\documentclass[preprint]{aastex}

\usepackage{longtable,lscape}
\usepackage{natbib}
\usepackage{graphics,graphicx}
\def\mdot{\.{M}}

\begin{document}

\title{Spectrum Synthesis Modeling of the X-ray Spectrum of GRO J1655-40
Taken During the 2005 Outburst.}

\author{T. R. Kallman\altaffilmark{1},M. A. Bautista\altaffilmark{2},
Stephane Goriely\altaffilmark{3},Claudio Mendoza\altaffilmark{4}, 
Jon M. Miller\altaffilmark{5},Patrick Palmeri\altaffilmark{6}, 
Pascal Quinet\altaffilmark{6},John Raymond\altaffilmark{7}}

\altaffiltext{1}{NASA Goddard Space Flight Center, Greenbelt, MD 20771}
\altaffiltext{2}{Department of Physics, Virginia Polytechnic and State University,  Blacksburg, VA 24061}
\altaffiltext{3}{Institut d'Astronomie et d'Astrophysique, Université Libre de Bruxelles, Campus de la Plaine CP 226, 1050 Brussels, Belgium}
\altaffiltext{4}{Centro de F\'{\i}sica, IVIC, Caracas 1020A, Venezuela}
\altaffiltext{5}{University of Michigan Department of Astronomy, 500 Church
Street, Ann Arbor, MI 48109}
\altaffiltext{6}{Astrophysique et Spectroscopie, Universit\'e de
Mons-Hainaut, B-7000 Mons, Belgium}
\altaffiltext{7}{Harvard-Smithsonian Center for Astrophysics, 60 Garden Street,
Cambridge, MA, 02138}

\begin{abstract}
The spectrum from the black hole X-ray transient GRO J1655-40.
obtained using the $Chandra$ High Energy Transmission Grating (HETG) 
in 2005 is notable as a laboratory for the study of warm absorbers, 
and for  the presence of many lines from odd-$Z$ elements between Na and Co (and Ti and Cr)  
not previously observed in X-rays.  We present synthetic spectral models which can be used to 
constrain these element abundances and other parameters describing the outflow 
from the warm absorber in this object.
We present results of fitting to the spectrum using various tools and 
techniques, including automated line fitting, phenomenological models, 
and photoionization modeling.  We show that the behavior of the curves of 
growth of lines from H-like and Li-like ions indicate that the lines are 
either saturated or affected by filling-in from scattered or a partially
covered continuum source.   We confirm the conclusion of previous work by \cite{Mill06}
and \cite{Mill08} which shows that the ionization conditions are not consistent with 
wind driving due to thermal expansion.  The spectrum provides the opportunity 
to measure abundances for several elements not typically observable 
in the X-ray band.  These show a pattern of enhancement for 
iron peak elements, and solar or sub-solar values for elements lighter 
than calcium.   Models show that this is consistent with enrichment by 
a core-collapse supernova.  We discuss the implications of 
these values for the evolutionary history of this system.
\end{abstract}

\section{Introduction.} 

X-ray spectra reveal that warm absorbers (absorption by partially 
ionized gas) are a common feature in compact objects.
Although warm absorbers were first detected in the 
spectra of active galaxies using the $Einstein$ satellite \citep{Halp84}, 
$Chandra$ and $XMM-Newton$ grating observations have shown that these occur in X-ray binaries 
as well, and information about the dynamics 
and other properties of this gas can have important implications 
for our understanding of accretion in these systems.
A notable example is the 900 ks observation using the $Chandra$  
High Energy Transmission Grating (HETG) 
of the Seyfert galaxy NGC 3783 \citep{Kasp02}, but this has been 
surpassed in signal-to-noise by the 2005 spectrum of the galactic black hole transient 
GRO J1655-40 \citep{Mill06}.   The statistical quality of this 
spectrum makes it one of the best available for testing and refinement of models
for warm absorber flows. 

GRO J1655-40 was discovered using BATSE onboard the Compton Gamma Ray Observatory \citep{Harm95}. 
Radio observations show apparently 
superluminal jets \citep{Hjel95}. Optical observations obtained 
during quiescence show  the companion star is a F3-5 
giant or sub-giant in a 2.62 day orbit around a 5-8 $M_\odot$ compact object 
\citep{Oros97}. 
Deep absorption dips have been observed \citep{Balu01} 
suggesting inclination $\sim 70^\circ$ \citep{Vand98}.  
GRO J1655-40 shows the highest-frequency 
quasi-periodic oscillations (QPOs) seen in a black hole \citep{Stro01}.  
Narrow X-ray absorption lines from highly ionized Fe were  detected 
by \citet{Yama01} and \citet{Ueda98}.
Similar features were detected from all the bright dipping low mass X-ray binaries (LMXBs) observed 
with XMM-Newton \citep{Diaz06}, and from the LMXB GX 13+1 \citep{Ueda01}. 
This suggests that ionized absorbers are a common feature of LMXBs,
although they may not be detected in objects viewed at lower inclination. 

The properties of the GRO J1655-40 warm absorber have been explored by 
\citet{Mill06} and \citet{Mill08}.  The spectrum resembles warm absorbers 
observed from other compact objects in that the lines are blueshifted, and that the 
inferred Doppler blueshift velocities are in the range 300 -- 1600 km/s.  The lines 
are identified primarily  as H- and He-like species of elements 
with nuclear charge $8\leq Z \leq 28$, and there is no clear evidence for absorption by 
low ionization material such as the iron M-shell UTA \citep{Beha03}.
The unprecedented signal-to-noise may account for the presence of lines from 
many trace elements previously not detected in X-rays, essentially all elements between Na and Co.
The detection of the 11.92 \AA\ line, arising from the $2p_{3/2}$ metastable level of Fe XXII,
implies a gas density $\geq 10^{14}$ cm$^{-3}$.  The spectrum is richer in features 
than another spectrum taken earlier in the same outburst, which showed only 
absorption by Fe XXVI L$\alpha$. The reason for this richness may be due to the 
higher column density of absorber  and to the softer continuum during this particular observation,
although this has not been tested quantitatively.

The observed line strengths can be used to infer the ionization balance, i.e. the 
ratio of abundances of H-like, He-like and lower ion stages for various elements.
Models for the ionization balance in the wind then yield the ionization 
parameter: the ratio of the ionizing flux to the gas density, and
the density is constrained by the Fe XXII line detection.
This, together with the observed luminosity and the observed outflow speed,
led \citet{Mill06} to conclude that the  radius 
where the outflow originates is too small to allow a wind driven by  thermal pressure.
That is, the likely ion thermal speeds in the gas are less than the escape velocity at the 
inferred radius.  On the other hand, the mass flux can be inferred from the line strengths, 
and the estimated rate exceeds  what is expected from outflows driven 
by radiation pressure.  This implies that the outflow must be driven by a different 
mechanism, such as magnetic forces, but this result has been controversial \citep{Netz06}.

The mass flux in the wind is important for our understanding of the mass and energy 
budget in accreting compact objects and it is clear that accurate models
for the ionization balance and synthetic spectrum are needed in order to reliably 
determine the properties of the wind.   It is the goal of this paper
to study this, and several issues which were not considered by \citet{Mill06, Mill08}:
(1) What element abundances are required to account for the lines from the 
many iron peak elements observed in the spectrum, and what might this tell us about 
the origin of the gas in the wind and accretion flow?  This is the only 
spectrum obtained from a warm absorber which clearly detects lines from odd-$Z$  elements with 
$Z \geq 10$ and from iron peak elements other than Fe and Ni.
The abundances are of interest since they
may contain clues to the evolutionary origin of the system; \citet{Isra99}
find evidence for enhanced O/H, Mg/H, Si/H, S/H, and relative to solar, but not Fe/H, and 
suggest that this may be due to enrichment by the supernova which produced the 
compact object.  (2) What is the possible influence of radiative transfer effects (line 
emission, partial covering, or finite energy resolution) on the inferred wind properties?
Analyses of X-ray warm absorber flows so far have not attempted to account in detail for these
processes (although hints to their importance in AGN come from the UV; \citet{Gabe05}).
They could systematically skew the derived column densities and ionization conditions.
It is straightforward, although time consuming, to construct models which will test 
these effects in various scenarios.
(3) Are there correlations between line shape or centroid and the ionization conditions where
that line is expected to dominate?   A flow with a predominantly  ordered velocity field 
and a central radiation source
is likely to show a gradient in ionization balance with position, and therefore with 
velocity.  This should be manifest as correlations between line width or offset with the 
ionization degree of its parent ion.  This effect is not found in  Seyfert galaxy  warm absorbers
\citep{Kasp02}, but examination of the GRO J1655-40 spectrum shows differences among the line 
profile shapes.  After careful attention to item (2) above, we can test this quantitatively.

In the remainder of this paper we present our model fits and interpretation of the 
$Chandra$ HETG spectrum of GRO J1655-40, attempting to address the above questions.  
This includes various fitting techniques, 
testing of radiative transfer effects, and discussion of the constraints on element abundances.
Finally, we discuss the implications of these results for the dynamics of the 
outflow, and possible evolutionary history of the source.

\section{Fitting:  Notch Models}  

The fits in this paper were performed using the same extraction of the 
HETG data as was used  by \cite{Mill08}.
$Chandra$ observed GRO J1655-40 for a total of 44.6 ks  starting at 
12:41:44 UT on 2005 April 1. 
Data was taken from the ACIS-S array dispersed by the High Energy Transmission 
Grating Spectrometer (HETGS). Continuous clocking mode was used in 
order to prevent photon pile-up.  As described by \cite{Mill08}, a gray 
filter was used in order to reduce the zero order counting rate.  Data was processed
using the CIAO reduction package, version 3.2.2.  The event file was filtered
to accept only standard event grades, good-time intervals, and to reject bad pixels.
Streaking was removed using the ``destreak'' tool, and spectra were 
extracted using ``tg\_resolve\_events'' and ``tg\_extract''.  Arfs were 
produced using the ``fullgarf'' tool along with canned rmfs.  First order 
HEG spectra and arfs and first-order MEG spectra and arfs were added using the 
``add\_grating\_spectra'' tool.  In this paper we do not fit to the RXTE data 
obtained simultaneously with the $Chandra$ HETG observation, but we do employ the continuum 
shape derived from the \cite{Mill08} fits which include the RXTE data.

Our procedure for analyzing the spectrum of  GRO J1655-40 consists of three separate 
parts.  First, we fit the spectrum using {\sc xspec}  together with simple analytic models 
describing the continuum and the lines.  This consists of a continuum shape which is 
the same as that used by \cite{Mill08}:  a power law plus a disk blackbody and cold absorption.  
Based on the fits by \cite{Mill08}, we use a power law index of 3.54  and a 
disk blackbody temperature with temperature  1.34 keV.  We allow the normalizations of 
the components to vary, and find a best fit normalization of  515.7$\pm$1.5 for the disk 
blackbody and $\leq$0.15 for the power law.
The flux is 2.02 $\times 10^{-8}$ ergs/cm$^2$/s 2-10 keV.  We ignore photons with wavelength greater than 
15 \AA, since there are few counts in this range and its inclusion has negligible effect
on the results at shorter wavelength.  We refer to this as model 1. 
The  fitting parameters and $\chi^2$ for this 
and the other fits discussed in this paper are  presented  in Table \ref{chi2table}.  
The various physical parameters for all the models we test are listed
in the first column.  Model 1, the best fit to the continuum only, gives $\chi^2/\nu=120618/8189$. 

\clearpage

\begin{deluxetable}{llllllll} 
\tabletypesize{\scriptsize}
\tablecolumns{8} 
\tablewidth{0pc} 
\tablecaption{\label{chi2table} Model results} 
\tablehead{ 
\colhead{fitting quantity}&\colhead{units}&\colhead{Model 1}&\colhead{Model 2}&\colhead{Model 3}&\colhead{Model 4}&\colhead{Model 5}&\colhead{Model 6}}
\startdata 
log($\xi_1$)   &erg cm s$^{-1}$&NA             &NA             &NA        &4.$^+_-0.1$         &3.8$^+_-0.1$         &4.0$^+_-0.1$        \\
log(N$_1$)     &cm$^{-2}$      &NA             &NA             &NA        &23.8$^+_-0.02$       &22.64$^+_-0.02$      &24.0$^+_-0.02$      \\
v$_{turb}$     &km s$^{-1}$    &NA             &NA             &NA        &50\tablenotemark{*}  &200\tablenotemark{*} &200\tablenotemark{*}\\
v$_{off}$      &km s$^{-1}$    &NA             &NA             &NA        &-375                 &-375                 &-375                \\
log($\xi_2$)   &erg cm s$^{-1}$&NA             &NA             &NA        &NA                   &4.6$^+_-$0.1          &NA                  \\
log(N$_2$)     &cm$^{-2}$      &NA             &NA             &NA        &NA                   &23.90                &NA                  \\
EW$_{fe26}$    &keV            &NA             &NA             &NA        &0.03                 &NA                   &0.03                \\
v$_{off,fe26}$ &km s$^{-1}$    &NA             &NA             &NA        &1451.               &NA                   &1451.              \\
v$_{turb,fe26}$&km s$^{-1}$    &NA             &NA             &NA        &0.03$_{-0.02}^{+0.01}$&NA                   &0.03$_{-0.02}^{+0.01}$\\
NH             &cm$^{-2}$      &21.49$^+_-0.01$&21.49$^+_-0.01$&NA        &21.49$^+_-0.01$      &21.48$^+_-0.01$      &21.48$^+_-0.01$     \\
$\gamma$       &               &3.54           &3.54           &3.54      &3.54\tablenotemark{*}&3.54\tablenotemark{*}&3.54\tablenotemark{*}\\
pl norm        &               &$\leq$0.01     &$\leq$0.01     &$\leq$0.15&$\leq$0.01           &$\leq$0.01         &$\leq$0.01             \\
diskbb norm    &               &513$^+_-$1   &538\tablenotemark{*}&534$+_-$1 &548$_{-1.5}^{+0.5}$&533$_{-1}^{+1.5}$&533$_{-1}^{+1.5}$ \\
$kT_{diskbb}$  &keV            &1.35           &1.35\tablenotemark{*}&1.35\tablenotemark{*}&1.35\tablenotemark{*}&1.35\tablenotemark{*}&1.35\tablenotemark{*}\\
$f_{scatt}$      &               &0              &0              &0         &0                    &0                  &0.37                \\
$\chi^2$       &               &120618         &18474          &20671     &33591                &34283              &24561               \\
dof            &               &8189           &8108           &8118      &8187                 &8189               &8184                \\
\enddata 
\tablenotetext{*}{These parameters were fixed during the fitting.}
\end{deluxetable} 

\clearpage

We then add the effects of absorption lines  by using negative Gaussian models for 
the lines, as was done by \cite{Mill08}.  The list of lines 
and their properties is the primary topic of the rest of this paper.  Initially we use the 
same lines as given in Table 1 of \cite{Mill08}.  There are 71 lines with well-determined wavelengths
and identifications in this list.  We allow the values of the wavelengths, widths, and line normalizations to be 
determined by the {\sc xspec} minimization.  This results in detections of essentially all the lines 
with parameters consistent with those of \cite{Mill08}.  In addition, we propose identifications for some of the 
lines which were not identified by those authors.   We point out that this procedure differs from 
those authors in that we fit to a single analytic global continuum, while they fit to a piece-wise
powerlaw.  The procedure used here is chosen for comparison with the fits
to photoionization models in the next section.  The best-fit to the continuum plus Gaussians gives
$\chi^2/\nu=$18474/8108.  We refer to this as model 2 (cf. Table \ref{chi2table}).  This fit is marginally 
acceptable based on standard arguments derived from $\chi^2$ statistics, and it 
is the best of the models presented in this paper.  This serves to illustrate the level 
of systematic errors, or errors in our continuum, 
which provide an effective limit to our ability to fit the spectrum.
Model 2, along with other models discussed below, is plotted in figures \ref{fita} -- \ref{fitn}.  
In this figure the vertical axis is the ratio of the model and data to the continuum-only model, model 1.
Successive models are offset by unity from each other. These figures 
show qualitatively the agreement between the model and the data which we achieve.  

The second approach to line fitting is to construct an automated line fitting 
program.   This takes the same continuum employed in the continuum-only fit, and then 
experiments with random placements of lines throughout the 1-15 \AA\ range.
These experiments begin with an initial  wavelength which is chosen randomly within this range
(but excluding a region within 2 Doppler widths of previously found lines), 
and then the line wavelength, width, and optical depth
are varied in an attempt to find a best fit (the wavelength is restricted to a region 
near the initial wavelength in this procedure).  The fit is considered valid if the $\chi^2$ improves 
by 3 with the inclusion of the line.  Line widths are limited to be less than 8 Doppler widths when compared
with a turbulent velocity of 100 km/s.  
The lines are assumed to be Doppler broadened only, and 
the absorption is a true Gaussian absorption profile.  This is in contrast to the {\sc xspec}
Gaussian line model, which treats absorption as a negative emission.   As we will show, many of our line fits
result in large optical depths, and in this case a Voigt profile fit would be preferable.
The pure Doppler profile likely results in an over-estimate of the line optical depth, since it cannot
produce as strong absorption line wings as would a Voigt profile.  However, the Voigt profile damping 
parameter value  depends on the line identification, and cannot be conveniently used as a fitting parameter.
In our synthetic spectral modeling, in the following section, we fit to Voigt profiles using 
accurate atomic rates for the damping parameters.
For convenience we refer to this as the notch model, although it assumes Gaussian absorption lines 
rather than true notches.   

This procedure yields a total of 
292 lines after a total of 20000 attempts at placing random lines.  
We neglect lines with equivalent widths less than 3.4 $\times 10^{-4}$ eV.  
We do not consider this to be necessarily 
an exhaustive list of lines in the GRO J1655-40 spectrum, but likely contains the strongest or least 
ambiguous features, and it is objective.  This procedure detects all the features in the 
\cite{Mill08} table, plus many others which are weaker or blended.  Some of these are undoubtedly artifacts 
of the shortcomings of our continuum model, and others may coincide with regions where bound-free continuum 
opacity is important.  These are discussed in turn in the following. This fit yields $\chi^2/\nu=20671/8118$
using the continuum from the {\sc xspec} Gaussian fit. We refer to this as model 3 (cf. Table \ref{chi2table}). 
The slightly worse $\chi^2$ value for this model when compared with model 2 is likely due to the 
limitations of our automated fitting procedure, particularly when two strong lines are close  
together or partially overlapping.  In addition to deriving wavelengths, line center optical depths, and 
widths ($\sigma$), we also derive errors on these quantities based on the $\Delta\chi^2$=3 criterion 
of \cite{Cash79}, and we calculate the line equivalent widths by integrating numerically over the best-fit model
profile.

\subsection{Line Identifications}

\begin{figure}[t!]
\includegraphics*[angle=90, scale=0.45]{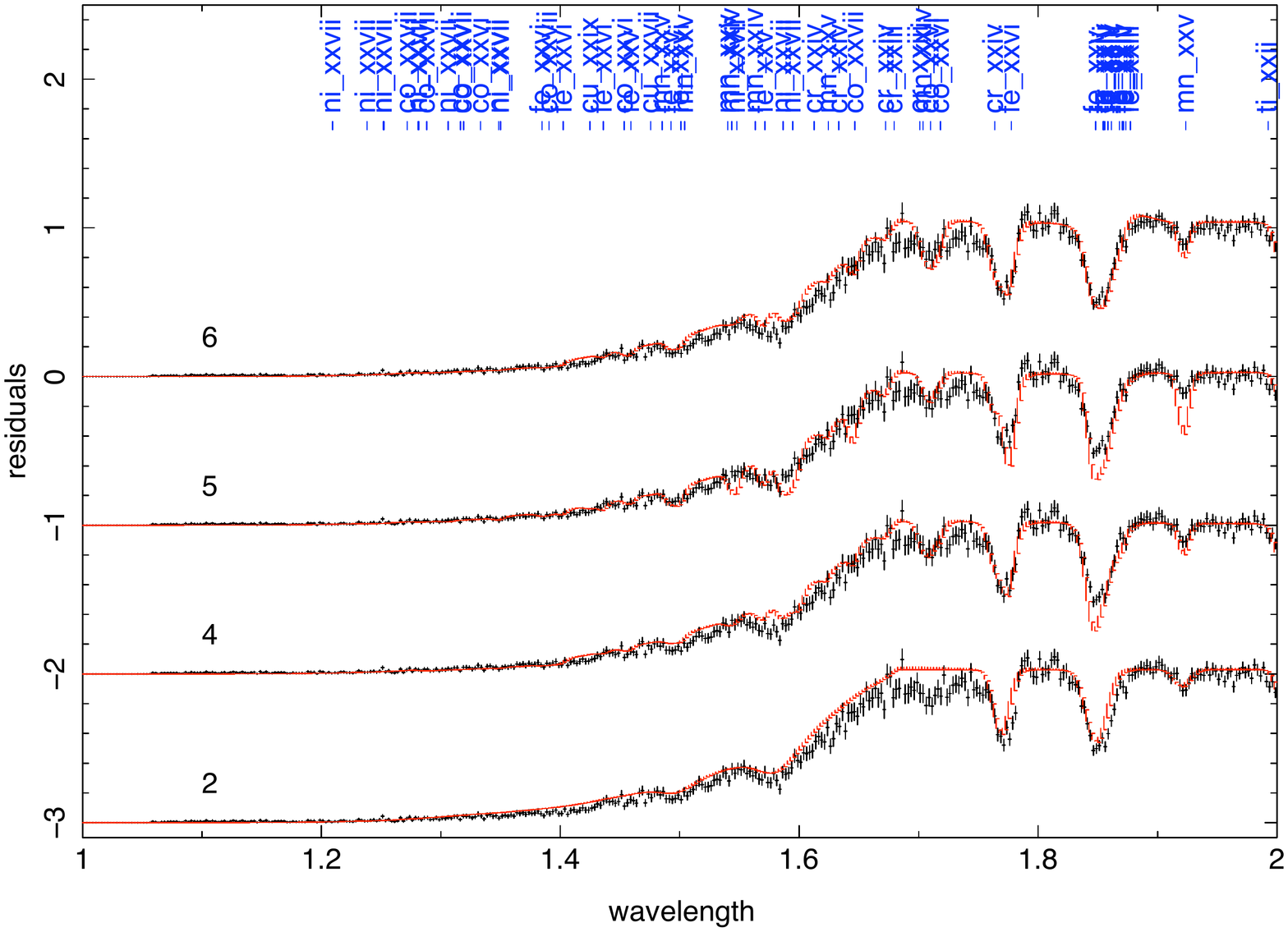}
\caption{\label{fita}Spectrum $\lambda\lambda$ 1 -- 2 \AA.  Spectrum is shown as ratio relative
to pure power law model (model 1).  Various models 2,4,5,6 are labeled. 
The vertical axis is the ratio of the model and data to the continuum-only model, model 1.
Successive models are offset by unity from each other.}
\end{figure}

\begin{figure}[b!]
\includegraphics*[angle=90, scale=0.45]{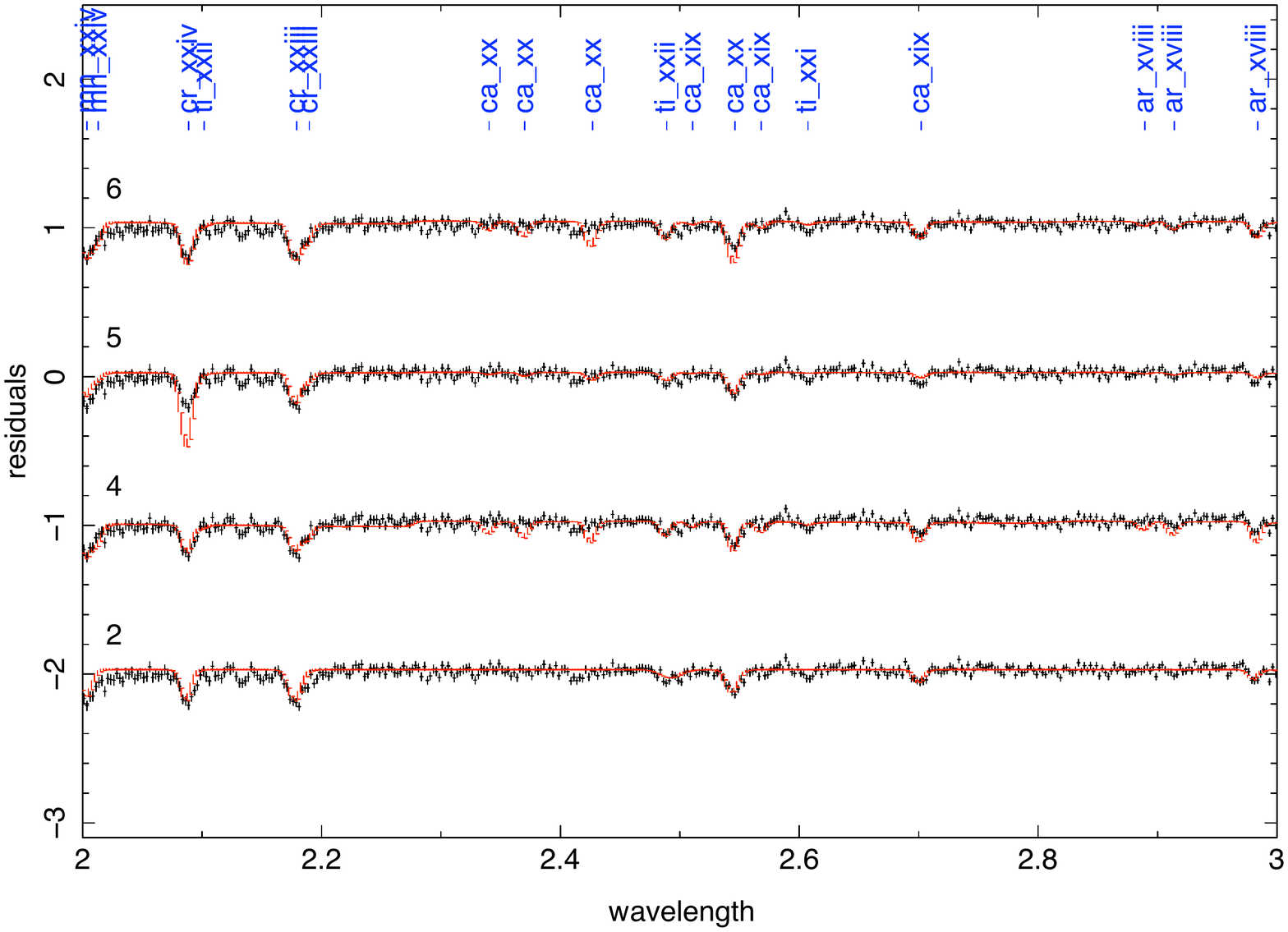}
\caption{\label{fitb}Spectrum $\lambda\lambda$  2 -- 3 \AA. Spectrum is shown as ratio relative
to pure power law model (model 1).  Various models 2,4,5,6 are labeled.
The vertical axis is the ratio of the model and data to the continuum-only model, model 1.
Successive models are offset by unity from each other.}
\end{figure}

\begin{figure}[t!]
\includegraphics*[angle=90, scale=0.45]{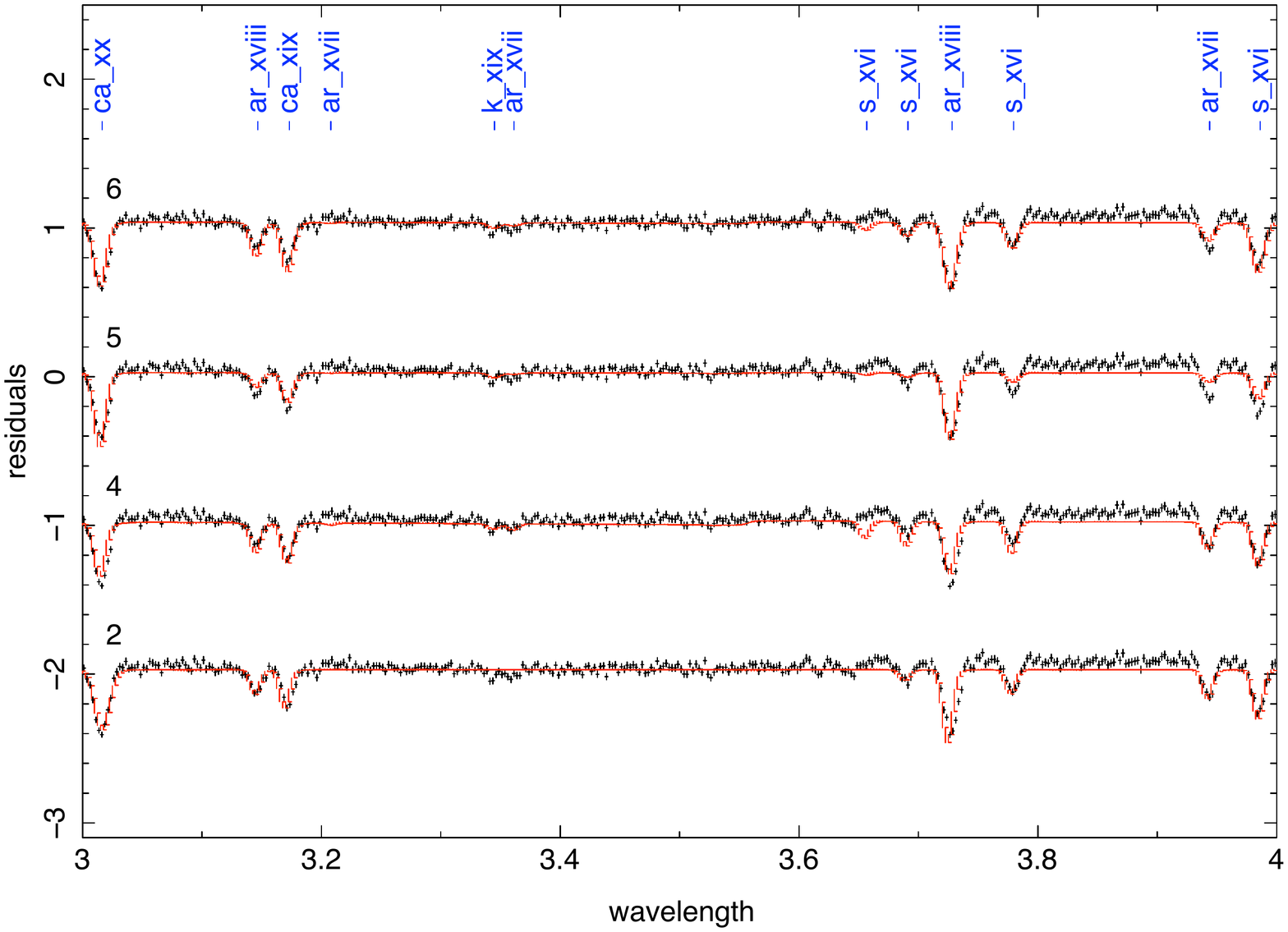}
\caption{\label{fitc}Spectrum $\lambda\lambda$ 3 -- 4 \AA. Spectrum is shown as ratio relative
to pure power law model (model 1).  Various models 2,4,5,6 are labeled.
The vertical axis is the ratio of the model and data to the continuum-only model, model 1.
Successive models are offset by unity from each other.}
\end{figure}

\begin{figure}[b!]
\includegraphics*[angle=90, scale=0.45]{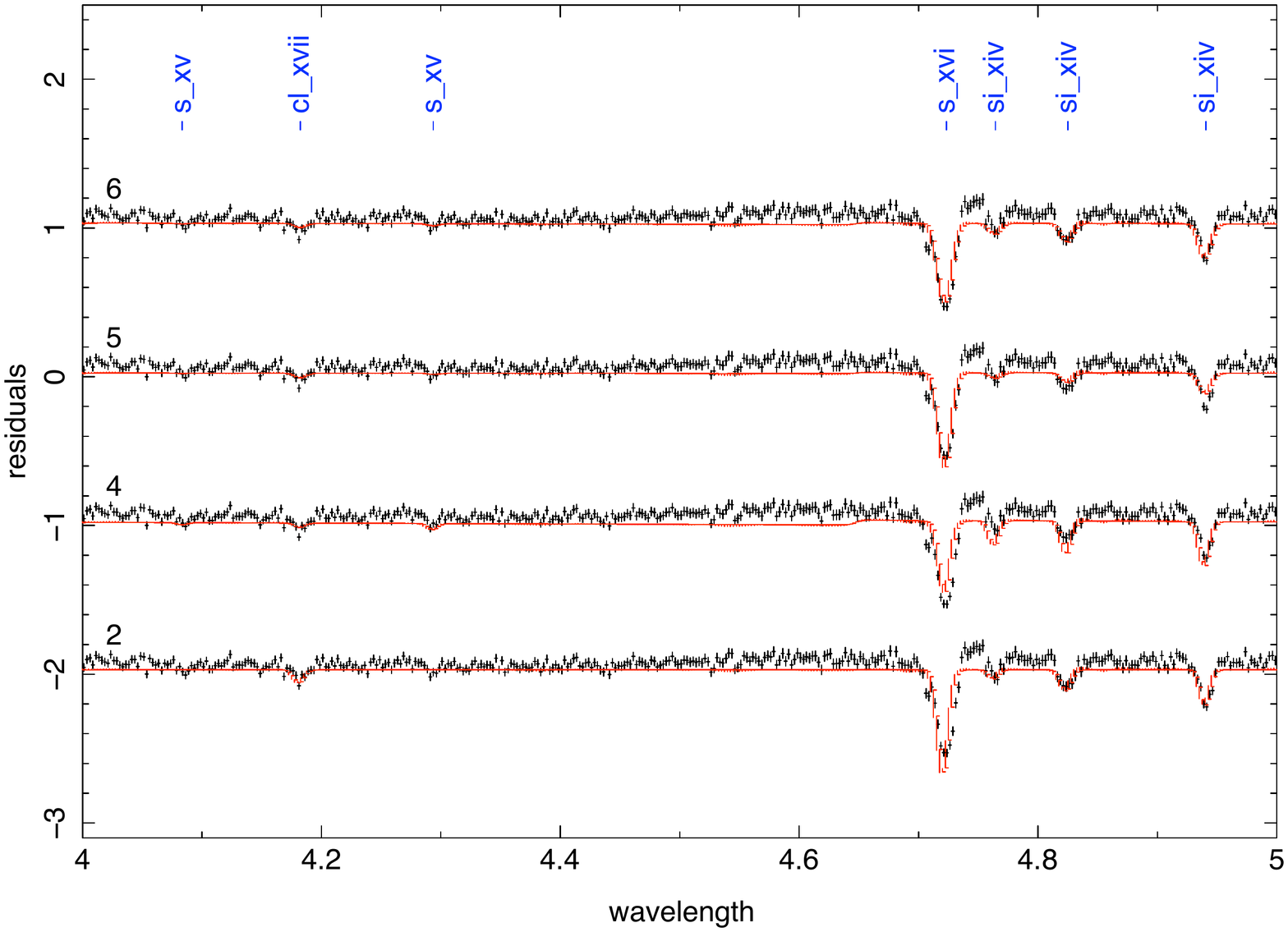}
\caption{\label{fitd}Spectrum $\lambda\lambda$ 4 -- 5 \AA. Spectrum is shown as ratio relative
to pure power law model (model 1).  Various models 2,4,5,6 are labeled.
The vertical axis is the ratio of the model and data to the continuum-only model, model 1.
Successive models are offset by unity from each other.}
\end{figure}

\begin{figure}[t!]
\includegraphics*[angle=90, scale=0.45]{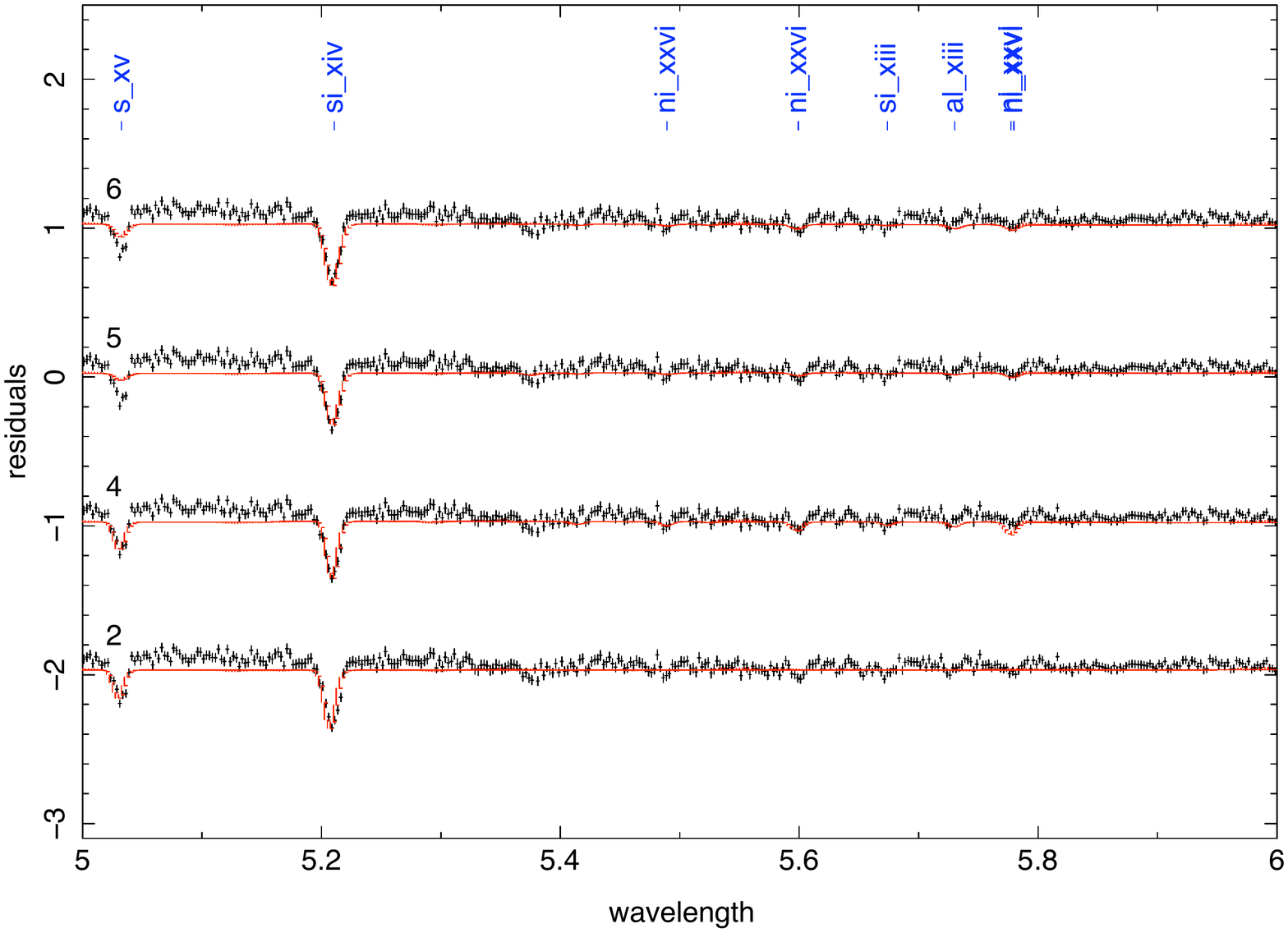}
\caption{\label{fite}Spectrum $\lambda\lambda$ 5 -- 6 \AA. Spectrum is shown as ratio relative
to pure power law model (model 1).  Various models 2,4,5,6 are labeled.
The vertical axis is the ratio of the model and data to the continuum-only model, model 1.
Successive models are offset by unity from each other.}
\end{figure}

\begin{figure}[b!]
\includegraphics*[angle=90, scale=0.45]{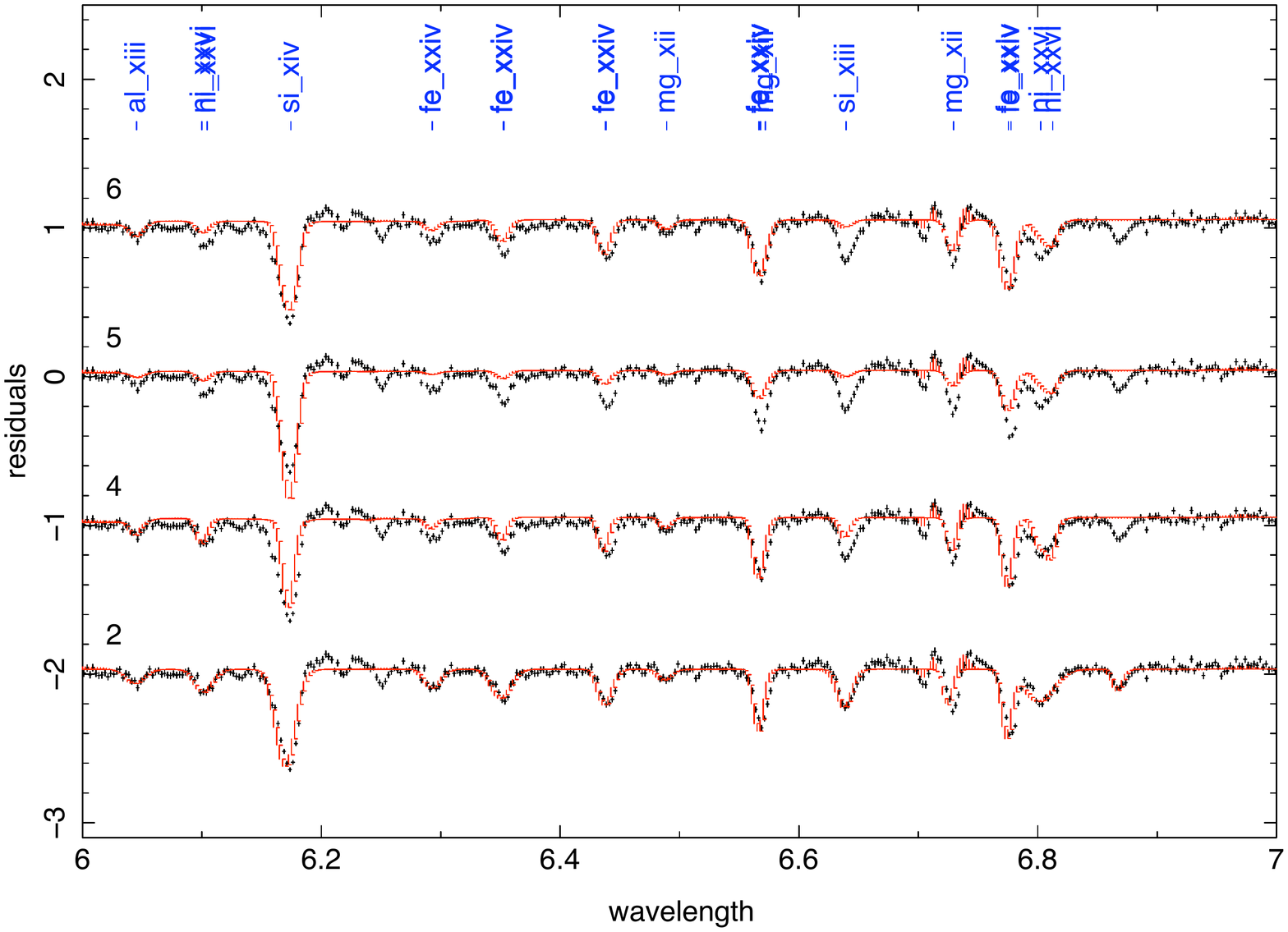}
\caption{\label{fitf}Spectrum $\lambda\lambda$ 6 -- 7 \AA. Spectrum is shown as ratio relative
to pure power law model (model 1).  Various models 2,4,5,6 are labeled.
The vertical axis is the ratio of the model and data to the continuum-only model, model 1.
Successive models are offset by unity from each other.}
\end{figure}

\begin{figure}[t!]
\includegraphics*[angle=90, scale=0.45]{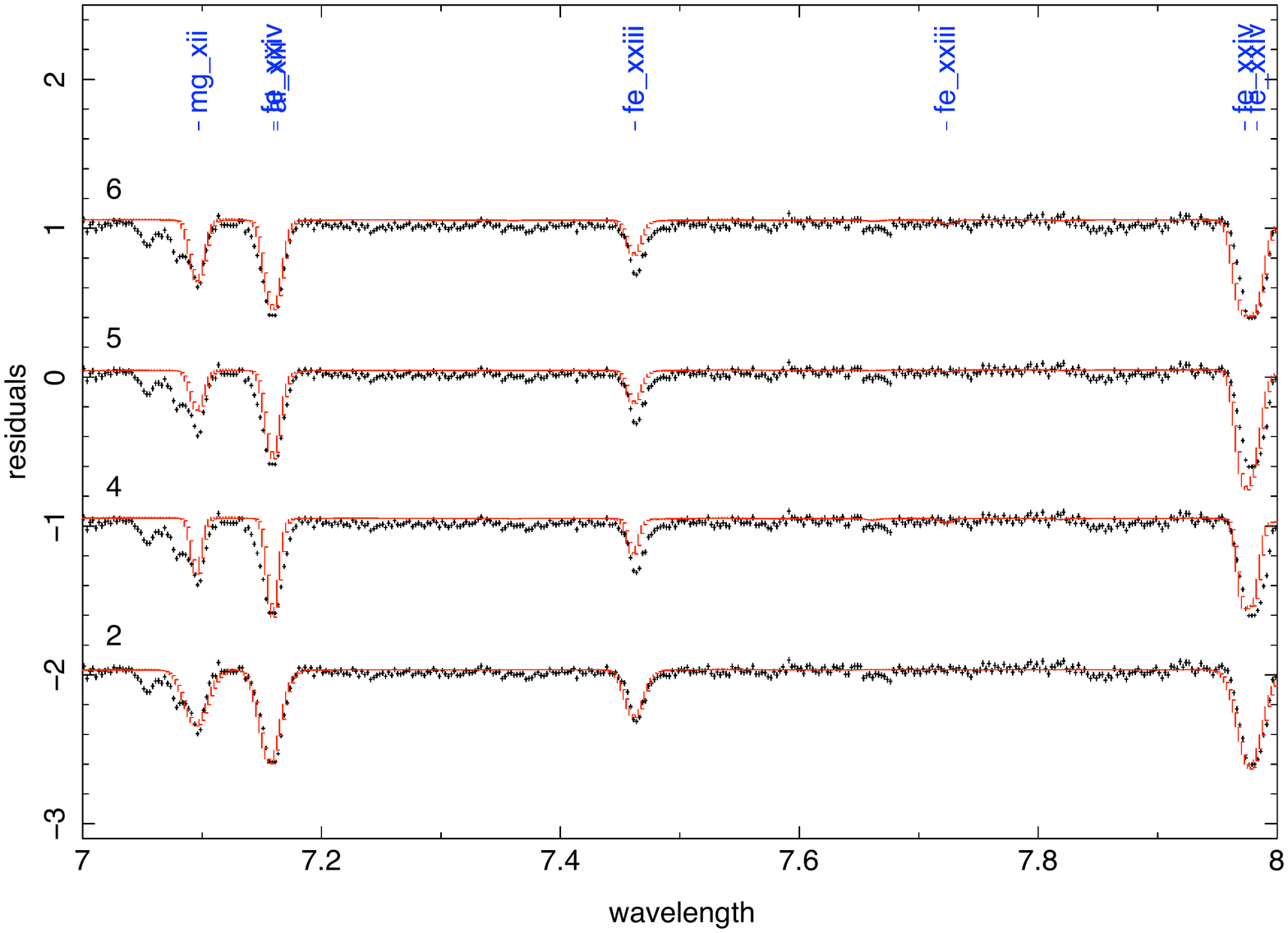}
\caption{\label{fitg}Spectrum $\lambda\lambda$ 7 -- 8 \AA. Spectrum is shown as ratio relative
to pure power law model (model 1).  Various models 2,4,5,6 are labeled.
The vertical axis is the ratio of the model and data to the continuum-only model, model 1.
Successive models are offset by unity from each other.}
\end{figure}

\begin{figure}[b!]
\includegraphics*[angle=90, scale=0.45]{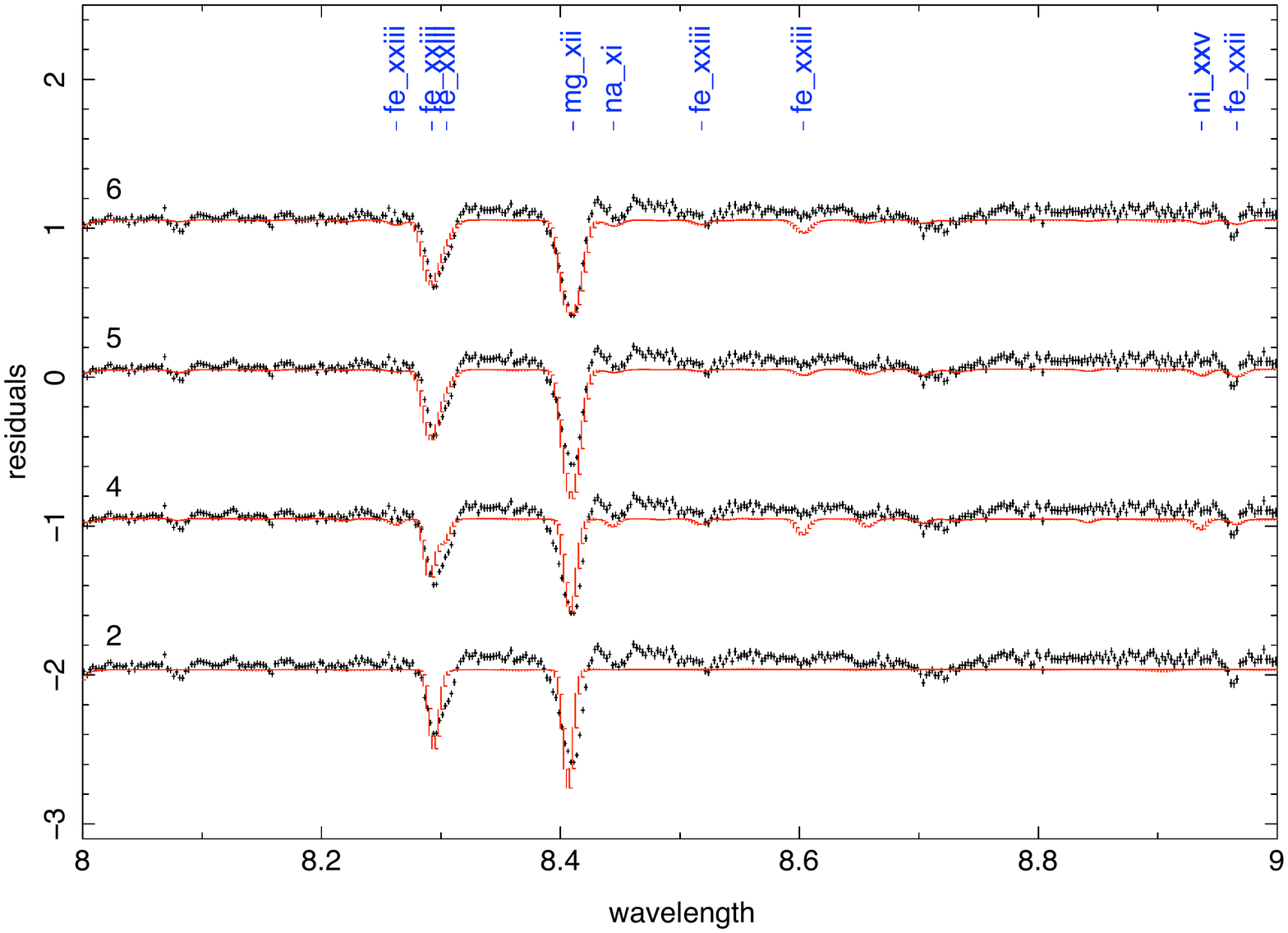}
\caption{\label{fith}Spectrum $\lambda\lambda$ 8 -- 9 \AA. Spectrum is shown as ratio relative
to pure power law model (model 1).  Various models 2,4,5,6 are labeled.
The vertical axis is the ratio of the model and data to the continuum-only model, model 1.
Successive models are offset by unity from each other.}
\end{figure}

\begin{figure}[t!]
\includegraphics*[angle=90, scale=0.45]{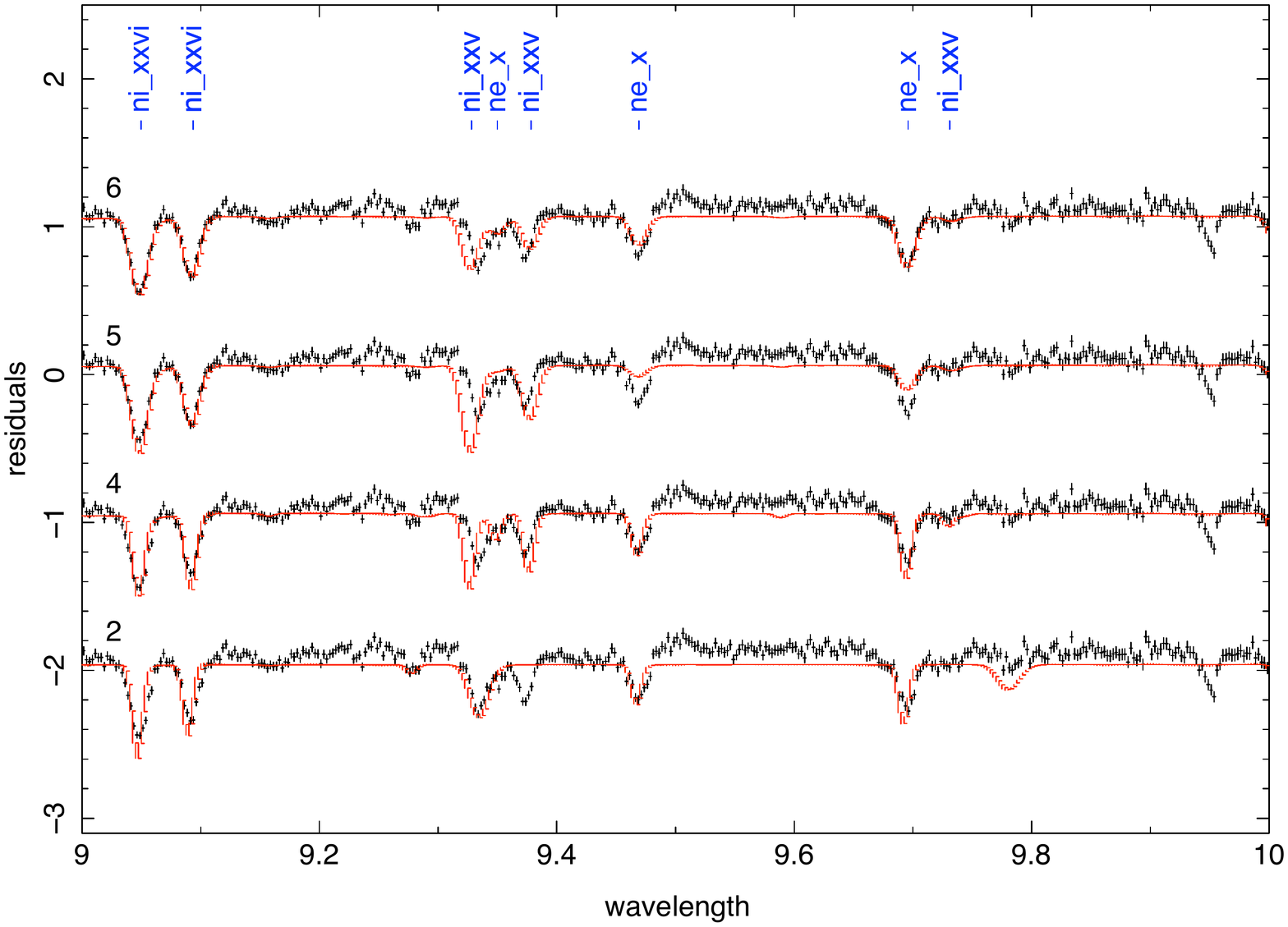}
\caption{\label{fiti}Spectrum $\lambda\lambda$ 9 -- 10 \AA. Spectrum is shown as ratio relative
to pure power law model (model 1).  Various models 2,4,5,6 are labeled.
The vertical axis is the ratio of the model and data to the continuum-only model, model 1.
Successive models are offset by unity from each other.}
\end{figure}

\begin{figure}[b!]
\includegraphics*[angle=90, scale=0.45]{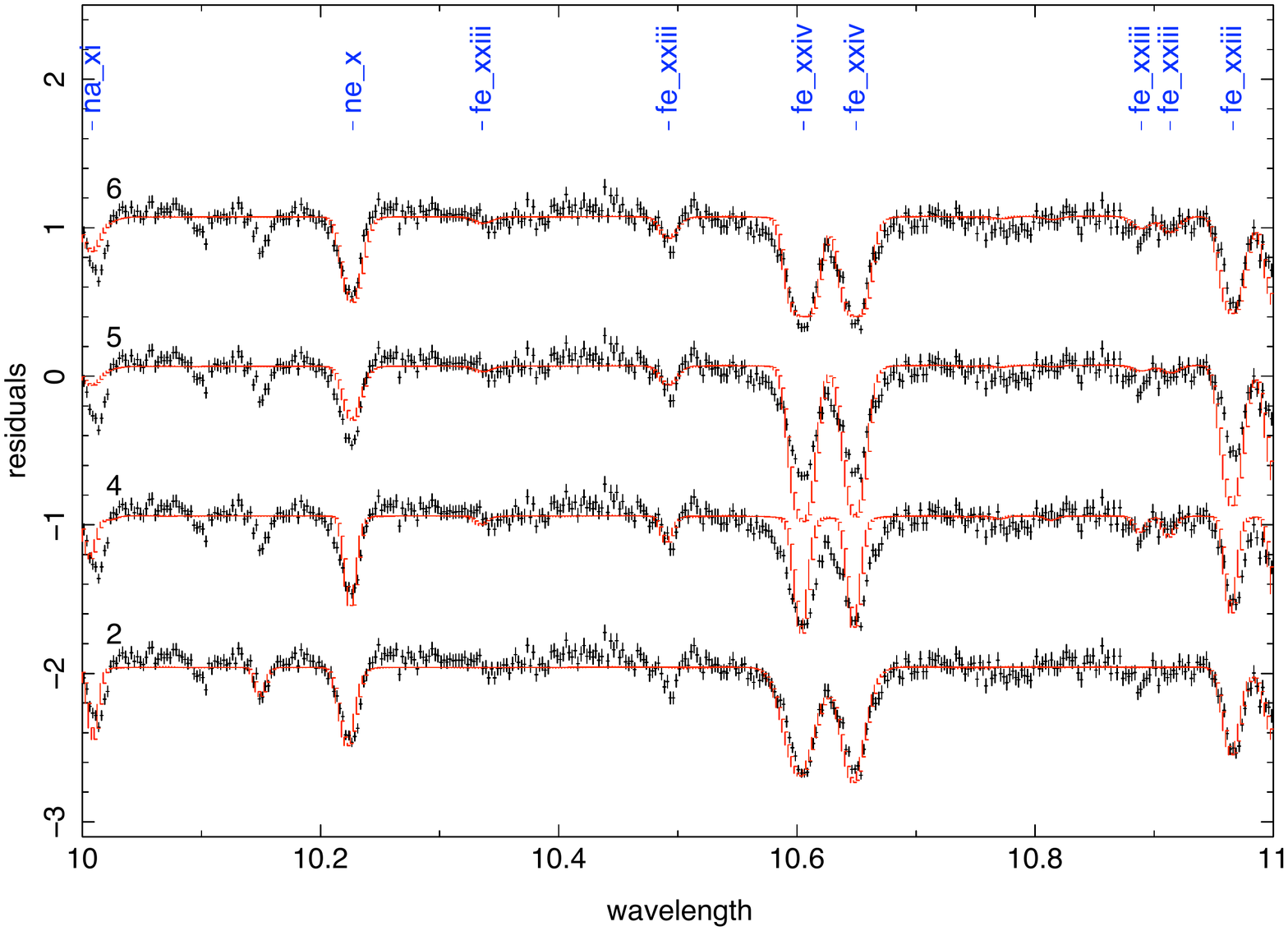}
\caption{\label{fitj}Spectrum $\lambda\lambda$ 10 -- 11 \AA. Spectrum is shown as ratio relative
to pure power law model (model 1).  Various models 2,4,5,6 are labeled.
The vertical axis is the ratio of the model and data to the continuum-only model, model 1.
Successive models are offset by unity from each other.}
\end{figure}

\begin{figure}[b!]
\includegraphics*[angle=90, scale=0.45]{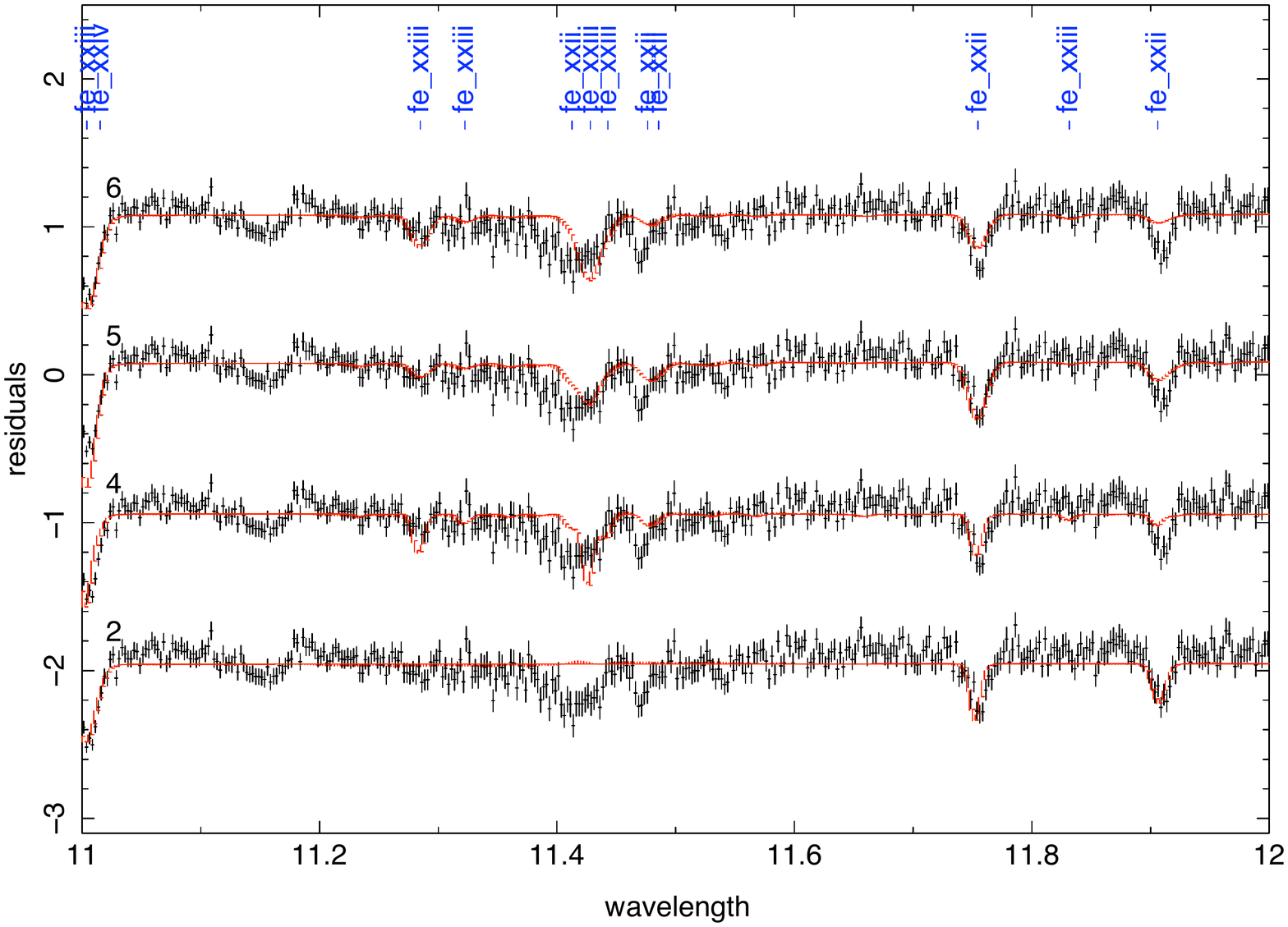}
\caption{\label{fitk}Spectrum $\lambda\lambda$ 11 -- 12 \AA. Spectrum is shown as ratio relative
to pure power law model (model 1).  Various models 2,4,5,6 are labeled.
The vertical axis is the ratio of the model and data to the continuum-only model, model 1.
Successive models are offset by unity from each other.}
\end{figure}

\begin{figure}[b!]
\includegraphics*[angle=0, scale=0.45]{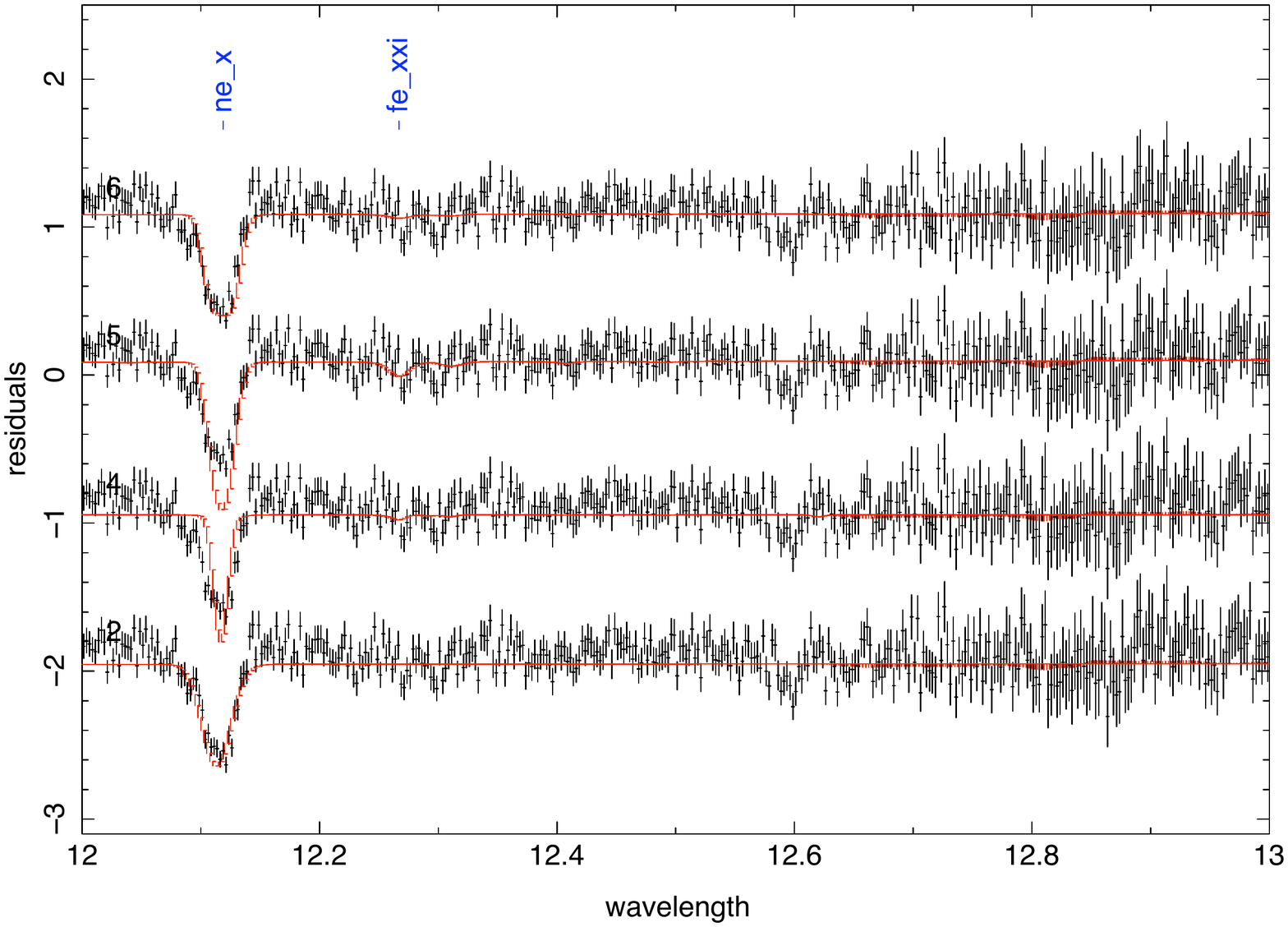}
\caption{\label{fitl}Spectrum $\lambda\lambda$ 12 -- 13 \AA. Spectrum is shown as ratio relative
to pure power law model (model 1).  Various models 2,4,5,6 are labeled.
The vertical axis is the ratio of the model and data to the continuum-only model, model 1.
Successive models are offset by unity from each other.}
\end{figure}

\begin{figure}[b!]
\includegraphics*[angle=0, scale=0.45]{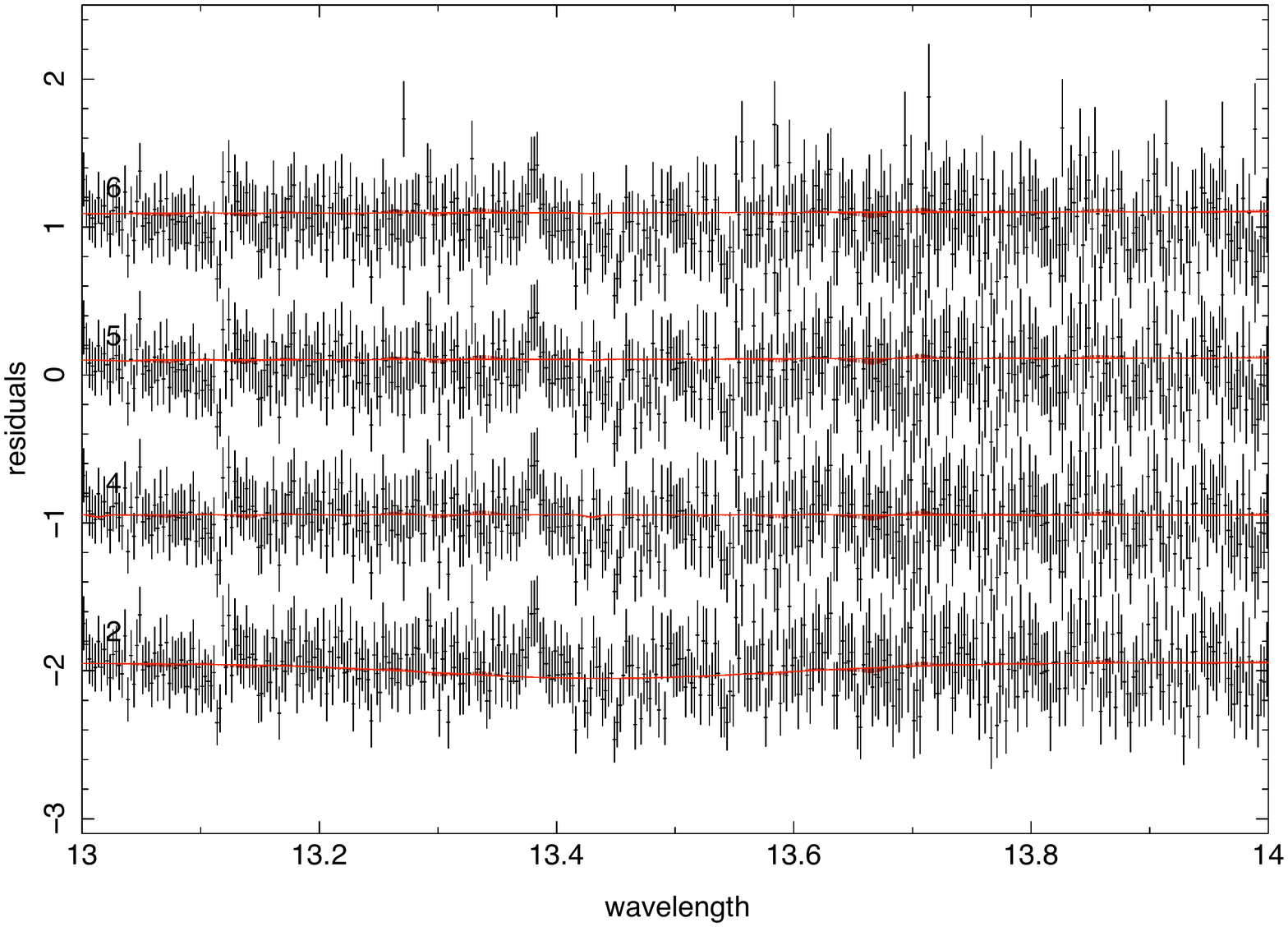}
\caption{\label{fitm}Spectrum $\lambda\lambda$ 13 -- 14 \AA. Spectrum is shown as ratio relative
to pure power law model (model 1).  Various models 2,4,5,6 are labeled.
The vertical axis is the ratio of the model and data to the continuum-only model, model 1.
Successive models are offset by unity from each other.}
\end{figure}

\begin{figure}[b!]
\includegraphics*[angle=0, scale=0.45]{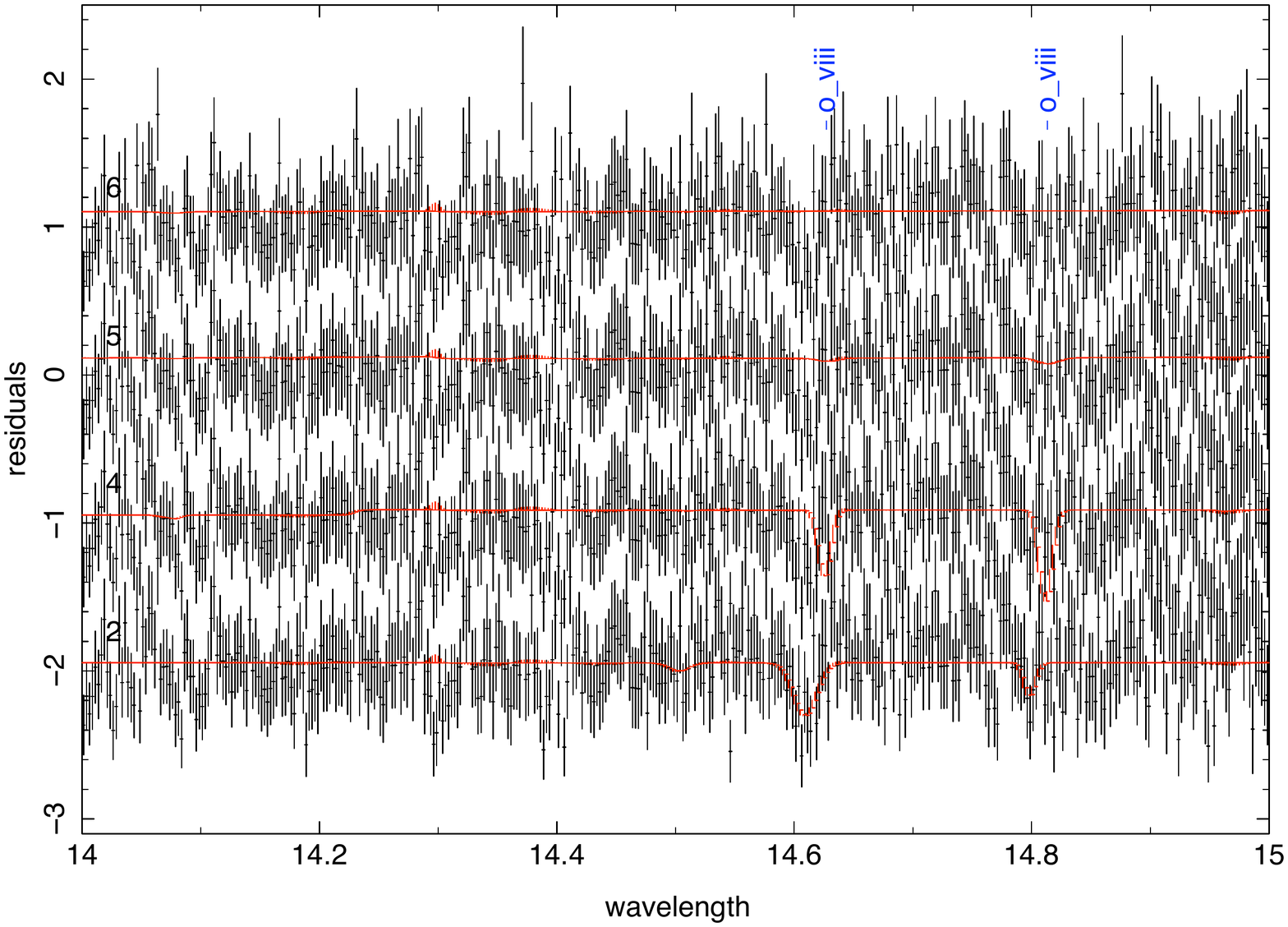}
\caption{\label{fitn}Spectrum $\lambda\lambda$ 14 -- 15 \AA. Spectrum is shown as ratio relative
to pure power law model (model 1).  Various models 2,4,5,6 are labeled.
The vertical axis is the ratio of the model and data to the continuum-only model, model 1.
Successive models are offset by unity from each other.}
\end{figure}

The list of lines we detect is given in 
Table \ref{linelist}.  This includes the wavelength, width  ($\sigma$) and
equivalent width  derived from the automated fitting leading to model 3.  We also provide identifications for 
the lines.  This is done by searching the linelist in the {\sc xstar} \citep{Kall01, Baut01} database and choosing the line
which has the greatest ratio of optical depth  to Doppler shift within a Doppler shift of 
$\leq^+_-$1500 km s$^{-1}$.  The optical depth used in this determination is calculated using 
the {\sc warmabs} analytic model, which is an implementation of  {\sc xstar} use as an analytic 
model within the {\sc xspec} X-ray spectral fitting package,  for the conditions
in model 4 described in the following section.   
We have updated both {\sc xstar} and {\sc warmabs} to include all the previously neglected elements
with $Z \leq 30$; this is described in the Appendix.
Note that the criterion for line identification is used to choose among the {\sc xstar} lines which fall within 
Doppler shifts of $^+_-$1500 km s$^{-1}$, but does not prevent a line from being included if there is 
no {\sc xstar} line within that
interval.  If there is an ID, then the parent ion, {\sc xstar} wavelength, and upper and lower level designations are 
also given in Table \ref{linelist}.  The identification, together with the optical depth derived from the model 3  fits
allows the equivalent hydrogen column density of the absorbing ion to be derived.  These 
are discussed in more detail in the following subsection, and are given in the table.  
The elemental abundances used in the calculation of equivalent hydrogen abundances are those of 
\citep{Grev96, Alle73}.
Figures \ref{fita} -- \ref{fitn} show the count spectrum observed by the HETG (relative to the 
continuum only model 1) together with the lines
and identifications from Table \ref{linelist}.  Also shown in these figures is our
fit to model 2 and to models 4, 5 and 6  which will be discussed in more detail below.

\clearpage

\clearpage

\clearpage

\clearpage

\clearpage

\clearpage

\clearpage

\clearpage

\begin{deluxetable}{rrrrrrrrrrrr} 
\tabletypesize{\scriptsize}
\tablecolumns{12} 
\tablewidth{0pc} 
\tablecaption{ \label{linelist} Line Properties} 
\tablehead{ 
\colhead{$\lambda_O$\tablenotemark{a}}& \colhead{index}&\colhead{$\lambda$\tablenotemark{a,b}}&\colhead{ion}&\colhead{v$_0$\tablenotemark{c}}&\colhead{ew\tablenotemark{d}}&\colhead{width\tablenotemark{c}}&\colhead{$\tau$\tablenotemark{e}}&\colhead{f\tablenotemark{f}}&\colhead{lower level}&\colhead{upper level}&\colhead{ion column\tablenotemark{g}}}
\startdata 
  1.544&    17264&  1.542&Mn XXV   & -357.5 & 0.0006&459.3& 0.71$^{+0.56}_{-0.41}$&  0.029 &1s.2S                &4p.2P$^o$             & 2.9$^{+2.3}_{-1.7}\times 10^{23}$\\
  1.567&   128239&  1.573&Fe XXV   & 1187.0 & 0.0029&772.1& 2.62$^{+1.44}_{-0.88}$&  0.152 &1s2.1S               &1s.3p.1P*            & 4.4$^{+2.4}_{-1.5}\times 10^{21}$\\
  1.582&   132917&  1.589&Ni XXVII & 1232.6 & 0.0084&506.7&23.06$^{+9.99}_{-8.75}$&  0.684 &1s2.1S               &1s.2p.1P             & 1.0$^{+0.7}_{-0.4}\times 10^{23}$\\
  1.601&         &       & no ID    &        & 0.0022&707.2& 1.92$^{+0.91}_{-0.62}$& & & &\\
  1.613&    17063&  1.615&Cr XXIV  &  297.6 & 0.0004&234.9& 1.18$^{+0.80}_{-0.60}$&  0.008 &1s.2S                &6p.2P$^o$             & 1.4$^{+0.9}_{-0.7}\times 10^{24}$\\
  1.626&    17263&  1.626&Mn XXV   &   77.5 & 0.0010&412.7& 1.22$^{+0.51}_{-0.40}$&  0.080 &1s.2S                &3p.2P$^o$             & 1.7$^{+0.7}_{-0.6}\times 10^{23}$\\
  1.646&   128543&  1.649&Co XXVII &  484.8 & 0.0006&407.0& 0.66$^{+0.34}_{-0.28}$&  0.422 &1s.2S                &2p.2P$^o$             & 7.4$^{+3.8}_{-3.1}\times 10^{22}$\\
  1.675&    17061&  1.674&Cr XXIV  & -107.5 & 0.0004&204.0& 1.22$^{+0.79}_{-0.63}$&  0.029 &1s.2S                &4p.2P$^o$             & 3.7$^{+2.4}_{-1.9}\times 10^{23}$\\
  1.693&         &       & no ID    &        & 0.0007&458.1& 0.58$^{+0.27}_{-0.23}$& & & &\\
  1.709&   128440&  1.712&Co XXVI  &  544.2 & 0.0015&539.4& 1.24$^{+0.40}_{-0.31}$&  0.693 &1s2.1S               &1s.2p.1P             & 8.2$^{+2.6}_{-2.1}\times 10^{22}$\\
  1.726&         &       & no ID    &        & 0.0012&701.0& 0.65$^{+0.24}_{-0.22}$& & & &\\
  1.742&         &       & no ID    &        & 0.0003&225.4& 0.71$^{+0.49}_{-0.36}$& & & &\\
  1.772&   128358&  1.780&Fe XXVI  & 1325.6 & 0.0053&453.6&10.56$^{+3.68}_{-2.72}$&  0.408 &1s.2S                &2p.2P$^o$             & 5.9$^{+2.1}_{-1.5}\times 10^{21}$\\
  1.838&         &       & no ID    &        & 0.0007&195.2& 2.35$^{+0.75}_{-0.55}$& & & &\\
  1.851&   128293&  1.850&Fe XXV   &  -81.0 & 0.0157&503.9&39.38$^{+9.99}_{-8.98}$&  0.775 &1s2.1S               &1s.2p.1P*            & 1.1$^{+0.3}_{-0.3}\times 10^{22}$\\
  1.865&   127741&  1.864&Fe XXIV  & -209.1 & 0.0006&179.5& 2.57$^{+0.96}_{-0.70}$&  0.149 &1s2.2s               &1s2s2p.2P0.5         & 3.7$^{+1.4}_{-1.0}\times 10^{21}$\\
  1.875&   127760&  1.873&Fe XXIV  & -320.0 & 0.0003&203.7& 0.41$^{+0.28}_{-0.24}$&  0.015 &1s2.2s               &1s2s2p.4P1.5         & 5.7$^{+3.9}_{-3.4}\times 10^{21}$\\
  1.923&    17262&  1.926&Mn XXV   &  480.5 & 0.0007&401.6& 0.44$^{+0.13}_{-0.12}$&  0.420 &1s.2S                &2p.2P$^o$             &10.0$^{+2.9}_{-2.7}\times 10^{21}$\\
  1.951&    16659&  1.948&Ti XXII  & -412.1 & 0.0002&299.7& 0.14$^{+0.14}_{-0.09}$&  0.014 &1s.2S                &5p.2P$^o$             & 4.1$^{+4.0}_{-2.6}\times 10^{23}$\\
  1.963&         &       & no ID    &        & 0.0003&212.7& 0.44$^{+0.23}_{-0.20}$& & & &\\
  1.993&    16658&  1.995&Ti XXII  &  313.1 & 0.0003&207.6& 0.36$^{+0.21}_{-0.19}$&  0.029 &1s.2S                &4p.2P$^o$             & 5.0$^{+2.9}_{-2.7}\times 10^{23}$\\
  2.004&    17159&  2.006&Mn XXIV  &  329.4 & 0.0018&566.0& 0.87$^{+0.15}_{-0.14}$&  0.711 &1s2.1S               &1s.2p.1P             & 1.1$^{+0.2}_{-0.2}\times 10^{22}$\\
  2.020&         &       & no ID    &        & 0.0007&493.5& 0.32$^{+0.10}_{-0.09}$& & & &\\
  2.036&         &       & no ID    &        & 0.0003&454.9& 0.17$^{+0.09}_{-0.09}$& & & &\\
  2.062&         &       & no ID    &        & 0.0003&442.3& 0.14$^{+0.08}_{-0.08}$& & & &\\
  2.088&    17059&  2.092&Cr XXIV  &  502.9 & 0.0014&416.2& 0.84$^{+0.13}_{-0.12}$&  0.421 &1s.2S                &2p.2P$^o$             & 1.4$^{+0.2}_{-0.2}\times 10^{22}$\\
  2.114&         &       & no ID    &        & 0.0002&161.4& 0.26$^{+0.20}_{-0.17}$& & & &\\
  2.153&         &       & no ID    &        & 0.0005&156.3& 0.73$^{+0.28}_{-0.26}$& & & &\\
  2.179&    16956&  2.182&Cr XXIII &  426.8 & 0.0020&526.2& 0.87$^{+0.11}_{-0.11}$&  0.721 &1s2.1S               &1s.2p.1P             & 8.2$^{+1.0}_{-1.0}\times 10^{21}$\\
  2.193&    17016&  2.193&Cr XXIII &  -68.4 & 0.0004&181.7& 0.52$^{+0.20}_{-0.18}$&  0.152 &1s2.1S               &1s.2p.3P             & 2.3$^{+0.9}_{-0.8}\times 10^{22}$\\
  2.205&         &       & no ID    &        & 0.0002&158.5& 0.22$^{+0.20}_{-0.14}$& & & &\\
  2.312&         &       & no ID    &        & 0.0003&158.4& 0.27$^{+0.19}_{-0.17}$& & & &\\
  2.361&         &       & no ID    &        & 0.0003&396.3& 0.10$^{+0.06}_{-0.05}$& & & &\\
  2.417&         &       & no ID    &        & 0.0004&403.5& 0.15$^{+0.06}_{-0.05}$& & & &\\
  2.491&    16656&  2.492&Ti XXII  &  150.5 & 0.0007&410.0& 0.22$^{+0.05}_{-0.05}$&  0.419 &1s.2S                &2p.2P$^o$             & 1.7$^{+0.4}_{-0.4}\times 10^{22}$\\
  2.502&    16183&  2.514&Ca XIX   & 1414.8 & 0.0003&151.4& 0.30$^{+0.16}_{-0.14}$&  0.027 &1s2.1S               &1s.5p.1P*            & 2.5$^{+1.3}_{-1.2}\times 10^{22}$\\
  2.545&    16282&  2.549&Ca XX    &  515.1 & 0.0011&136.9& 1.69$^{+0.29}_{-0.26}$&  0.078 &1s.2S                &3p.2P$^o$             & 4.8$^{+0.8}_{-0.7}\times 10^{22}$\\
  2.555&         &       & no ID    &        & 0.0002&140.0& 0.20$^{+0.17}_{-0.13}$& & & &\\
  2.701&    16143&  2.705&Ca XIX   &  488.7 & 0.0008&400.9& 0.21$^{+0.05}_{-0.05}$&  0.152 &1s2.1S               &1s.3p.1P*            & 2.9$^{+0.7}_{-0.7}\times 10^{21}$\\
  2.857&         &       & no ID    &        & 0.0003&117.3& 0.18$^{+0.13}_{-0.11}$& & & &\\
  2.877&         &       & no ID    &        & 0.0003&124.1& 0.23$^{+0.16}_{-0.15}$& & & &\\
  2.982&    13532&  2.987&Ar XVIII &  548.3 & 0.0007&131.4& 0.48$^{+0.15}_{-0.13}$&  0.029 &1s.2S                &4p.2P$^o$             & 1.5$^{+0.5}_{-0.4}\times 10^{22}$\\
  3.016&    16262&  3.020&Ca XX    &  427.7 & 0.0050&415.2& 1.24$^{+0.08}_{-0.07}$&  0.411 &1s.2S                &2p.2P$^o$             & 5.7$^{+0.4}_{-0.3}\times 10^{21}$\\
  3.145&    13528&  3.151&Ar XVIII &  534.2 & 0.0016&304.3& 0.40$^{+0.06}_{-0.05}$&  0.078 &1s.2S                &3p.2P$^o$             & 4.3$^{+0.6}_{-0.6}\times 10^{21}$\\
  3.172&    16146&  3.177&Ca XIX   &  491.8 & 0.0024&203.9& 1.08$^{+0.12}_{-0.12}$&  0.770 &1s2.1S               &1s.2p.1P*            & 2.5$^{+0.3}_{-0.3}\times 10^{21}$\\
  3.355&    13463&  3.365&Ar XVII  &  938.9 & 0.0009&476.4& 0.11$^{+0.03}_{-0.03}$&  0.153 &1s2.1S               &1s.3p.1P*            & 5.5$^{+1.8}_{-1.7}\times 10^{20}$\\
  3.690&    11876&  3.696&S XVI    &  479.7 & 0.0012&107.1& 0.49$^{+0.14}_{-0.15}$&  0.014 &1s.2S                &5p.2P$^o$             & 8.0$^{+2.4}_{-2.4}\times 10^{21}$\\
  3.727&    13535&  3.733&Ar XVIII &  474.9 & 0.0072&351.1& 1.08$^{+0.07}_{-0.06}$&  0.412 &1s.2S                &2p.2P$^o$             & 1.9$^{+0.1}_{-0.1}\times 10^{21}$\\
  3.779&    11875&  3.784&S XVI    &  429.5 & 0.0019& 86.5& 1.18$^{+0.30}_{-0.24}$&  0.029 &1s.2S                &4p.2P$^o$             & 9.0$^{+2.3}_{-1.8}\times 10^{21}$\\
  3.943&    13446&  3.949&Ar XVII  &  441.3 & 0.0025& 85.2& 1.70$^{+0.33}_{-0.28}$&  0.766 &1s2.1S               &1s.2p.1P*            & 1.5$^{+0.3}_{-0.2}\times 10^{21}$\\
  3.985&    11878&  3.991&S XVI    &  466.8 & 0.0044&173.4& 1.20$^{+0.14}_{-0.12}$&  0.079 &1s.2S                &3p.2P$^o$             & 3.2$^{+0.4}_{-0.3}\times 10^{21}$\\
  4.183&    12039&  4.188&Cl XVII  &  329.9 & 0.0012& 90.5& 0.41$^{+0.16}_{-0.15}$&  0.416 &1s.2S                &2p.2P$^o$             & 7.4$^{+3.0}_{-2.7}\times 10^{21}$\\
  4.722&    11871&  4.729&S XVI    &  457.4 & 0.0184&407.6& 1.27$^{+0.08}_{-0.07}$&  0.413 &1s.2S                &2p.2P$^o$             & 5.4$^{+0.3}_{-0.3}\times 10^{20}$\\
  4.764&     9813&  4.770&Si XIV   &  403.0 & 0.0008& 72.9& 0.20$^{+0.22}_{-0.12}$&  0.008 &1s.2S                &6p.2P$^o$             & 2.2$^{+2.4}_{-1.4}\times 10^{21}$\\
  4.824&     9810&  4.831&Si XIV   &  441.5 & 0.0031&128.4& 0.48$^{+0.14}_{-0.13}$&  0.014 &1s.2S                &5p.2P$^o$             & 2.9$^{+0.9}_{-0.8}\times 10^{21}$\\
  4.941&     9811&  4.947&Si XIV   &  358.2 & 0.0053& 86.8& 1.98$^{+0.51}_{-0.37}$&  0.029 &1s.2S                &4p.2P$^o$             & 5.6$^{+1.4}_{-1.1}\times 10^{21}$\\
  5.032&    11734&  5.039&S XV     &  401.2 & 0.0044& 66.2& 2.07$^{+0.46}_{-0.37}$&  0.761 &1s2.1S               &1s.2p.1P*            & 4.5$^{+1.0}_{-0.8}\times 10^{20}$\\
  5.209&     9817&  5.217&Si XIV   &  472.2 & 0.0115&237.9& 0.92$^{+0.09}_{-0.08}$&  0.079 &1s.2S                &3p.2P$^o$             & 9.1$^{+0.9}_{-0.8}\times 10^{20}$\\
  5.375&         &       & no ID    &        & 0.0022&244.9& 0.11$^{+0.05}_{-0.05}$& & & &\\
  5.600&   132830&  5.606&Ni XXVI  &  332.2 & 0.0023& 62.5& 0.46$^{+0.22}_{-0.20}$&  0.007 &1s2.2s               &1S2\_7p              & 6.2$^{+2.9}_{-2.7}\times 10^{22}$\\
  6.004&         &       & no ID    &        & 0.0010& 78.3& 0.12$^{+0.09}_{-0.07}$& & & &\\
  6.018&         &       & no ID    &        & 0.0025& 64.8& 0.37$^{+0.13}_{-0.13}$& & & &\\
  6.045&     8043&  6.053&Al XIII  &  392.0 & 0.0063&414.3& 0.14$^{+0.02}_{-0.02}$&  0.079 &1s.2S                &3p.2P$^o$             & 1.8$^{+0.3}_{-0.3}\times 10^{21}$\\
  6.065&         &       & no ID    &        & 0.0014& 55.7& 0.23$^{+0.14}_{-0.13}$& & & &\\
  6.077&         &       & no ID    &        & 0.0031& 50.3& 0.66$^{+0.23}_{-0.21}$& & & &\\
  6.103&   132826&  6.108&Ni XXVI  &  226.1 & 0.0088&407.7& 0.22$^{+0.02}_{-0.02}$&  0.024 &1s2.2s               &1s2.5p               & 7.4$^{+0.8}_{-0.7}\times 10^{21}$\\
  6.135&         &       & no ID    &        & 0.0020& 62.9& 0.28$^{+0.12}_{-0.12}$& & & &\\
  6.154&     9816&  6.182&Si XIV   & 1374.7 & 0.0034&164.6& 0.17$^{+0.04}_{-0.04}$&  0.414 &1s.2S                &2p.2P$^o$             & 2.7$^{+0.7}_{-0.6}\times 10^{19}$\\
  6.172&     9816&  6.182&Si XIV   &  495.8 & 0.0414&419.4& 1.35$^{+0.04}_{-0.04}$&  0.414 &1s.2S                &2p.2P$^o$             & 2.1$^{+0.1}_{-0.1}\times 10^{20}$\\
  6.295&         &       & no ID    &        & 0.0075&332.6& 0.20$^{+0.03}_{-0.02}$& & & &\\
  6.352&   128163&  6.360&Fe XXIV  &  392.0 & 0.0138&425.8& 0.28$^{+0.02}_{-0.02}$&  0.002 &1s2.2s               &1S2\_9p              & 7.8$^{+0.6}_{-0.6}\times 10^{21}$\\
  6.375&         &       & no ID    &        & 0.0025& 58.5& 0.35$^{+0.13}_{-0.13}$& & & &\\
  6.413&         &       & no ID    &        & 0.0020& 58.8& 0.27$^{+0.13}_{-0.12}$& & & &\\
  6.440&   128161&  6.446&Fe XXIV  &  270.2 & 0.0139&274.1& 0.41$^{+0.03}_{-0.03}$&  0.004 &1s2.2s               &1S2\_8p               & 6.5$^{+0.5}_{-0.5}\times 10^{21}$\\
  6.489&     7856&  6.497&Mg XII   &  389.7 & 0.0043&162.7& 0.19$^{+0.04}_{-0.04}$&  0.008 &1s.2S                &6p.2P$^o$             & 1.7$^{+0.4}_{-0.4}\times 10^{21}$\\
  6.550&        &        & no ID   &        & 0.0019& 67.4& 0.20$^{+0.10}_{-0.10}$& & & &\\
  6.569&   128159&  6.575&Fe XXIV  &  260.3 & 0.0204&242.7& 0.74$^{+0.04}_{-0.04}$&  0.007 &1s2.2s               &1S2\_7p               & 6.7$^{+0.3}_{-0.4}\times 10^{21}$\\
  6.590&     7853&  6.580&Mg XII   & -455.2 & 0.0013& 59.5& 0.15$^{+0.11}_{-0.09}$&  0.014 &1s.2S                &5p.2P$^o$             & 6.9$^{+5.4}_{-4.4}\times 10^{20}$\\
  6.640&     9691&  6.648&Si XIII  &  361.4 & 0.0163&291.7& 0.42$^{+0.03}_{-0.03}$&  0.748 &1s2.1S               &1s.2p.1P*            & 3.4$^{+0.3}_{-0.3}\times 10^{19}$\\
  6.729&     7852&  6.738&Mg XII   &  401.2 & 0.0100& 54.7& 3.40$^{+0.92}_{-0.59}$&  0.029 &1s.2S                &4p.2P$^o$             & 7.7$^{+2.1}_{-1.3}\times 10^{21}$\\
  6.778&   128157&  6.784&Fe XXIV  &  256.7 & 0.0269&289.2& 0.76$^{+0.03}_{-0.03}$&  0.012 &1s2.2s               &1S2\_6p               & 3.7$^{+0.2}_{-0.1}\times 10^{21}$\\
  6.805&   132821&  6.811&Ni XXVI  &  264.5 & 0.0198&416.0& 0.34$^{+0.02}_{-0.02}$&  0.032 &1s2.2s               &1s2.4p               & 7.7$^{+0.5}_{-0.4}\times 10^{21}$\\
  6.842&         &       & no ID    &        & 0.0013& 75.2& 0.11$^{+0.07}_{-0.07}$& & & &\\
  6.869&         &       & no ID    &        & 0.0090&271.0& 0.20$^{+0.03}_{-0.02}$& & & &\\
  6.956&         &       & no ID    &        & 0.0015& 73.4& 0.12$^{+0.08}_{-0.07}$& & & &\\
  7.013&         &       & no ID    &        & 0.0026& 58.3& 0.26$^{+0.10}_{-0.10}$& & & &\\
  7.058&         &       & no ID    &        & 0.0144&445.5& 0.20$^{+0.02}_{-0.02}$& & & &\\
  7.093&     7855&  7.106&Mg XII   &  557.0 & 0.0426&495.7& 0.61$^{+0.02}_{-0.02}$&  0.079 &1s.2S                &3p.2P$^o$             & 4.8$^{+0.2}_{-0.2}\times 10^{20}$\\
  7.159&   128021&  7.169&Fe XXIV  &  419.1 & 0.0556&392.7& 1.17$^{+0.02}_{-0.02}$&  0.026 &1s2.2s               &1s2.5p               & 2.5$^{+0.1}_{-0.1}\times 10^{21}$\\
  7.223&         &       & no ID    &        & 0.0022& 48.2& 0.25$^{+0.12}_{-0.11}$& & & &\\
  7.242&         &       & no ID    &        & 0.0058&312.1& 0.11$^{+0.01}_{-0.04}$& & & &\\
  7.265&         &       & no ID    &        & 0.0024& 68.3& 0.18$^{+0.07}_{-0.06}$& & & &\\
  7.283&         &       & no ID    &        & 0.0030& 60.1& 0.27$^{+0.08}_{-0.08}$& & & &\\
  7.373&         &       & no ID    &        & 0.0051& 47.7& 0.64$^{+0.14}_{-0.13}$& & & &\\
  7.387&         &       & no ID    &        & 0.0029& 65.1& 0.22$^{+0.07}_{-0.07}$& & & &\\
  7.400&         &       & no ID    &        & 0.0020& 61.6& 0.15$^{+0.07}_{-0.07}$& & & &\\
  7.415&         &       & no ID    &        & 0.0029& 61.9& 0.23$^{+0.08}_{-0.07}$& & & &\\
  7.432&         &       & no ID    &        & 0.0026& 54.2& 0.24$^{+0.09}_{-0.09}$& & & &\\
  7.464&   126341&  7.472&Fe XXIII &  321.5 & 0.0272&348.1& 0.48$^{+0.02}_{-0.02}$&  0.056 &2s2                  &2s.5p                & 4.7$^{+0.2}_{-0.2}\times 10^{20}$\\
  7.490&         &       & no ID    &        & 0.0036& 58.0& 0.31$^{+0.09}_{-0.09}$& & & &\\
  7.513&         &       & no ID    &        & 0.0023& 52.0& 0.21$^{+0.11}_{-0.11}$& & & &\\
  7.533&         &       & no ID    &        & 0.0034& 59.9& 0.27$^{+0.10}_{-0.09}$& & & &\\
  7.555&         &       & no ID    &        & 0.0020& 59.7& 0.15$^{+0.08}_{-0.08}$& & & &\\
  7.575&         &       & no ID    &        & 0.0042& 53.8& 0.38$^{+0.11}_{-0.10}$& & & &\\
  7.617&         &       & no ID    &        & 0.0022& 43.5& 0.23$^{+0.15}_{-0.14}$& & & &\\
  7.664&         &       & no ID    &        & 0.0067&278.2& 0.10$^{+0.03}_{-0.02}$& & & &\\
  7.700&   126899&  7.733&Fe XXIII & 1285.7 & 0.0027& 48.3& 0.25$^{+0.12}_{-0.11}$&  0.035 &2s.2p                &2s.5d                & 3.6$^{+1.8}_{-1.7}\times 10^{20}$\\
  7.722&   126899&  7.733&Fe XXIII &  427.3 & 0.0018& 46.4& 0.17$^{+0.12}_{-0.11}$&  0.035 &2s.2p                &2s.5d                & 2.5$^{+1.8}_{-1.6}\times 10^{20}$\\
  7.743&   126899&  7.733&Fe XXIII & -387.5 & 0.0030& 72.0& 0.18$^{+0.07}_{-0.06}$&  0.035 &2s.2p                &2s.5d                & 2.6$^{+1.0}_{-0.9}\times 10^{20}$\\
  7.850&         &       & no ID    &        & 0.0056& 48.5& 0.55$^{+0.13}_{-0.13}$& & & &\\
  7.863&         &       & no ID    &        & 0.0037& 53.6& 0.29$^{+0.10}_{-0.10}$& & & &\\
  7.875&         &       & no ID    &        & 0.0037& 57.2& 0.27$^{+0.09}_{-0.09}$& & & &\\
  7.890&         &       & no ID    &        & 0.0028& 51.8& 0.22$^{+0.10}_{-0.10}$& & & &\\
  7.904&         &       & no ID    &        & 0.0035& 59.5& 0.24$^{+0.09}_{-0.08}$& & & &\\
  7.946&   127946&  7.983&Fe XXIV  & 1396.9 & 0.0019& 51.1& 0.14$^{+0.09}_{-0.08}$&  0.062 &1s2.2s               &1s2.4p               & 1.2$^{+0.7}_{-0.6}\times 10^{20}$\\
  7.980&   127946&  7.983&Fe XXIV  &  112.8 & 0.0829&409.0& 1.24$^{+0.04}_{-0.02}$&  0.062 &1s2.2s               &1s2.4p               & 1.0$^{+0.0}_{-0.0}\times 10^{21}$\\
  8.081&   120327&  8.090&Fe XXII  &  348.9 & 0.0043& 52.5& 0.33$^{+0.10}_{-0.09}$&  0.050 &2s2.2p               &2s2.5d               & 3.3$^{+1.0}_{-0.9}\times 10^{20}$\\
  8.296&   125709&  8.303&Fe XXIII &  249.5 & 0.0431&311.2& 0.67$^{+0.03}_{-0.03}$&  0.144 &2s2                  &2s.4p                & 2.3$^{+0.1}_{-0.1}\times 10^{20}$\\
  8.393&     7883&  8.421&Mg XII   & 1000.8 & 0.0042& 50.3& 0.29$^{+0.12}_{-0.11}$&  0.414 &1s.2S                &2p.2P$^o$             & 3.6$^{+1.5}_{-1.4}\times 10^{19}$\\
  8.409&     7883&  8.421&Mg XII   &  428.1 & 0.0704&285.7& 1.28$^{+0.04}_{-0.04}$&  0.414 &1s.2S                &2p.2P$^o$             & 1.6$^{+0.0}_{-0.0}\times 10^{20}$\\
  8.705&   119538&  8.715&Fe XXII  &  344.6 & 0.0061& 54.4& 0.36$^{+0.11}_{-0.11}$&  0.062 &2s2.2p               &2s.2p(3P).4p         & 2.7$^{+0.8}_{-0.8}\times 10^{20}$\\
  8.721&   119538&  8.715&Fe XXII  & -206.4 & 0.0057& 57.7& 0.31$^{+0.10}_{-0.10}$&  0.062 &2s2.2p               &2s.2p(3P).4p         & 2.3$^{+0.8}_{-0.7}\times 10^{20}$\\
  8.963&   119527&  8.977&Fe XXII  &  478.6 & 0.0077& 41.5& 0.61$^{+0.19}_{-0.18}$&  0.122 &2s2.2p               &2s2.4d               & 2.2$^{+0.7}_{-0.6}\times 10^{20}$\\
  9.048&   132824&  9.061&Ni XXVI  &  431.0 & 0.0602&244.6& 0.88$^{+0.05}_{-0.04}$&  0.248 &1s2.2s               &1s2.3p               & 1.9$^{+0.1}_{-0.1}\times 10^{21}$\\
  9.092&   132807&  9.105&Ni XXVI  &  428.9 & 0.0409&180.2& 0.76$^{+0.06}_{-0.05}$&  0.130 &1s2.2s               &1s2.3p               & 3.2$^{+0.2}_{-0.2}\times 10^{21}$\\
  9.158&     7744&  9.169&Mg XI    &  353.8 & 0.0038& 36.1& 0.28$^{+0.21}_{-0.18}$&  0.738 &1s2.1S               &1s.2p.1P             & 1.8$^{+1.3}_{-1.1}\times 10^{19}$\\
  9.336&   132741&  9.340&Ni XXV   &  118.9 & 0.0414&304.6& 0.42$^{+0.03}_{-0.03}$&  0.462 &2s2                  &2s.3p                & 4.7$^{+0.3}_{-0.4}\times 10^{20}$\\
  9.373&   132694&  9.390&Ni XXV   &  534.5 & 0.0289&164.9& 0.48$^{+0.05}_{-0.05}$&  0.231 &2s2                  &2s.3p                & 1.1$^{+0.1}_{-0.1}\times 10^{21}$\\
  9.469&     6241&  9.481&Ne X     &  380.2 & 0.0299&181.8& 0.42$^{+0.04}_{-0.04}$&  0.014 &1s.2S                &5p.2P$^o$             & 1.9$^{+0.2}_{-0.2}\times 10^{21}$\\
  9.695&     6218&  9.708&Ne X     &  402.3 & 0.0403&189.3& 0.54$^{+0.05}_{-0.05}$&  0.029 &1s.2S                &4p.2P$^o$             & 1.1$^{+0.1}_{-0.1}\times 10^{21}$\\
  9.723&     6218&  9.708&Ne X     & -462.8 & 0.0039& 58.1& 0.14$^{+0.12}_{-0.09}$&  0.029 &1s.2S                &4p.2P$^o$             & 3.0$^{+2.5}_{-1.9}\times 10^{20}$\\
  9.783&         &       & no ID    &        & 0.0037& 40.3& 0.19$^{+0.18}_{-0.12}$& & & &\\
 10.010&     6380& 10.021&Na XI    &  323.7 & 0.0531&183.2& 0.72$^{+0.07}_{-0.06}$&  0.416 &1s.2S                &2p.2P$^o$             & 9.6$^{+0.9}_{-0.8}\times 10^{20}$\\
 10.100&         &       & no ID    &        & 0.0122& 59.5& 0.43$^{+0.16}_{-0.14}$& & & &\\
 10.150&         &       & no ID    &        & 0.0234& 88.0& 0.60$^{+0.12}_{-0.11}$& & & &\\
 10.210&     6207& 10.240&Ne X     &  881.5 & 0.0047& 48.1& 0.18$^{+0.16}_{-0.11}$&  0.079 &1s.2S                &3p.2P$^o$             & 1.3$^{+1.2}_{-0.8}\times 10^{20}$\\
 10.220&     6207& 10.240&Ne X     &  587.1 & 0.0879&250.4& 0.87$^{+0.05}_{-0.05}$&  0.079 &1s.2S                &3p.2P$^o$             & 6.5$^{+0.4}_{-0.4}\times 10^{20}$\\
 10.340&   124604& 10.349&Fe XXIII &  255.3 & 0.0086& 57.1& 0.28$^{+0.16}_{-0.14}$&  0.006 &2s2                  &2p.3d                & 1.8$^{+1.1}_{-0.9}\times 10^{21}$\\
 10.490&   124935& 10.505&Fe XXIII &  443.3 & 0.0233& 66.5& 0.77$^{+0.21}_{-0.16}$&  0.021 &2s2                  &2p.3s                & 1.4$^{+0.4}_{-0.3}\times 10^{21}$\\
 10.570&   127980& 10.619&Fe XXIV  & 1390.7 & 0.0309&479.7& 0.12$^{+0.03}_{-0.03}$&  0.247 &1s2.2s               &1s2.3p               & 1.8$^{+0.5}_{-0.5}\times 10^{19}$\\
 10.610&   127980& 10.619&Fe XXIV  &  254.5 & 0.2323&490.0& 1.30$^{+0.05}_{-0.06}$&  0.247 &1s2.2s               &1s2.3p               & 2.0$^{+0.1}_{-0.1}\times 10^{20}$\\
 10.650&   127947& 10.663&Fe XXIV  &  366.2 & 0.3370&716.3& 1.13$^{+0.05}_{-0.05}$&  0.129 &1s2.2s               &1s2.3p               & 3.3$^{+0.1}_{-0.2}\times 10^{20}$\\
 10.780&   124988& 10.785&Fe XXIII &  128.0 & 0.0346&478.1& 0.12$^{+0.04}_{-0.03}$&  0.002 &2s2                  &2s.3d                & 2.5$^{+0.7}_{-0.7}\times 10^{21}$\\
 10.890&   124606& 10.903&Fe XXIII &  352.6 & 0.0402&435.8& 0.15$^{+0.04}_{-0.03}$&  0.082 &2s.2p                &2p.3p                & 6.8$^{+1.6}_{-1.5}\times 10^{19}$\\
 10.930&   125174& 10.980&Fe XXIII & 1372.3 & 0.0079& 41.0& 0.31$^{+0.25}_{-0.19}$&  0.414 &2s2                  &2s.3p                & 2.7$^{+2.2}_{-1.7}\times 10^{19}$\\
 10.970&   125174& 10.980&Fe XXIII &  273.5 & 0.1490&406.7& 0.74$^{+0.05}_{-0.04}$&  0.414 &2s2                  &2s.3p                & 6.6$^{+0.4}_{-0.4}\times 10^{19}$\\
 11.010&   125372& 11.018&Fe XXIII &  218.0 & 0.1051&221.2& 0.99$^{+0.07}_{-0.08}$&  0.254 &2s2                  &2s.3p                & 1.4$^{+0.1}_{-0.1}\times 10^{20}$\\
 11.150&         &       & no ID    &        & 0.0350&382.1& 0.15$^{+0.04}_{-0.04}$& & & &\\
 11.280&   125155& 11.299&Fe XXIII &  508.0 & 0.0289&253.5& 0.15$^{+0.06}_{-0.06}$&  0.751 &2s.2p                &2s.3d                & 7.3$^{+2.8}_{-2.8}\times 10^{18}$\\
 11.310&   125155& 11.299&Fe XXIII & -289.1 & 0.0224&170.2& 0.18$^{+0.08}_{-0.08}$&  0.751 &2s.2p                &2s.3d                & 8.5$^{+4.0}_{-3.9}\times 10^{18}$\\
 11.350&   124975& 11.337&Fe XXIII & -348.9 & 0.0504&481.3& 0.16$^{+0.06}_{-0.05}$&  0.552 &2s.2p                &2s.3d                & 1.0$^{+0.4}_{-0.3}\times 10^{19}$\\
 11.420&   124980& 11.442&Fe XXIII &  585.8 & 0.1575&500.7& 0.52$^{+0.08}_{-0.07}$&  0.615 &2s.2p                &2s.3d                & 3.0$^{+0.4}_{-0.4}\times 10^{19}$\\
 11.470&   124991& 11.457&Fe XXIII & -337.4 & 0.0621&253.4& 0.34$^{+0.08}_{-0.07}$&  0.110 &2s.2p                &2s.3d                & 1.1$^{+0.3}_{-0.2}\times 10^{20}$\\
 11.540&         &       & no ID    &        & 0.0357&437.0& 0.11$^{+0.04}_{-0.04}$& & & &\\
 11.760&   119502& 11.769&Fe XXII  &  242.3 & 0.0710&192.6& 0.51$^{+0.09}_{-0.08}$&  0.673 &2s2.2p               &2s2.3d               & 2.6$^{+0.5}_{-0.4}\times 10^{19}$\\
 11.910&   120311& 11.921&Fe XXII  &  277.1 & 0.0546&145.6& 0.50$^{+0.12}_{-0.11}$&  0.597 &2s2.2p               &2s2.3d               & 2.9$^{+0.7}_{-0.6}\times 10^{19}$\\
 12.120&     6204& 12.134&Ne X     &  344.1 & 0.2336&358.4& 1.09$^{+0.08}_{-0.08}$&  0.415 &1s.2S                &2p.2P$^o$             & 1.3$^{+0.1}_{-0.1}\times 10^{20}$\\
 12.300&    89984& 12.282&Fe XXI   & -431.7 & 0.0226& 66.1& 0.40$^{+0.26}_{-0.21}$&  1.305 &2s2.2p2              &2s2.2p.3d            &10.0$^{+6.6}_{-5.2}\times 10^{18}$\\
 12.600&         &       & no ID    &        & 0.0550&163.0& 0.36$^{+0.12}_{-0.11}$& & & &\\
 12.640&         &       & no ID    &        & 0.0319&208.2& 0.15$^{+0.09}_{-0.09}$& & & &\\
 12.820&         &       & no ID    &        & 0.0703&500.8& 0.15$^{+0.12}_{-0.09}$& & & &\\
 12.870&         &       & no ID    &        & 0.0566&213.6& 0.25$^{+0.20}_{-0.16}$& & & &\\
 13.110&         &       & no ID    &        & 0.0700&175.5& 0.38$^{+0.19}_{-0.17}$& & & &\\
 13.240&         &       & no ID    &        & 0.0744&300.9& 0.21$^{+0.14}_{-0.12}$& & & &\\
 13.310&         &       & no ID    &        & 0.0464&118.2& 0.35$^{+0.29}_{-0.22}$& & & &\\
 13.450&         &       & no ID    &        & 0.2285&553.4& 0.41$^{+0.13}_{-0.11}$& & & &\\
 13.540&         &       & no ID    &        & 0.1563&564.8& 0.25$^{+0.12}_{-0.11}$& & & &\\
 13.580&         &       & no ID    &        & 0.0313& 30.1& 1.27$^{+9.99}_{-0.80}$& & & &\\
 13.660&         &       & no ID    &        & 0.0443& 29.2& 3.28$^{+9.99}_{-2.07}$& & & &\\
 13.680&         &       & no ID    &        & 0.0365& 26.1& 2.42$^{+9.99}_{-1.52}$& & & &\\
 13.770&         &       & no ID    &        & 0.1267&268.8& 0.39$^{+0.19}_{-0.17}$& & & &\\
 13.820&         &       & no ID    &        & 0.0393& 29.6& 1.92$^{+9.99}_{-1.21}$& & & &\\
 13.880&         &       & no ID    &        & 0.1279&450.4& 0.24$^{+0.13}_{-0.12}$& & & &\\
 13.930&         &       & no ID    &        & 0.0883&150.1& 0.49$^{+0.29}_{-0.23}$& & & &\\
 14.010&         &       & no ID    &        & 0.1935&482.0& 0.35$^{+0.14}_{-0.13}$& & & &\\
 14.090&    16073& 14.097&Ca XVIII &  149.0 & 0.1243&302.6& 0.40$^{+0.17}_{-0.15}$&  0.062 &1s2.2s               &1s2.4p               & 2.6$^{+1.1}_{-1.0}\times 10^{21}$\\
 14.160&         &       & no ID    &        & 0.0626& 72.3& 0.77$^{+0.73}_{-0.49}$& & & &\\
 14.190&         &       & no ID    &        & 0.1305&246.3& 0.40$^{+0.21}_{-0.18}$& & & &\\
 14.250&         &       & no ID    &        & 0.0399& 27.8& 1.85$^{+9.99}_{-1.17}$& & & &\\
 14.300&         &       & no ID    &        & 0.1415&160.0& 0.77$^{+0.34}_{-0.27}$& & & &\\
 14.360&         &       & no ID    &        & 0.0404& 27.8& 1.81$^{+9.99}_{-1.14}$& & & &\\
 14.390&         &       & no ID    &        & 0.1554&266.4& 0.43$^{+0.19}_{-0.17}$& & & &\\
 14.430&         &       & no ID    &        & 0.0576& 78.8& 0.56$^{+0.59}_{-0.36}$& & & &\\
 14.470&         &       & no ID    &        & 0.0527&112.3& 0.32$^{+0.38}_{-0.20}$& & & &\\
 14.550&         &       & no ID    &        & 0.0577& 26.3& 7.71$^{+9.99}_{-4.86}$& & & &\\
 14.610&     4675& 14.645&O VIII   &  724.9 & 0.1763&189.3& 0.75$^{+0.30}_{-0.25}$&  0.008 &1s.2S                &6p.2P$^o$             & 1.8$^{+0.7}_{-0.6}\times 10^{20}$\\
 14.770&     4634& 14.832&O VIII   & 1255.2 & 0.0501& 26.7& 3.02$^{+9.99}_{-1.90}$&  0.014 &1s.2S                &5p.2P$^o$             & 4.1$^{+9.9}_{-2.6}\times 10^{20}$\\
 14.820&     4634& 14.832&O VIII   &  238.9 & 0.1715&454.9& 0.26$^{+0.15}_{-0.13}$&  0.014 &1s.2S                &5p.2P$^o$             & 3.5$^{+2.0}_{-1.8}\times 10^{19}$\\
\enddata 
\tablenotetext{a}{Wavelength in $\AA$}
\tablenotetext{b}{Wavelength taken from the {\sc xstar} database.}
\tablenotetext{c}{Velocity is in units of km sec$^{-1}$}
\tablenotetext{d}{Equivalent width in units of eV.}
\tablenotetext{e}{Line center optical depth.}
\tablenotetext{f}{Oscillator strength}
\tablenotetext{g}{Column density is in units of cm$^{-2}$}
\end{deluxetable} 

\clearpage

The range between 1-2.5 \AA\ is dominated by the H-like L$\alpha$ and 
He-like allowed $n=$1 -- 2 lines from the elements Fe
(1.77, 1.85 \AA\ ), Mn (1.92, 2.01 \AA\ ) and Cr (2.08, 2.18 \AA\ ).
Wavelengths shortward of the Fe XXVI L$\alpha$ line at 1.77 \AA\ contain the K lines of elements heavier 
than Fe, in addition to the higher series lines of Fe.  Although there are marginal detections 
of some of these  lines, there are too few counts in this region to strongly 
constrain the abundances of the heavier elements Co, Cu, Zn.  Ni is an exception to this, since it also has lines from 
Li-like Ni XXVI at longer wavelength.  It is important to emphasize that, although Ly$\alpha$ lines are not clearly 
detected for Co, Cu and Zn, we cannot set meaningful upper limits to their strength because our model atoms 
for these elements do not include $n=$2 -- 3 lines from the Li-like and lower ionization stages.

The 2 -- 3 \AA\ region includes the He-like resonance line of Mn XXIV, the  K lines from H- and He-like  
Cr (2.088 and 2.179 \AA\ ), He-like Sc (2.877 \AA\ ), H-like Ti (2.491 \AA\ ). 
The apparent absence of lines from V, H-like Sc and He-like Ti provides 
relatively secure upper limits on the abundances of these elements.  We point out that 
the existence of evaluated wavelengths for these lines from NIST provides added validity to this 
conclusion.  In this range are also
$n=$1 -- 3 lines from Ca, which are stronger than the $n=$1 -- 2 lines from Sc, Ti, Cr and Mn.
In this range are also $n=$1 -- 3 lines from Ca, Ar, S, Si.  For Si XIV Rydberg series lines are detected up to $n=6$.  

At 4.18 \AA\ we marginally detect the L$\alpha$ line from H-like Cl XVII.  The He-like line from this element
is not detected, but we note the absence of evaluated wavelength for this line.  
Similar comments apply to the H-like line from P, near 5.38 \AA.
The ratio of the He and H-like lines from S near 4.72 and 5.04 \AA\ follows the same behavior as 
for Ca and Ar.  The 6 --7\AA\ region is dominated by lines from Si XIII and XIV and  Li-like Fe and Ni; for Fe these are detected up 
to $1s^22s - 1s^210p$.  At 10.67 and 10.88\AA\ are the lines from Fe XXII which provide the density diagnostics.  These lines
together extend the range of ionization of observed species to include ions which are indicative of lower ionization parameter 
gas than the hydrogenic and helium-like ions which predominate.

Table \ref{linelist} includes 175 lines of which 100 have IDs.  The table of \cite{Mill08} includes 102 lines, 
of which 15 have no IDs.  
Our table has 44 lines which are not in the \cite{Mill08} table, although we note that they used a 
5$\sigma$ criterion for line detection, while ours is 3$\sigma$.
Of the unidentified lines in \cite{Mill08} we propose IDs for 5.6 \AA\ Ni XXVI 2s -- 7p and 9.372 \AA\ Ni XXV 2s$^2$ -- 2s3p.
We have no IDs for the unidentified lines from \cite{Mill08} at $\lambda\lambda$ 6.25, 7.0555, 7.0851, 9.9509 \AA.  
These latter 3 are crudely consistent with lines 
arising from the 2p excited levels of Ni XXVI (rest wavelengths 7.048, 7.0950 and 9.6383).  But the results of 
{\sc warmabs} suggest optical depths for these lines which are smaller than the resonance lines by factors $\geq 10^4$ assuming
a density of 10$^{15}$ cm$^{-3}$ and no trapping of line radiation.  
All the 75 lines from our notch model for which we have no identifications, have small 
equivalent widths, within a factor of 2 of our cutoff of 3.5 $\times 10^{-4}$.  We speculate that some of them could be 
due to the notch
algorithm attempting to correct for discrepancies between the model continuum and the data, perhaps 
due to bound-free continuum absorption.  In addition, visual inspection of the 
spectrum shows lines which do not appear in our table.  Many of these have possible IDs as lines from excited 
levels, but none appears in the {\sc warmabs} models at sufficient strength to be included in our table.
Other lines in this category include:  2.60 \AA\  
possibly Ti XXI He-like 1 -- 2,  2.15 \AA\ possibly  Sc XX  He-like 1 -- 3,  features at 3.61, 5.375, 5.62, 5.65 \AA\ with 
no obvious ID,  and 5.72 \AA\ possibly Al XIII 1 -- 4.  Features at 6.22 and 6.24 \AA\ could be due to Si XIII 1s2s -- 2s2p.
Discernible features at $\lambda\lambda$ 6.7, 6.87, 7.06, 9.24, 9.78, 9.96, 10.1, 10.15 and 11.15 \AA\,  
have no obvious IDs.  Many of these 
correspond to lines in our list which are from excited levels, or from ion stages which are too low to 
coexist with the dominant ions in the {\sc warmabs} models.  We also note
an emission line at 13.38 \AA\ which coincides with the $\lambda$ 13.387 \AA\
2s -- 3p transition in Ti XXII listed in \citet{Shir00}.  If this is correct, it is hard understand why this 
line should appear in emission when other analogous lines from Fe XXIV and Ni XXVI, for instance, 
appear in absorption.    Other interesting lines not identified
by \cite{Mill08} are the Fe XXIII lines between 11.28 \AA\/
and 11.47 \AA , since they arise from the metastable 2s2p$^3P$
level.  Their presence corroborates the density estimate from the
Fe XXII lines.



\begin{figure}[h]
\includegraphics*[angle=270, scale=0.6]{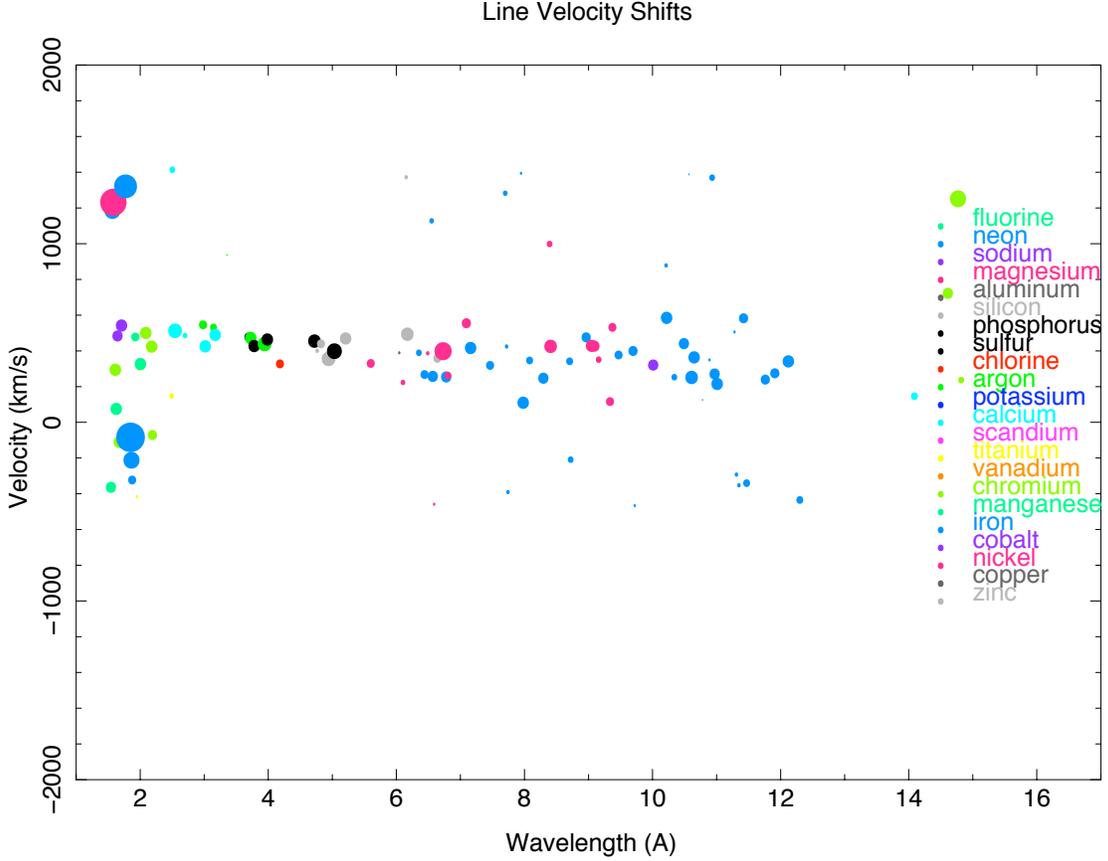}
\caption{\label{vofffig}Doppler velocities obtained by comparing measured notch wavelengths 
with lab wavelengths from {\sc xstar} database.  Element color coding is given in the legend.
The dot size is proportional to the line optical depth.}
\end{figure}

The Doppler shift of the lines relative to the lab wavelength are shown graphically in figure \ref{vofffig}.
In this figure, the color corresponds to the element, and the size of the dot corresponds to the line 
optical depth.
This shows that the lines cluster in a range  around  400 km s$^{-1}$ from zero offset
(here and in what follows we quote blueshifted velocities as positive, and conversely).  
This corresponds to $\sim$2 -- 10 m\AA\ for 
most lines.  Many of the laboratory wavelengths are uncertain at this level.  Given this, the most notable aspect of 
the line shifts is the large velocity offset of the Fe XXVI L$\alpha$  and the Ni XXVII lines.  
These are among the strongest lines in the spectrum, and are blueshifted by $\simeq$1300 km/s, which is significantly 
greater than any other lines in the spectrum.  This, together with the fact that these are the two highest-ionization lines
in the spectrum, suggests that these lines are partly formed in a separate component of the flow.  If so,  this component would 
have higher ionization, and higher velocity, than the component responsible for the rest of the lines.  On the other 
hand, we consider the possibility that this is an artifact of shortcomings in the HETG calibration when 
applied to fitting absorption of the  steeply sloping continuum under these lines.  In what 
follows we will examine alternative explanations for this in our fits to the spectrum.

The line widths in Table \ref{linelist} range from 1 -- 10 m\AA.  We have searched and found no significant 
correlation between this quantity and simple quantities characterizing the parent ion, such as the ionization potential
or isoelectronic sequence. 
There is a weak tendency for the largest widths to be associated with lower ionization potential ions.
An example is the L$\alpha$ line from Si XIV.  However, there are also narrower lines from ions with 
comparable ionization potential, so this does not constitute a statistically significant correlation.

\subsection{Ion Column Densities}

\begin{figure}[h]
\includegraphics*[angle=270, scale=0.6]{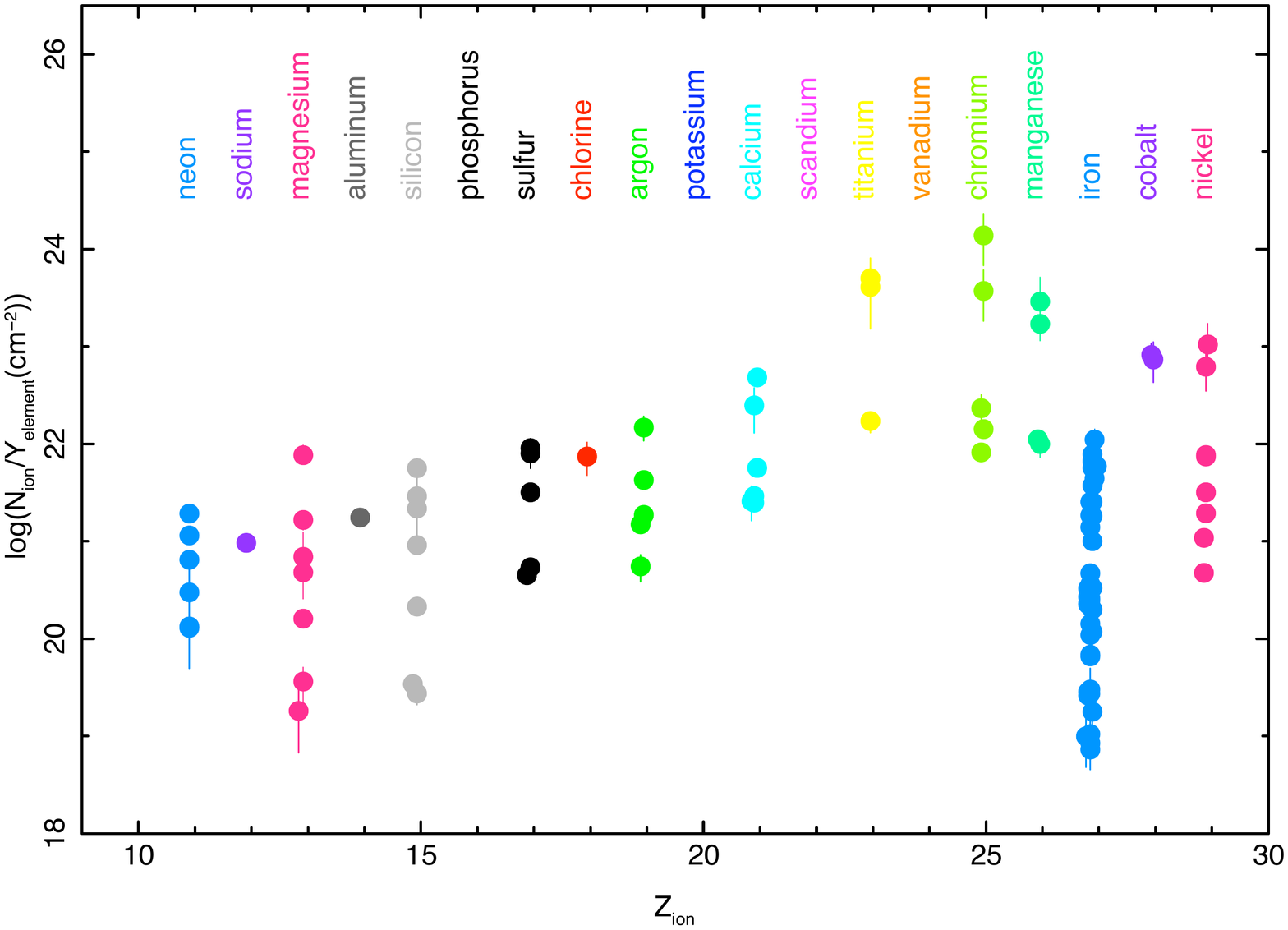}
\caption{\label{colunitfig}Element column densities obtained assuming linear 
curve of growth, cosmic element abundances, and unit ion fractions for 
all lines in Table \ref{linelist}.  Each data point represents one line.  
The horizontal axis is the quantity $Z_{ion}$ 
where we distinguish between contributions of various ions by assigning
$Z_{ion}=Z_{element}+1-0.1*(ion stage)$, where $(ion stage)$=1 for hydrogenic, 2 for He-like, etc.
Element color coding is the same as in figure \ref{vofffig}.}
\end{figure}

Ion column densities can be derived from the line identifications in Table \ref{linelist}, assuming that the equivalent widths 
lie on the linear part of the curve of growth.  That is, we calculate

\begin{equation}
N_{ion}=\frac{\tau_{line}}{\frac{\pi e^2}{m_e c} \frac{f}{\Delta\nu_D}}
\end{equation}

\noindent where $\Delta\nu_D$ is the total Doppler width in frequency units, including both thermal and 
turbulent contributions. $\tau_{line}$ is the line depth measured from the notch fit to the spectrum, 
and since it is based on Gaussian line fits its accuracy diminishes when values become large.
In Table \ref{linelist}  we list the value of the ratio $N_{ion}/Y_{element}$, 
where $Y_{element}$ is the elemental solar \citep{Grev96, Alle73}  abundance relative to hydrogen.
$N_{ion}/Y_{element}$, is the equivalent hydrogen column density implied by a given line if its ion fraction 
were unity, if the elemental abundance were solar, and if it lies on the linear
part of the curve of growth.  In figure \ref{colunitfig} we plot this quantity.  The horizontal 
axis is related to the  atomic number of the parent element by 
$Z_{ion}=Z_{element}+1-0.1*(ion stage)$, where $(ion stage)$=1 for hydrogenic, 2 for He-like, etc.  
Error bars are plotted and, in most cases, are small compared with the 
the interval between points.
If the assumptions of unit ionization fraction, cosmic abundances,  and linear curve of growth
were correct, then all the points in this diagram would lie along one horizontal line.
Since most elements have lines from more than one ion, the assumption of unit ionization fraction
cannot be correct; for these the true column should be the sum of the columns for 
various ions if the other assumptions were correct.  However, this neglects the 
possible contributions from ions which do not produce 
observed lines, such as fully stripped species, which are likely to be important 
for the lower-$Z$ elements.  

It is clear from figure \ref{colunitfig} that there is a greater dispersion 
of column densities within elements than can be accounted for by ionization effects.
This will be discussed in greater detail in the next section, and suggests the limitation 
of our second assumption, that of the linear curve of growth.  Figure \ref{colunitfig}
also shows  departures from solar abundance ratios, in the sense of 
apparently enhanced abundances for most elements between  Sc and Mn, relative to elements 
with $Z \leq 16$.  This conclusion is dependent quantitatively on the excitation and ionization conditions 
in the absorber, and in the following section we will attempt to quantify the abundances 
for a variety of assumptions about the state of the gas.  

We expect 
qualitatively that the lower-$Z$ elements will be more highly ionized than the higher-$Z$ elements, 
for most plausible ionization mechanisms.  This would predict that the apparent elemental abundances 
would be systematically greater for the low-$Z$ elements than for the high-$Z$ elements in figure 
\ref{colunitfig}, since the low-$Z$ elements would have a greater fraction of their ions in 
the unobservable fully stripped stage.  This is the opposite behavior to what we observe, and
so reinforces the conclusion that the elements with $Z \geq 16$  have enhanced abundances
relative to those with lower $Z$.

\subsection{Curve of Growth}

\begin{figure}[h]
\includegraphics*[angle=270, scale=0.6]{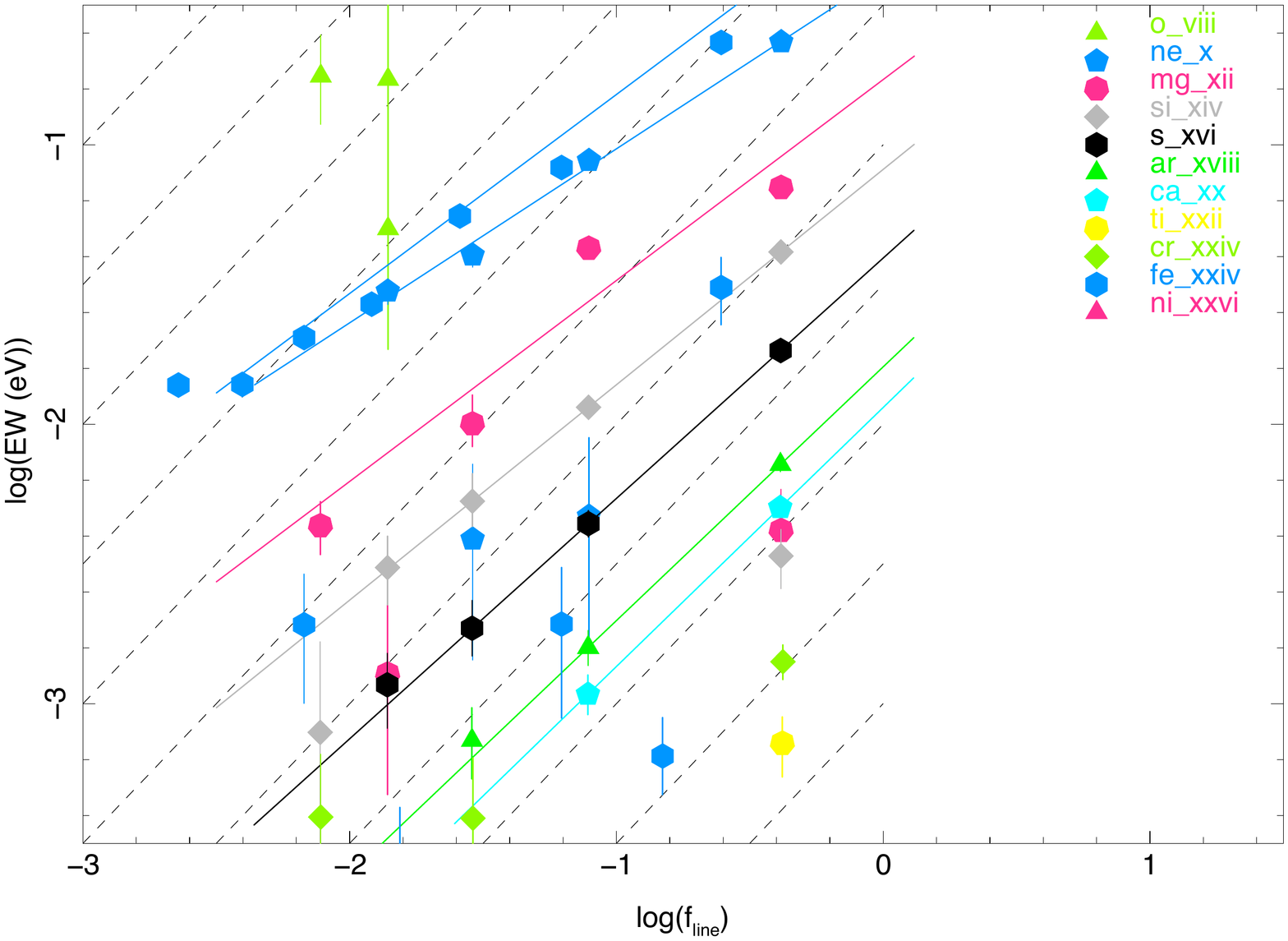}
\caption{\label{coghlike}Curve of growth for H-like and Li-like ions.  
Element color coding is the same as in figure \ref{vofffig}.  Dashed diagonal 
lines show simple proportionality 
behavior expected for unsaturated lines.  Solid lines are 
linear regression fits to data with errors less than 0.1.}
\end{figure}

The validity of the procedure used to derive the columns in figure \ref{colunitfig} depends on 
the assumption that the lines lie on the linear part of the curve of growth.  If this is correct, then
the line equivalent widths should be proportional to the transition oscillator strengths, and 
comparison of various lines from the same ion should show this proportionality.
In figure \ref{coghlike} we test this procedure using the Lyman series lines 
from H-like ions of Ne, Mg, Al, Si, S, Ar, Ca, Cr, and Mn.  We also include the $2s-np$ lines 
from Li-like Fe XXIV and Ni XXVI.  Solid lines are 
linear regression fits to data with errors less than 0.1. Also shown on this figure 
are diagonal lines (dashed) corresponding to the proportionality expected for 
linear curve of growth.  This shows that, although some ions appear to follow the 
linear trend, the strongest lines in particular grow more slowly than linearly.
This is particularly apparent for the lines of Ne X, Fe XXIV and Ni XXVI.  Each ion has at least 5 lines,
and the trend is apparent across the line strengths.  This shows that the simple analysis 
provided in figure \ref{colunitfig} and Table \ref{linelist} 
are likely not adequate for the purposes of inferring abundances and the lines may be saturated.  
On the other hand, weaker lines such as those of Si XIV do apparently follow the linear trend.  

\begin{figure}[h]
\includegraphics*[angle=270, scale=0.6]{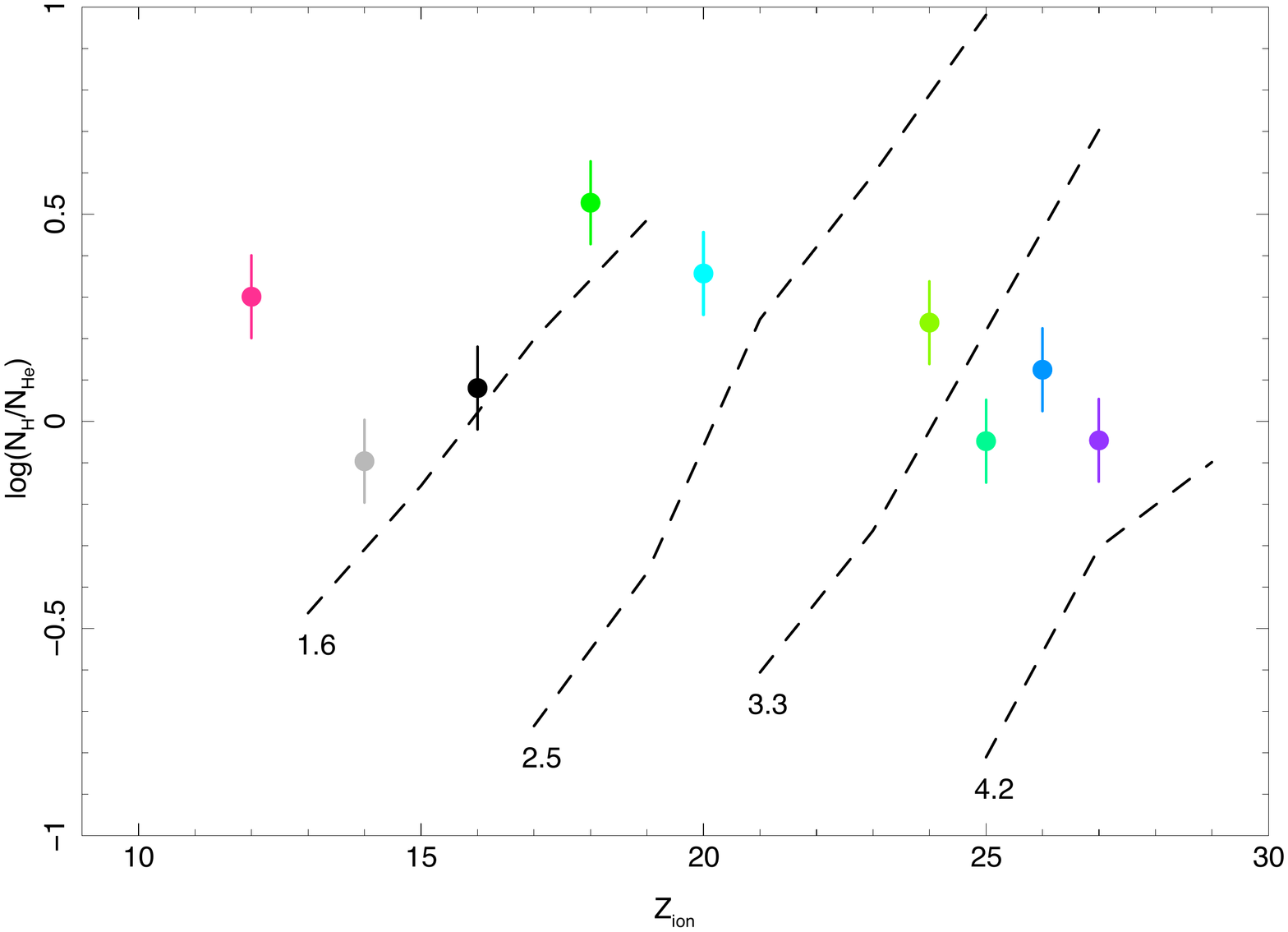}
\caption{\label{hheplot}Ratio of column densities 
derived from linear curve of grow analysis for H-line to He-line ions, 
versus the nuclear charge.  Element color coding is the same as in 
figure \ref{vofffig}.}
\end{figure}

Another illustration of the effects of saturation is shown in figure 
\ref{hheplot}.  This shows the ratios of He-like to H-like ion abundances inferred 
from the line equivalent widths and identifications in Table \ref{linelist}.  These ratios are
independent of element abundance, owing to the fact that each ratio is taken between
ions from the same element.  This shows that the ratios are all in the range 
between 10$^{-1}$ and 10$^{+1}$ for $ 12 \leq Z \leq 26$, 
but that there is no systematic trend with $Z$.  We might expect the ratio to increase with $Z$, 
as higher $Z$ elements would be generally less highly ionized, for plausible ionization mechanisms.
Also shown in figure \ref{hheplot} are the contours traced by 
an {\sc xstar} photoionization model described in more detail
in the Appendix.  These are labeled by the value of log($\xi$), where $\xi=L/(nR^2)$ is the 
ionization parameter as defined by \cite{Tart69}; $L$ is the source energy luminosity integrated 
from 1 -- 1000 Ry, $n$ is the gas number density, and $R$ is the distance from the continuum 
source to the absorber.  These show that a given 
value of ionization parameter predicts that the He/H abundance ratio should 
increase between adjacent elements by a factor $\sim$5.  The figure 
clearly shows that a single ionization parameter
cannot account for the ratios displayed by  all the elements.  This could indicate
the existence of a broad range of ionization parameters in the source, spanning values
indicated by this figure. On the other hand, it could be associated with radiative transfer effects,
such as saturation, which make ion fractions inferred from a linear curve of growth unreliable.

Possible explanations for the departures 
from the linear curve of growth include the influence of saturation which causes
the curve to flatten when the lines become optically thick in the Doppler core.
Other possibilities include filling-in of the lines by an additional continuum
emission component which is not seen in transmission through the warm absorber, 
and also radiation transfer effects in the absorber itself.  The latter includes 
forward scattering, which may also depend on the relative size of the continuum 
source and the absorber.  Also, thermal emission can fill in the lines.  In the following 
subsection we discuss these in turn.

\subsection{Radiation Transfer and Curve of Growth}

\begin{figure}[h]
\includegraphics*[angle=270, scale=0.6]{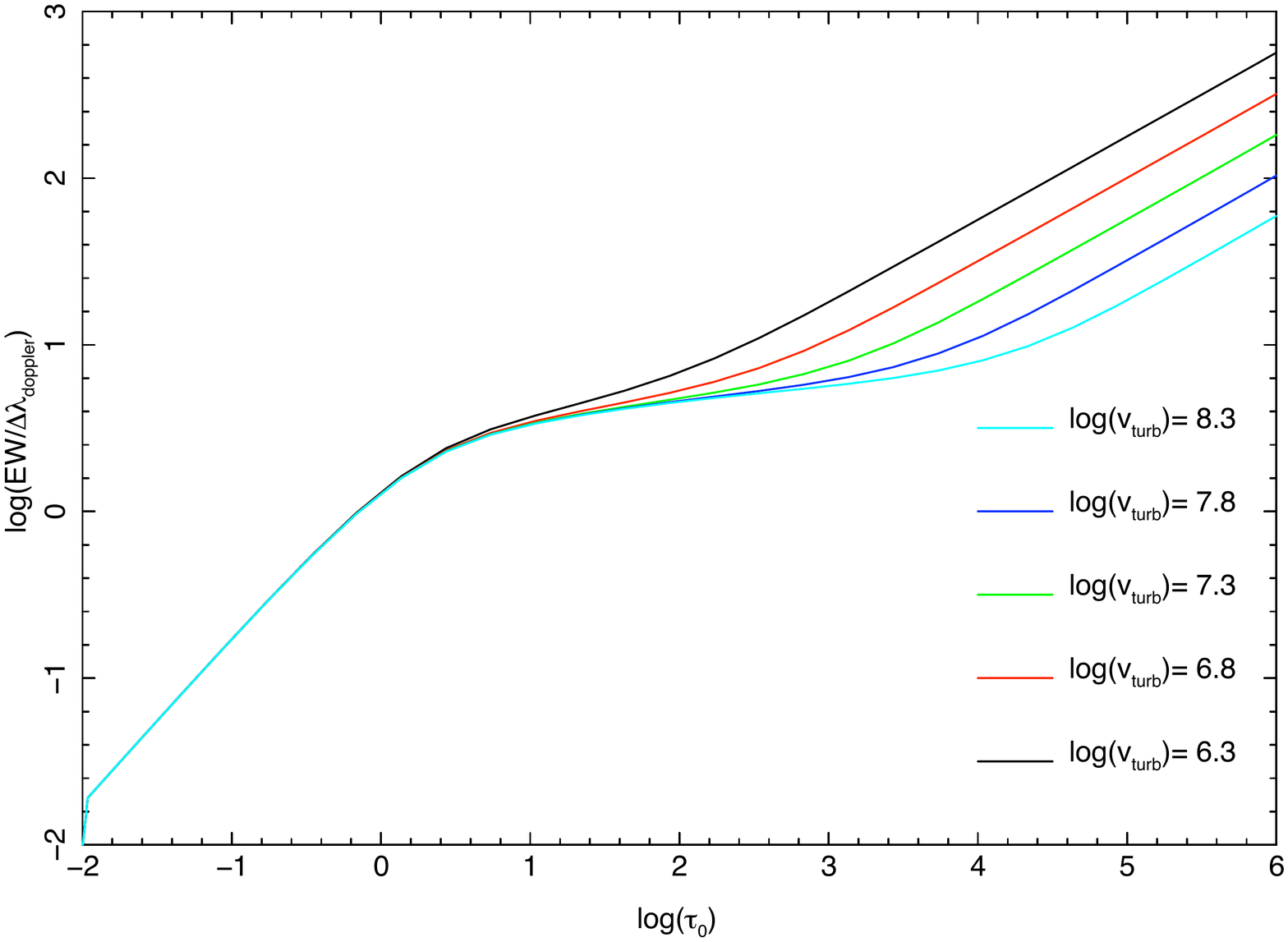}
\caption{\label{figcog}Curve of growth for various choices of the velocity characterizing the 
Doppler broadening $v_{turb}$, and for a damping parameter corresponding to that for the 
Si XIV L$\alpha$ line, at a Doppler broadening velocity of 20 km s$^{-1}$. }
\end{figure}

The standard curve of growth for a resonance line can be written:

\begin{equation}
\label{curgrowth}
EW=\int{d\varepsilon \left(1 - e^{-\tau(\varepsilon)}\right)}
\end{equation}

\begin{equation}
\tau(\varepsilon)=\frac{\pi e^2}{m_ec} f x_{ij}Y_j N \frac{\lambda}{v_{turb}} 
\phi(\frac{\varepsilon}{\Delta\varepsilon_{turb}})
\end{equation}

\noindent where $\varepsilon$ is the photon energy, $f$ is the oscillator strength,  $x_{ij}$ is the ion 
fraction, $Y_j$ is the element abundance, $N$ is the total column density,  $\lambda$ is the line wavelength,
$v_{Turb}$ is the turbulent velocity (including the thermal ion speed), 
and $\phi$ is the profile function, which includes both a Doppler core and damping wings.
This is shown in figure \ref{figcog}, for various choices of the velocity characterizing the 
Doppler broadening $v_{turb}$, and for a natural width corresponding to that for the Si XIV L$\alpha$ line.  
This shows that, for a given equivalent width, 
the effects of saturation are more likely to  be important when the Doppler broadening is smallest.
That is, the equivalent width where the curve flattens is approximately proportional to the 
Doppler width.  As we will show, the widths of many lines from GRO J1655-40 are not 
constrained from below by the HETG spectrum.  Thus, a possible 
explanation for the slower than linear curve of growth is that the line Doppler widths 
are small enough such that the strongest lines in figure \ref{coghlike} are affected by saturation.

Filling in by continuum from a separate source which is not absorbed by the warm 
absorber would preferentially affect lines with larger optical depths.  This could
qualitatively explain the flattening of the curve of growth which is observed. 
However, the filling in would also have the wavelength dependence of the 
additional continuum component.  Although this is unknown, the simplest assumption 
would be that it is the same as the primary continuum.  We have tested this possibility
quantitatively using direct fitting, and will discuss this further in the following section.


The standard curve of growth in equation (\ref{curgrowth}) assumes the simplest possible
geometry and microphysics affecting the line: (i)The continuum source has negligible size;
(ii) The absorber exists only in a narrow region along the line of sight to the 
continuum source; (iii) The excitation and deexcitation of the transition responsible 
for the line is affected only by the radiative excitation and spontaneous decay
connecting the two atomic levels.  In addition, the gas is assumed to have a velocity field
which can be characterized by a Maxwellian distribution with a well-defined thermal or turbulent 
velocity. 

If the continuum source has finite size, or if the warm absorber exists outside the line 
of sight to the source, then the line profiles will be affected by photons which scatter 
into our line of sight.  Although there is no simple general expression for the intensity observed 
from an extended scattering atmosphere in this case, it is straightforward to show that in the limit of 
small optical depth, the scattered emission contribution scales proportional to $(D/R)^2$, 
where $D$ is the characteristic size of the continuum source and $R$ the characteristic distance
from the continuum source to the absorber.  This contribution is independent of optical depth
in this limit, and so will not affect the shape of the curve of growth at small $\tau$.

Departures from assumption (iii), concerning the population kinetics, can be expected 
if the radiative decay of the upper level is suppressed or 
if the upper level can be populated by another process such as recombination or 
collisional excitation.  Some of these possibilities were discussed by \cite{Masa04}. 
If the upper level can decay via alternate channels, such as 
branching to other levels radiatively, collisional deexcitation, or autoionization, 
the basic absorption profile properties will be unchanged (although these processes can affect 
the damping parameter).  Suppression of the upper level decay by collisional deexcitation
could occur if the product of line optical depth and electron density, $n_e$,
is large, but this requires optical depth $\sim 10^{18}/n_e$ for iron.  
This would affect our limits on possible filling-in of the line by collisional or 
recombination radiation which we present in the next section.  Populating the upper level 
by collisions or recombination requires suitable temperature and ionization conditions.  
Efficient population by collisional excitation 
requires temperatures greater than can be accounted for 
by photoionization heating and radiative cooling, and this in turn will affect the ionization 
balance.  It also implies the existence of an additional heating mechanism.   
We have tested these possibilities quantitatively using direct fitting, 
and discuss the results below.

\section{Direct Fitting}

Many of the issues discussed in the previous section can be tested using 
direct fitting.  These include:  detector resolution, counting 
statistics, and scattered and thermal (either collisional or 
recombination) emission.  In order to do so, we use the {\sc warmabs} analytic model which 
interfaces with the {\sc xspec} spectral fitting package.  {\sc warmabs} makes use of a stored,
precomputed table of level populations for a family of photoionization models. It 
uses these to calculate a synthetic spectrum `on the fly' within {\sc xspec}.  The advantage
of this is that it calculates the synthetic spectrum automatically for the 
energy grid of the data such that it matches the energy resolution of the instrument.  Interpolation
can introduce significant numerical errors when applied to absorption spectra, 
since the absorption coefficient can change rapidly over a narrow range in energy.
{\sc Warmabs} calculates all absorption lines using a Voigt function including damping 
due to both radiative  and Auger decays where applicable.  Line profiles are calculated 
using sub-gridding, on an energy scale which is a fraction of a Doppler width, and 
the opacity and transmittivity then mapped onto the detector grid.  Bound-free absorption
is also included.  {\sc Warmabs} uses the full database and computational routines from the {\sc xstar}
package \citep{Kall01, Baut01}, and differs from {\sc xstar} in that the level populations 
are pre-calculated rather than calculated simultaneously with the spectrum.  The level populations
are calculated using a full `collisional-radiative' calculation, albeit with relatively simple model ions
in most cases.  Thus they include the effects of upper level depletion due to thermalization and photoionization
automatically.   We have 
extended the {\sc xstar} database to include all the trace elements seen in the GRO J1655-40 spectrum.
A description is provided in the Appendix.

{\sc Warmabs} does not calculate the flux transmitted by a fully self-consistent slab of 
gas, as {\sc xstar} does.  Rather, it calculates the opacity from the illuminated face of 
such a slab, and then calculates optical depth and transmitted flux analytically by assuming
the opacity is uniform throughout the slab.  A real slab will shield its interior, 
which will then have lower ionization and greater opacity than a slab whose opacity is 
assumed to be uniform.  Thus, the results of {\sc warmabs} fitting will be absorbing columns which 
may be greater than would be produced by a real slab.  The importance of this effect depends 
sensitively on the assumed slab structure, i.e. its geometrical thickness, and so 
its importance cannot be estimated in general.  Also, this effect is negligible
at high ionization parameters.  As we will show, our fits require large ionization parameters.
In this case, the advantages of the energy resolution provided by {\sc warmabs} outweigh
its limitations.

\subsection{Model 4: Narrow Lines}

Our basic fit assumes that the absorption is provided by a single component 
of photoionized gas.  We adopt the ionizing continuum used by \cite{Mill08} which
consists of a power law plus disk blackbody, as described in the previous section.  
The power law photon index is 3.54 and the disk inner temperature is kT=1.35 keV.  
In our fit we account for the curve of growth 
by adopting a small turbulent velocity, 50 km s$^{-1}$, so that the ratios of 
these lines are on the saturated part of the curve of growth.  
We then search for the single ionization parameter which  most nearly accounts for 
all the lines in the spectrum.  We do this using {\sc warmabs} and the {\sc xspec} package and stepping 
through ionization parameter in intervals $\Delta$log($\xi$)=0.1.  
{\sc Warmabs} accept the turbulent velocity as an input, but the actual line 
width is calculated including both this turbulent velocity and the thermal ion 
velocity corresponding to the equilibrium temperature.  As shown in the 
Appendix figures \ref{ionbala} -- \ref{ionbald}, in order to simultaneously  produce the ions of iron ranging 
from B-like (Fe XXII) to H-like, an ionization parameter in the range 3$\leq$log($\xi$)$\leq$4
is needed.  We find the best fit at log($\xi$)=4.0$^+_-$0.1, 
log(N)=23.8  and a blueshift for the absorber of 375 km s$^{-1}$.
This can be compared with the radial velocity of the system of 141$^+_-$1  km s$^{-1}$  and the 
orbital semi-amplitude of the secondary star of 215.5 $^+_-$2.4  km s$^{-1}$ \citep{Shah02}.
This suggests that the outflow is not moving fast when compared with the maximum
velocities characterizing the orbital motion in the system.  We have also performed equivalent 
fits for this set of assumptions using the {\sc isis} fitting package \citep{Houc00}, and have
verified that the results are independent of which fitting package is used.

We refer to this single component {\sc warmabs} with $v_{turb}=50 $ km s$^{-1}$ 
as model 4. We assume a gas density of 10$^{15}$ cm$^{-3}$ in calculating the level populations used 
by {\sc warmabs}.  We have not extensively tested our results at lower densities, and we point out 
that it is likely that densities as low as $10^{13.8}$ cm$^{-3}$ can produce the 
Fe XXII lines.  We employ the same ionizing continuum as derived in the model 1 fit when 
calculating the populations and gas temperature using {\sc xstar}.
In our fitting for model 4 and subsequent models we allow the normalizations 
of the two continuum components to vary.
The best fit flux is 1.8 $\times 10^{-8}$ erg cm$^{-2}$ s$^{-1}$ in the 2 -- 10 keV band.
In Figures \ref{fita}-\ref{fitn} we show the observed count rate spectrum (black) 
together with all the models discussed in this section as red curves, labeled 
according to the model number. 
Strong lines are marked in blue along with parent ion.

\begin{figure}[h]
\includegraphics*[angle=270, scale=0.6]{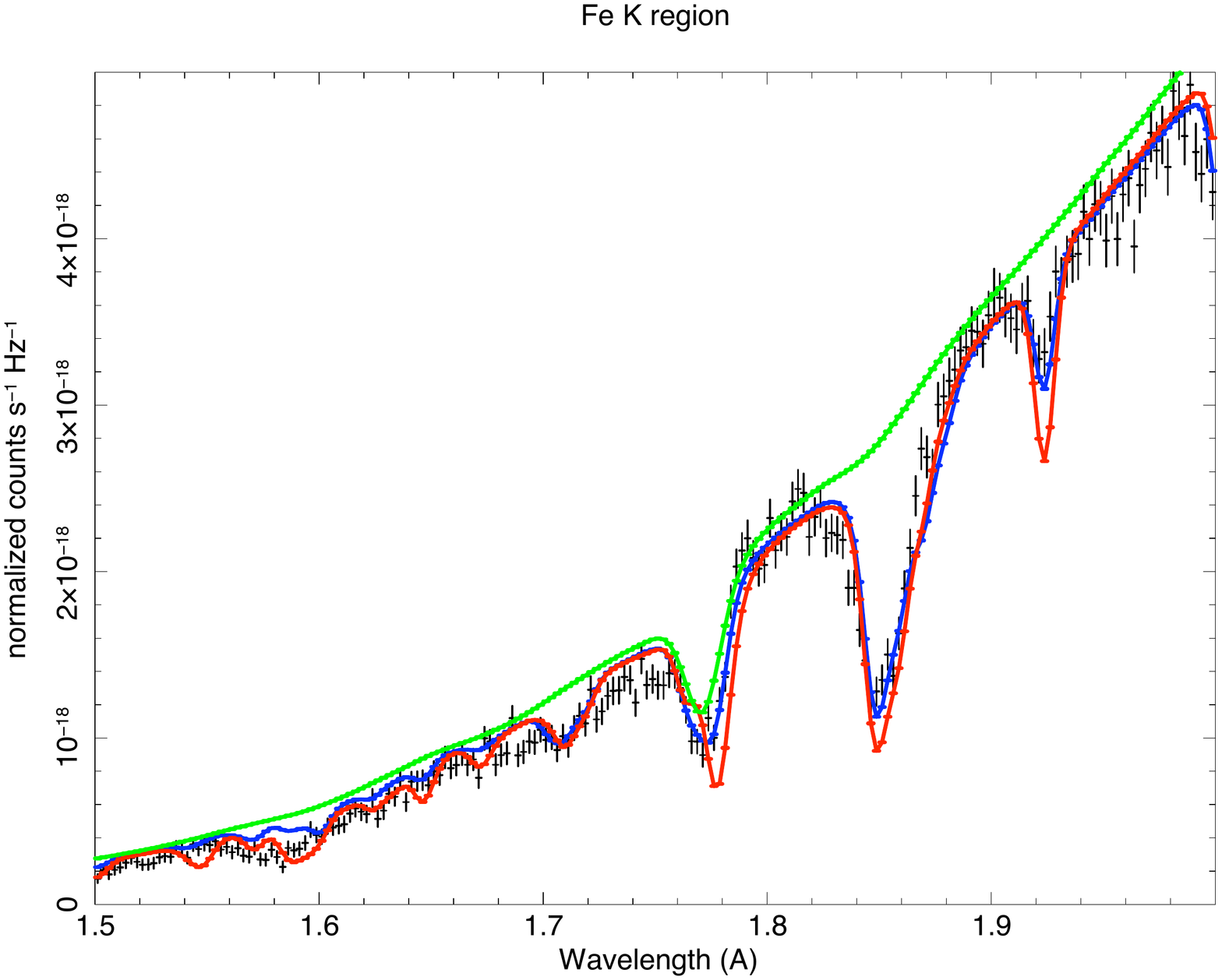}
\caption{\label{fitfeblowup}detail of spectrum in the 1.5 -- 2 \AA\ region.  Crosses are data, blue is model 4, red is model 5,
green is the Fe XXVI L$\alpha$ Gaussian contribution to model 4.}
\end{figure}

As discussed above, and shown in figure \ref{vofffig},
the majority of lines have centroid wavelengths which are consistent 
with a single Doppler blueshift, approximately 400 km s$^{-1}$ with respect to their laboratory values.  
The notable exception is the Fe XXVI L$\alpha$ line near 1.77 \AA\, which has a centroid wavelength corresponding
to a Doppler blueshift of $\simeq$1300 km s$^{-1}$.  In model 4, we adopt the hypothesis that 
this line arises in a separate velocity component of the flow, which must have a much 
greater ionization parameter so that it does not show up in the other lines.  Since the 
{\sc xstar} grid of model level populations does not extend to values where Fe XXVI is the only ion 
of significant abundance, we do this by adding a single Gaussian component for the high 
velocity part of this line.  A consequence of this is that the 375 km s$^{-1}$ component of the 
flow, which we model with {\sc warmabs} and an ionization parameter log($\xi$)=4,
accurately fits to the red part of the Fe XXVI L$\alpha$ profile, while the Gaussian 
accounts for the rest.  This is shown in figure \ref{fitfeblowup}, which shows the region surrounding 
the line.  The data is shown as the black bars, and model 4 is shown as the dark blue 
points.  The green points show the continuum + Fe XXVI L$\alpha$ Gaussian model 
alone.  The equivalent width of the Fe XXVI Gaussian is 29.6 eV.
The $\chi^2$ for this model is 34211 for 8187 degrees of freedom.
These parameters are summarized in Table \ref{chi2table}.

An alternative possibility is that the Doppler blueshift of the Fe XXVI line is affected by instrumental 
effects and shortcomings in the available HETG response function.  This could be due 
to the fact that the continuum count spectrum is steeply decreasing in the line region, as shown 
in figure \ref{fitfeblowup}. Internal scattering could cause photons from the continuum adjacent to 
the longer wavelength wing of the line to scatter into the line core, and this effect would 
be stronger on the long wavelength side than on the short wavelength side.  The width of the 
line spread function is comparable to the line blueshift.  In addition, the wavelength calibration 
may be affected by the use of continuous clocking mode.  On the other hand, other lines in 
the 1.5 -- 2 \AA\ region, although at slightly longer wavelength, do not show this effect, 
as exemplified by the Fe XXV $1s^2-1s2p$ line.  Furthermore, we have confirmed that the response matrix we use 
reproduces the line spread function obtained from the $Chandra$ calibration database 
\footnote{http://cxc.harvard.edu/caldb/calibration/gratings.html}.  So, although there is no obvious shortcoming
in the response matrix which would explain the discrepancy between the Fe XXVI L$\alpha$ wavelength shift 
and those of other lines, we will examine both hypotheses: that 
the high velocity part of the Fe XXVI line is associated with a separate kinematic component 
(as in  model 4), and that it is not.  The latter case also corresponds to the assumptions of \cite{Mill08}.
The primary conclusions of this paper, with respect to dynamics, abundances, turbulence, and geometry,
are not dependent on this assumption. 

The model 4 fit accurately reproduces the strength and shape of the Fe XXVI and Fe XXV lines, although we predict 
a slightly stronger red wing on the Fe XXV than is observed.  This is the contribution of Fe K$\alpha$ 
from lower ion stages of iron, Fe XXIV and XXIII.  The strengths of these lines provides a lower bound on the ionization
of iron.  We also slightly overestimate the strength of the Mn XXV L$\alpha$ line, relative to the Mn He-like K$\alpha$ line.
This may be due to an error in the ionization balance for this element; as shown in figure \ref{hheplot}, there are apparent 
departures from a simple monotonic trend in the H/He ratio  with nuclear charge.  
The region longward of 3\AA\ includes the lines  from Ca (3.02, 3.18 \AA) and Ar (3.73, 3.94 \AA).  
Our single component ionization balance is too low for both these elements, adequately accounting for the 
He-like lines but under-predicting the H-like lines.  We also point out that our assumed line 
width of 50 km s$^{-1}$ adequately fits the profiles of essentially all the lines shortward of 6 \AA.

Model 4 is based on a physical model for the absorber in GRO J1655-40, and it represents the best 
fit that we can obtain to the spectrum using a single {\sc warmabs} component.  As shown in figure \ref{fita}-\ref{fitn},
this accounts for the depths and positions of essentially all the lines in the spectrum.  Discrepancies
between the model and the data fall in three categories: errors in the continuum, errors in line widths, and errors
in line strengths (including lines which are missing).  In what follows we will explore these in turn.  As we will 
show, the best fits we obtain to a physical model have $\chi^2/\nu$ $\sim$ 2.9.  This can be compared with the 
phenomenological notch fit, model 3, which has reduced $\chi^2/\nu$ $\sim$ 2.2.  The difference likely 
represents the ability of the phenomenological notch fit to account for features which may not be physical lines
but rather part of the continuum or consequences of calibration uncertainties.  This, we think, 
represents the ultimate limit in our ability to fit the spectrum.   In what follows, we proceed 
and attempt to interpret the $\chi^2$ values from physical models when compared with each other, but without 
relying on standard interpretations of these values and their relation to probability of random occurrence, etc.
That is, we acknowledge that our fits are not acceptable based on these standard arguments about $\chi^2$, but 
nevertheless interpret confidence intervals on the fitting parameters by evaluating $\Delta\chi^2$ as if 
they were.

Model 4 is designed to fit to the curves of growth of lines from 
Fe XXIV and Ni XXVI by having a relatively small turbulent velocity, so that the lines 
from these ions are at least partially saturated.  This works for these ions, as is apparent 
from figure \ref{fitf}, and it also is consistent with the observed widths of the lines which are 
not resolved.  However, it does not account for the observed widths of the broader 
lines, such as Si XIII and Si XIV, and also for the 2s -- 3p lines of Fe XXIV near
10.6 \AA.  \cite{Mill08} adopt a line width of 300 km s$^{-1}$, which is a better fit to the 
observed widths for many lines.  We conclude that saturation with a single profile component, although
it accounts for the curves of growth of the lines of Fe XXIV, is not consistent with 
the curves of growth of lines from Si XIV, nor is it consistent with observed widths of some of the strong lines.
A possible explanation is that the lines consist of  multiple narrow 
components closely spaced in velocity so they appear blended together in the HETG, 
and where the components furthest from 
line center are unsaturated and so appear only in the low members of the Li-like 2s -- np series.
UV warm absorber lines in AGN appear to follow this behavior \citep{Gabe05}.

\subsection{Model 5: Two components plus broad lines}

Another way to fit the Fe XXVI L$\alpha$ line is to add a higher ionization parameter {\sc warmabs} component
to the fit.  We do this in model 5, which has a {\sc warmabs} component with the same ionization parameter
as model 4, but with the addition of a second component at an ionization parameter which can make 
the Fe XXVI  line sufficiently strong without over-producing the lower ionization lines.  
This model also has a turbulent velocity of 200 km s$^{-1}$.  We vary
the column densities and element abundances of both components in order to find the best 
fit.  In doing so, we force both components to have the same bulk outflow velocity, the same
turbulent velocity, and the same elemental abundances.  Thus, the two components are allowed 
to differ only in their ionization parameters and column densities.  This produces
a fit with $\chi^2/\nu=36281/8189$.  This fit is also shown in figures \ref{fita} -- \ref{fitn}.
In figure \ref{fitfeblowup} we show a blowup of the iron K region, comparing this model with 
model 4.  Model 4 is decomposed into the part accounted for by the Gaussian (green) and the total (blue).  
Model 5 is shown as the red points.  This shows that
in spite of the nearly equivalent $\chi^2$, model 5 does less well than model 4 in fitting the iron lines; it
generally over-predicts their strengths.  This is particularly true in the case of the Fe 
XXVI L$\alpha$  line.   Since this model has the same 375 km s$^{-1}$ outflow velocity as model 4, the best fit 
accounts for the blue edge of the Fe XXVI L$\alpha$ by over-predicting the line near 
line center and the red edge.   Thus, the conclusions from this model differ from model 4 (and subsequent
models) in that they do not depend on the existence of two kinematic components.

In other ways, the comparison between models 5 and 4 reflects the fact that model 5 has on average
a higher ionization parameter.  As a result, model 5 tends to over-predict the ratio of 
H-like to He-like lines, compared with both model 4 and with the observation.  Also, 
model 5 has a larger turbulent velocity, and therefore predicts a steeper curve 
of growth for most lines.  This fact is apparent from the Fe XXIV lines 
in the 6 -- 7 \AA\ region.  On the other hand, model 5 fits better than model 4 to lines 
which appear to follow the linear curve of growth, such as the L$\beta$/L$\alpha$ lines of 
Si XXIV, and also to lines which are detectably broadened, such as the 10.6 \AA\ doublet 
of Fe XXIV.

\subsection{Model 6: Partial Covering}

An additional possible reason for the apparent saturation of the curves of growth could be due to
systematic effects or calibration uncertainties associated with the $Chandra$ telescope, grating 
and detector.  This could lead to 
apparent scattering of photons in the continuum into the cores of absorption lines which is not 
accounted for by the detector response matrix, thus preventing
the residual flux in the lines from going to zero.  This effect cannot be evaluated accurately without
calibration data which includes narrow absorption line features, but we can get an 
indication of its plausibility by fitting to a model 
which includes some `leakage' of photons into lines from adjacent continuum regions.  We do this 
by fitting to a partial covering model in which the {\sc warmabs} component is partially diluted, i.e. 
$model=((1-C)+C \times {\sc warmabs}) \times continuum$ where $1-C$ is the fraction of scattered 
continuum at each energy.  We also adopt a turbulent velocity of 200 km s$^{-1}$ for this model, 
since we do not need to account for saturated curves of growth; the dilution of the {\sc warmabs} lines
tends to flatten the curves of growth.  
We show this as model 6 in Table \ref{chi2table} and in 
figures \ref{fita} -- \ref{fitn}.  This model has $\chi^2/\nu=24561/8184$,  
and so is the best of the fits using the {\sc warmabs} model.  
It also employs a turbulent velocity which is sufficient to account for the widths 
of resolved lines while at the same time fitting to the curves of 
growth for Fe XXIV and Ni XXVI.  

The best-fit value of $C$ is 0.37.  We can interpret this as being due to true partial 
covering in the object, i.e. if the continuum source is more extended than the absorber, or 
as being due to instrumental scattering which is not accounted for by the response matrix we used.  
The latter can be further subdivided into shortcomings in the response in accounting for the line spread
function (LSF) for the HETG which occur in the core region of the LSF, and shortcomings which occur in the wings.
The core region of the LSF \footnote{http://cxc.harvard.edu/proposer/POG/html/chap8.html\#fg:hetg-heglrf}.  
can be approxmiately represented by a Gaussian with full-width-half-maximum (FWHM) of $\simeq$1300 km s$^{-1}$ 
at the wavelength of the iron line, 1.77 \AA.  So two intrinsically narrow lines 
of equal intensity will fill the region between them to 30$\%$ of their peak intensities 
if they are separated by  $\simeq$ 900 km s$^{-1}$.    This is comparable to the typical line 
widths we find in our fits.  However, as discussed previously, the response matrix 
we use does accurately reproduce the LSF in the core region, so it would require a large error in the 
calibration files, the LSF and the resulting response matrix, in order to account for the filling in 
of the lines in this way.  

Another possibility is that there exists significant scattering in the line wings which is 
not accounted for by the calibration.  This is difficult to evaluate quantitatively, except to point out that the 
uncertainty in the LSF at the extremes of the wings appears to be at most $\sim$2 -- 4 $\%$ of the maximum value. 
In order to make up the covering fractions we require, the integrated area under the wings would
have to be $\sim$30 -- 40 $\%$ of the area in the core of the line.  The wings would have to 
extend to $\simeq$10 times the $\sigma$ of the core, or $\simeq$7600 km s$^{-1}$.    
We cannot evaluate the likelihood of this possibility reliably,
and cannot conclusively rule it out, although such a large departure from the calibration would 
be surprising.  This observation differs from typical observations in its 
use of continuous clocking mode.  This prevents the use of detector regions adjacent to the 
readout strip for background subtraction, but the counting rate in this case is high so 
that background should not be important at the levels considered here.  The fact that the 
response matrix agrees with the LSF in the core of the line is further evidence that 
the use of continuous clocking is not the source of the apparent partial covering.

We conclude that true partial covering in the source is more likely than instrumental 
effects, and thus plays a role
in determining the line curves of growth and the overall quality of the fit.  
Owing to its superior $\chi^2$ we adopt this model as the one which most 
nearly accounts for the observed spectrum and consider the physical assumptions
to be most nearly correct for the GRO J1655-40 outflow. 
We emphasize that the partial covering 
does not account for the high velocity component in Fe XXVI L$\alpha$, and model 6 includes
the same high velocity Gaussian contribution to this line as model 4.

\subsection{Other Models}

We also have examined the possible influence of thermal emission filling in lines 
on the one-component {\sc warmabs}  fit, i.e. model 4.  That is, we have included an additional 
emission component, which emits due to `thermal' (i.e. collisional and recombination) processes
rather than resonance scattering.  This component is calculated using the {\sc photemis} model, 
and is the emission analog of {\sc warmabs}.  It is assumed to have the same 
ionization parameter, abundances, redshift, and turbulent velocity as the warm absorber 
component.  We allow the normalization of the emission component, which is proportional 
to its emission measure, to vary, along with the optical depth of the warm absorber. 
In doing so, we find the best fit to be that with zero thermal emission, and 
the statistical upper limit on the thermal emission component corresponds to 
an emission measure $EM \leq 1.2 \times 10^{56}$ cm$^{-3}$.  
We will discuss the implications of this in the next section.

We have also examined the possibility that the absorption is produced in a plasma which 
is in coronal equilibrium instead of photoionized.  This is what might  be expected 
if the gas is heated mechanically, and if the outflow is due to thermal expansion 
of such a mechanically heated wind.   This is done using the {\sc xspec} analytic model 
{\sc hotabs}, which calculates the absorption spectrum of partially ionized gas if the 
ionization is due to electron impact.  The free parameter describing the ionization balance
in this gas is the electron temperature.   A key difference between a photoionized gas and 
a gas in coronal ionization equilibrium is that the ionization abundance distributions from 
photoionized gas have more overlap in parameter space than does a coronal gas.  That is, 
at a given temperature each element  in a coronal gas is most likely to exist in a pure ionization 
state, while at a given ionization parameter in a photoionized gas, each element is likely to
have a mixture of two or more ionization states.  In addition, the 
ions which can coexist in a coronal plasma all tend to have similar ionization potential.
 A consequence is that it is impossible for coronal equilibrium to allow the coexistence of
H-like or He-like ions of elements with very different nuclear charge, eg. Fe XXV and S XVI. 
In contrast, a photoionized plasma at log($\xi$)=4 does allow this.  For this reason, 
coronal equilibrium models do not fit the GRO J1655-40 spectra as well as photoionized models. 


\subsection{Abundances}

\begin{figure}[h]
\includegraphics*[angle=270, scale=0.6]{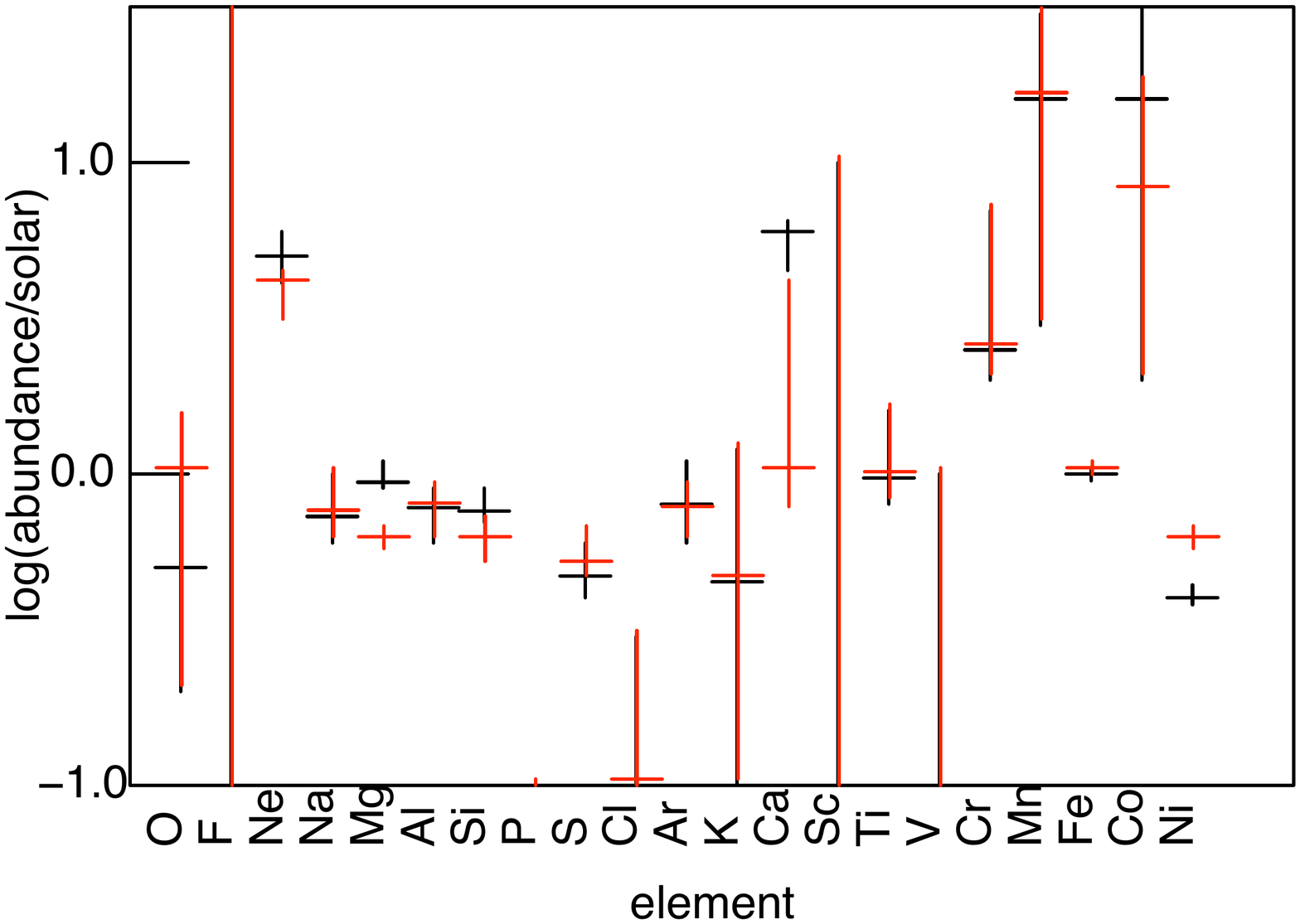}
\caption{\label{abundfig}Element abundances:  model 4(black) and model 6 (red).
log(abundance)=0 corresponds to solar \cite{Grev96} values.}
\end{figure}

We have also varied the elemental abundances and explored the limits for these allowed by the 
$\chi^2$ statistical criterion.  In doing this, we rely on the criterion of \citet{Cash79}, where
a 99$\%$ confidence interval is defined by the values of the parameter which 
fall within $\Delta\chi^2 \leq 10$ of the best fit value.  We display these results graphically 
in figure \ref{abundfig} for the elements O -- Ni.  
The abundances here are taken relative to the solar values of \citet{Grev96} for 
abundant elements, and \citet{Alle73} for the new elements added for this calculation.
In all of our models we fix the abundance of Fe at 1 relative to solar.  In Figure \ref{abundfig} the results 
of model 4 are shown in black, and the results of model 6 are shown in red.  This shows that 
the models agree on the abundances of elements 
Sc through Co, such that  the abundances of Cr, Mn and Co are all enhanced relative to Fe by 
at least 50 $\%$.  
For Ca  model 4  predicts values greater than solar by 0.5-1 dex while model 
6 is consistent with solar.  
As discussed above, we conclude that model 6 most nearly 
fits the overall properties of the spectrum, and in what follows we discuss 
the implications of the abundance pattern from this model. 
Limits on the abundance of O come from the O VIII 1s -- 5p line at 14.8 \AA\ and 1s -- 6p at
14.64 \AA.  This results in $Y_O$ values in the range 0.2 -- 1.5.

The abundance patterns shown in figure \ref{abundfig} are crudely consistent between models
4 and 6, and show enhanced abundances of Cr and Mn relative to Fe.  Results for Ca-V are ambiguous,
and suggest no strong enhancement, while the abundances of Na-Cl are smaller than the 
solar ratio, relative to Fe.  

\begin{figure}
\includegraphics[angle=0.,scale=0.6]{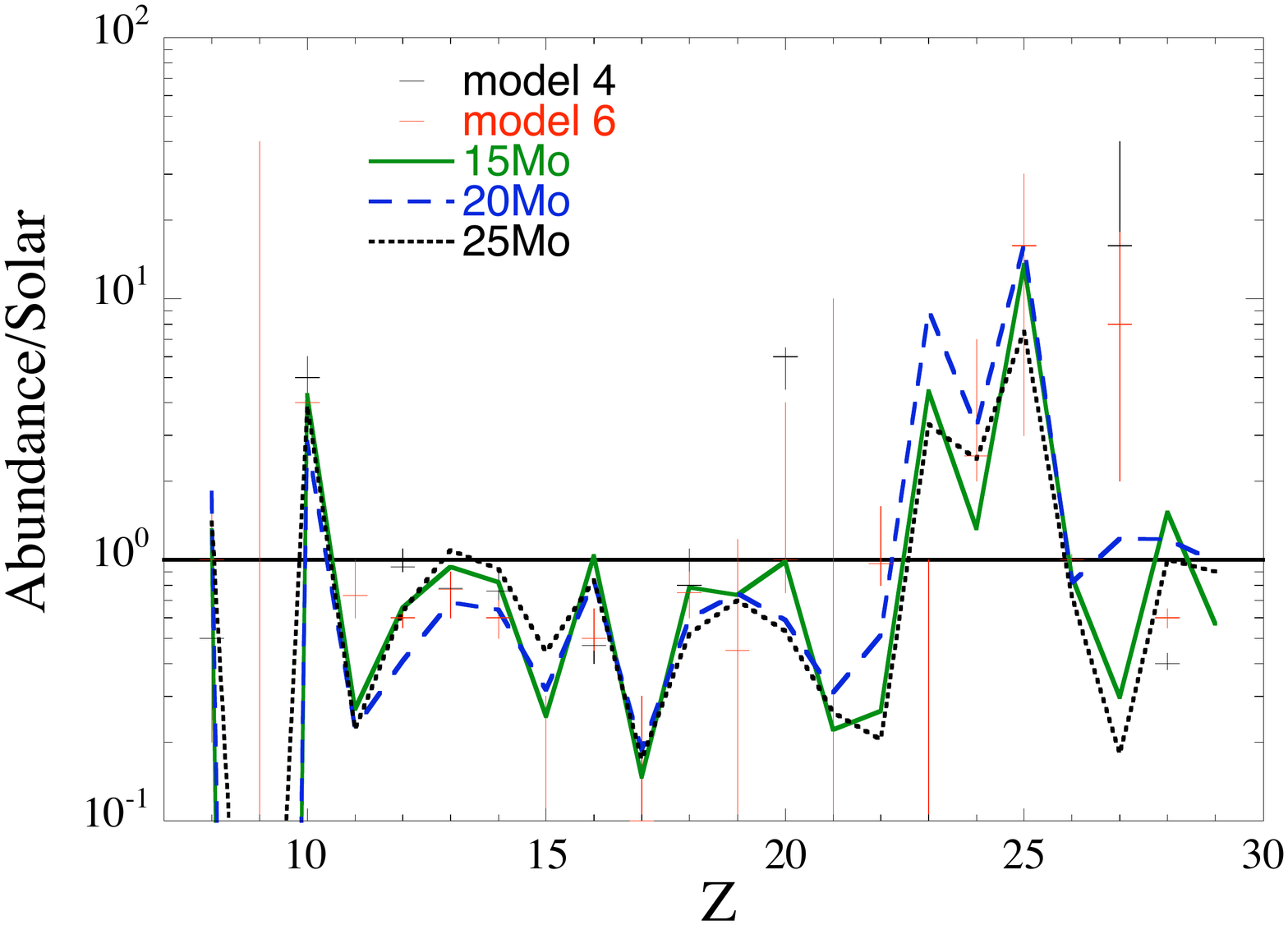}
\caption{Comparison of observed and predicted overabundances. The symbols with errors bars correspond to the 1-component (black) and 2-component (red) models. The lines correspond to model calculations in a 15~M$_{\odot}$ (solid), 20~M$_{\odot}$ (dashed) and 25~M$_{\odot}$ (dotted) star, as described in the text. }
\label{fig_abth1}
\end{figure}

The overabundances of Fe-group elements strongly suggest that the observed 
matter has been subject to high-temperature burning conditions. For this 
reason, we have compared the observed abundances against nucleosynthesis 
models in massive stars, and more particularly resulting from the hydrostatic 
C-, O- and Si-burning phases. We use the presupernova nucleosynthesis yields 
calculated in model stars of 15, 20, and 25 solar masses  with solar metallicity 
\citep{Limo00}. Each of the C-, O- or Si-burning zone has been weighted by 
the relative masses needed to best reproduce the observations. The relative 
mass ratio (C:O:Si) obtained is (6:1:1) for the three model stars. As shown 
in Fig. ~\ref{fig_abth1}, the predicted abundances (relative to solar) are 
in rather close agreement with the observed pattern for models 4 and 6. 

As far as the light elements are concerned, the agreement between theory 
and observation is rather good. The O overabundance predicted reaches 1.3 
(1.8 for the 20 M$_{\odot}$ star) in agreement with the 1.5 upper limit determined 
in the 2-component model. Note, however, that the O abundance is not well 
constrained by the observed spectrum owing to strong interstellar absorption. 
Furthermore, we consider it possible that the low $Z$ elements are affected 
by a separate ionizing continuum component. This could suppress 
absorption from O even if the abundance were greater than solar, although 
this is contrived given the absence of observations of such radiation. 
We also point out  the possibility that the Ne and O lines are at least 
partly of interstellar origin (e.g. \citet{Juet04}).
This would require a coincidence in velocity between the GRO J1655 outflow 
and the intervening gas.
It would also place more severe constraints on the 
nucleosynthetic models, since it would decrease the inferred abundances 
of these elements intrinsic to the source.

Concerning 
the underproduction of Na, the disagreement may be due to uncertainties affecting 
the nuclear reaction rates or the thermodynamical conditions in the combustion 
zones.  For the heavier elements above Ca, the agreement is still good though 
discrepancies can be seen in particular for Ti, V, and Co. It should be mentioned 
here that, in the nucleosynthesis simulations, all unstable nuclei produced 
have been assumed to $\beta$-decay (except the long-lived $^{53}$Mn).


It is interesting to compare with the results of the optical abundance determination by \citet{Isra99}.
These authors find evidence for enhanced O/H, Mg/H, Si/H, S/H relative to solar, but Fe/H and 
Cr/H are approximately solar.  This differs from our results for 
models 4 and 6, insofar as they can 
be compared, since \citet{Isra99} do not measure abundances for Mn, Co, and Ni.  The conclusions 
of \citet{Isra99} have not been confirmed in an investigation by \cite{Foel07}.  
This also underscores the fact that the abundances we measure 
are relative to each other, and all are for elements heavier than O.  We have 
assumed that iron is solar and quote other abundances relative to that, but 
we have no constraints on abundances for light elements, or for any abundances
relative to hydrogen.

\section{Discussion}

Other implications of our models include the fact that the curves of growth, in the absence of 
partial covering, indicate a velocity structure which has a  small turbulent velocity, $\simeq$50 km s$^{-1}$ 
at the same time as a larger bulk velocity, $\simeq$400 km s$^{-1}$ .  Since the radial velocity 
of the GROJ1655-40 system is  141$^+_-$1  km s$^{-1}$, this implies an outflow consisting 
of material which is cold or still compared with the bulk flow.  This is unusual when compared with flows which 
are well studied in stars and non-thermal outflows from compact objects.

All of our models fit to the iron K region by assuming the existence of a separate, 
higher ionization component which is primarily responsible for producing the Fe XXVI L$\alpha$ line.  
This is because essentially all the other lines in the spectrum are consistent with a single ionization 
parameter and outflow velocity.  Models 4 and 6 fit the Fe XXVI L$\alpha$ line using an ad hoc 
Gaussian, while model 5 fits it using a separate {\sc warmabs} component.  Although the latter approach 
is more physically consistent, it does not fit the feature as well because for this model we force the 
two components to have the same outflow velocity.

In either case, we find that the majority of lines fit to a single ionization component, 
log($\xi$)$\simeq$4.  In addition to this, we have the observed luminosity $L\simeq 5 \times 10^{37}$ erg s$^{-1}$
\citep{Mill08}  and density constraints from the Fe XXII metastable line implying $n \geq 10^{14} {\rm cm}^{-3}$.  Taken together, these determine the location of the absorber:  
$R=\sqrt{L/n/\xi} \simeq 10^9 L_{37}^{0.5} n_{15}^{-0.5} \xi_4^{-0.5}$ 
cm, and its size:  $\Delta R=N/n \simeq 10^9 N_{24} n_{15}^{-1}$ cm.  In these 
equations $L_{37}$ is the 
luminosity in units of $10^{37}$ erg s$^{-1}$,  $n_{15}$ is the density in units of 10$^{15}$ cm$^{-3}$
and $\xi_4=\xi/10^4$.  
That is, for plausible values, $R \leq 7 \times 10^9$ cm, $\Delta R \leq 10^{10}$ cm.
This location can be compared with the `Compton radius' within which photoionized 
gas cannot escape the gravity of the black hole \citep{Bege83}, which is 
$R_{IC}=10^{10} (M/M_\odot) T_{IC8}^{-1}$, where $T_{IC8}$ is the temperature of the photoionized 
gas in units of 10$^8$K, and for the GRO J1655-40 spectrum this value is $\leq$0.03 
and the mass of the black hole is  5 -- 8 $M_\odot$. 
If so, $R_{IC} \simeq  2 \times 10^{12}$  cm.
Thus the inferred wind location is well within the Compton radius
and is inconsistent with an outflow driven by thermal expansion,
even though weak flows are possible at $\sim 0.1 R_{IC}$ as stated by  \cite{Wood96}.  
This conclusion  is consistent with  that of \cite{Mill08}.   

On the other hand, we note that the virial radius corresponding
to the outflow speed we measure is 1.3 $\times 10^{12}$ cm, considerably greater than the position 
we infer from the X-ray ionization balance.  
This, together with the fact that the outflow speed is comparable to the orbital speed 
in the system raise the possibility that the outflow could be associated with the secondary star or a region 
of the binary which is at comparable distance from the black hole.
If so, then the radius we infer from the ionization balance arguments is an underestimate.
This is difficult to understand in view of the density constraints from the Fe XXII lines; 
it would require that the metastable level of Fe XXII were populated by radiative pumping 
rather than by collisions.   This, in turn, would require that the intensity of the radiation
in the $\sim 100 \AA$ region to be close to LTE, and there is no known source for such radiation
far from the black hole.  In addition, the observed outflow speed does not appear to vary during the 
observation \citep{Mill08}, which spanned $\simeq$20$\%$ of an orbital period, which argues against an origin 
associated with the companion star or accretion stream. Thus, we consider this possibility to be unlikely.

Another astrophysical system which appears to show a comparable contrast  between line turbulent width 
and outflow velocity are  FU Orionis stars.  In these systems the terminal
velocity of the wind is 300-400 km/s, and the rotational broadening (seen
in absorption) of the wind is about 50 km/s \citep{Calv93,Hart95}.  The intrinsic turbulent velocity 
needed to produce the lines is likely much smaller.  If so, the 
contrast between outflow and turbulent velocities is comparable to 
the contrast between the observed width and the virial speed at the inferred 
position for the GRO J1655-40 outflow. 

We note also that the flow timescale is 
$t_{dyn}=R/v_{out}\simeq 27$s.  This can be compared with the recombination timescale 
$t_{rec}=(n \alpha)^{-1} \simeq 10^{-4}$s, showing that the assumption of ionization equilibrium
is valid for these simple assumptions about the density and location of the outflow.  This 
does not change qualitatively if the absorber is located at the virial radius.

We can also explore the implications of the emission measure upper limit 
derived in the previous section, $EM \leq 1.2 \times 10^{56}$ cm$^{-3}$.   The emission measure
expected from a constant density shell with size derived from the ionization parameter, density and 
thickness is $EM=10^{57} \Omega L_{37} N_{24} \xi_4^{-1} {\rm cm}^{-3}$, where
$\Omega$ is the solid angle of the warm absorber/emitter.  So we infer 
$\Omega \leq 0.12 L_{37}^{-1} N_{24}^{-1} \xi_4$ or $\Omega\leq 0.024$ for 
the most likely value of $L_{37}$.  This can be compared with the constraint derived 
by \cite{Mill06} which is $\Omega\leq 1.4$ based on the observed limit on the Fe XXIV 2p -- 3s 
emission line at 11.43 \AA.  We can also then estimate the 
mass loss rate in the outflow:  
\mdot$= \Omega R^2 n m_H v_{out} = 2 \times 10^{15} N_{24}^{-1} v_7$ gm s$^{-1}$
where $v_7$ is the outflow speed in units of 100 km s$^{-1}$; so 
\mdot$ \simeq 8 \times 10^{15}$ gm s$^{-1}$ for our fits.  This is small compared 
with the mass accretion rate required to fuel the X-ray source, 
\mdot$_{acc} \simeq L/(\eta c^2)\simeq 6 \times 10^{17}$ gm $^{-1}$.
The \cite{Mill06} limit on the solid angle allows a considerably larger mass 
loss rate, $\leq$10$\%$ of the accretion rate.

Another conclusion of our work is the existence of two velocity components, one with 
ionization parameter log($\xi$)=4 and speed $v$=375 km s$^{-1}$ relative to the Earth, 
and the other with higher speed and much greater ionization parameter.
We find for this second component a speed $\simeq$1400 km s$^{-1}$, and infer that the 
ionization parameter be high enough that Fe XXVI and Ni XXVIII are the only ions with 
appreciable abundance, say log($\xi)\geq 6$.  Then the mass flux is 
\mdot= $10^{17} \xi_{6}^{-1} v_8 \frac{\Omega}{4 \pi}$ gm s$^{-1}$, where we have assumed
a source luminosity 5 $\times 10^{37}$ erg s$^{-1}$ and $v_8$ is the speed in units of 
1000 km s$^{-1}$.  It is more difficult to constrain the fractional solid angle of this 
component than it is for the lower ionization component because any emission may be masked 
by the lower velocity component, and we have not attempted to test this.  It is clear that
this component can carry more mass than the low ionization component if its solid 
angle exceeds 0.02.  However, the radial location  of this component cannot 
be reliably established, since the features in the spectrum which provide density 
constraints do not share the outflow velocity of this component.

The results found here agree qualitatively with those of \cite{Mill08} 
in the sense that the ionization parameter and density of the flow containing most 
of the lines imply a wind location which is too close to the black hole to be
easily explained by a driving thermal mechanism.  We favor a scenario in which 
there are two components to the flow, and the lower ionization component is somewhat 
less ionized and the higher ionization component is more ionized than the 
single component found by \cite{Mill08}.  Our spectral fitting allows us to constrain
elemental abundances, and we find enhanced abundances for several elements in the iron 
peak.  We do not find a pattern of systematic enhancements for $\alpha$-capture elements.

\acknowledgements This work was funded in part by a grant from the Chandra theory program.
We thank the referee, Frits Paerels, for many constructive comments.  

\appendix

\section{Appendix:  Photoionization Models}

In order to model the spectrum of GRO J1655-40 we have made modifications and 
enhancements  of the  {\sc xstar} \footnote{http://heasarc.gsfc.nasa.gov/docs/software/xstar/xstar.html}
photoionization code
\citep{Kall01}.  This  
code calculates the transfer of X-rays and other ionizing radiation, and 
the ionization balance, opacity and reprocessed emission from gas under 
a variety of physical conditions.  It contains a relatively complete and up-to-date 
collection of atomic cross sections, rate coefficients and atomic energy levels. 
The code and atomic database, along with the `{\sc warmabs/photemis}' analytic models for {\sc xspec},
are freely available, distributed 
as part of the ftools package, and have been widely used in interpreting X-ray spectra.

The standard {\sc xstar} distribution includes all abundant even-$Z$ elements (plus N), i.e.:
H, He, C, N, O, Ne, Mg, Si, S, Ar, Ca, Fe and Ni.  In order to model GRO J1655-40 
we have added all the other elements with $Z \leq 30$, with the exception of Li, Be, and B.
The atomic data needed to do this, in addition to the resulting ionization balance for 
an optically thin gas, are discussed in this appendix.  
These data  include: energy levels, line wavelengths, oscillator
strengths, photoionization cross sections, recombination rate coefficients,
and electron impact excitation and ionization rates.
Distorted wave dielectronic and radiative recombination rate coefficients,
photoionization cross sections, and photoexcitation-autoionization rates
have been calculated  \citep{Badn03} for ions of these elements
belonging to the H-like through Na-like isoelectronic sequences and are
also available in the ORNL ADAS 
database \citep{Schu00}. Concerning the other
atomic parameters, a survey of atomic databases  \citep{Kall07}
reveals that they are far from complete for any of these elements.
Therefore we rely heavily on hydrogenic scaling 
for many of these quantities.

In order to model the spectrum we have modified the {\sc xspec} analytic model {\sc warmabs} 
in order to account for the new elements.  {\sc warmabs} uses the optically thin ionization balance
calculated by {\sc xstar}, along with the same atomic database as {\sc xstar}, to calculate
the synthetic spectrum expected for a gas seen in transmission with a given ionization parameter,
column density, turbulent velocity and abundance set.  This takes into account all of 
the absorption processes, i.e. lines and photoionization cross sections,  included in {\sc xstar}.
A complementary model, {\sc photemis}, calculates the emission spectrum expected from such a gas, 
but that is not used in the calculations presented here.

Rates and atomic structure for elements H, He, C, N, O, Ne, Mg, Si, S, Ar, Ca, Fe and Ni are the 
same as those described in \citep{Kall01, Baut01}, and available in version 2.1l of the {\sc xstar} code
(http://heasarc.gsfc.nasa.gov/docs/software/xstar/xstar.html).  For the odd-$Z$ elements, and the iron 
peak elements not included in the release version of {\sc xstar}, we use atomic structure scaled according 
to the following prescription:  for the H- and He-like isoelectronic sequences we scale 
all energies and rates according to that expected for hydrogenic quantities:  energies scale 
according to $Z^2$, photoionization cross sections scale according to $Z^{-2}$,  transition 
probabilities scale according to $Z^4$.  For these isosequences we scale the full atomic structure and 
rate set from {\sc xstar}, including all levels through $n=6$ for hydrogenic and through $n=5$ for helium-like ions.
For other isoelectronic sequences we adopt hydrogenic scaling of a highly simplified hydrogen-like 
ion, consisting of two spectroscopic levels, 1s and 2p, along with a super-level and continuum.  
The energies, transition probabilities, and other rates are scaled according to a hydrogenic prescription
with an effective $Z=\sqrt{E_{th}/13.6{\rm eV}}$, where $E_{th}$ is the first ionization potential
taken from \cite{Alle73}.  Following this, we correct the energies of the n=2 levels using the evaluated 
wavelengths of the 2 -- 1 transitions from the NIST database.  These are available for 
F VIII, Al XII, Al XIII, Sc XX, Sc XXI, Ti XXI, Ti XXII, V XXII, V XXIII, Cr XXIII, Cr XXIV, 
Mn XXIV, Mn XXV, Co XXVI, Co XXVII, Cu XXVIII, Cu XXIX. 
We do not have evaluated wavelengths for the K lines of Na, P, Cl, and K.

For ions not previously included in {\sc xstar}, ground state photoionization cross sections are taken from the 
calculations of \cite{Vern95}. 

Recombination rate coefficients were taken from the calculations reported 
in the series of the papers by Badnell and coworkers \citep{Badn03} and available from the 
website \footnote{http://amdpp.phys.strath.ac.uk/tamoc/DATA/}.  These are available for all elements He -- Zn and 
for isoelectronic sequences H -- Mg-like for both dielectronic (DR) and radiative
(RR) recombination.  For isoelectronic sequences Al-like through Ni-like, we make 
use of rates from \cite{Aldr73, Arna92, Shul82}, which cover ions of 
Si, S, Ar, Ca, Fe, and Ni in these isosequences.  For ions of other elements in these isosequences
we interpolate along isoelectronic sequence.   {\sc xstar} does not directly use 
the total recombination rate, but rather calculates rates onto a set of 
spectroscopic levels, typically with principle quantum numbers $n\leq$6, using photoionization 
cross sections and the Milne relation.  Then it calculates a photoionization cross section for
one or more fictitious superlevels such that the total recombination rate for the 
superlevel(s) plus the spectroscopic levels adds to the desired total rate taken 
from one of the above compilations.  The super levels are generally assumed to decay 
directly to ground without the emission of any observable cascade radiation.  The
exception is the decay of the H- and He- isoelectronic sequences, for which we have explicitly calculated
the decay of the superlevels to the spectroscopic levels using a full cascade calculation
\citep{Kall01, Baut01}.  

\subsection{Ionization Balance}

A key input affecting the ionization balance is the spectral energy distribution (SED)
assumed for the source.  We take the spectrum adopted by \citet{Mill08}, consisting of 
a disk black body with inner temperature kT=1.35 keV, plus a power law with photon index 
$\Gamma=3.54$.  The normalizations are chosen such that the low-energy cross-over between the 
two components is at 1 keV.  Below this energy we assume that the spectrum is flat (in photons).
We assume that the spectrum incident on the gas in the GROJ1655 system is not affected by 
interstellar absorption.  This spectrum is much steeper in the X-ray band than the conventional 
$\Gamma\simeq$2 power law applied to AGN.  The ionization balance is correspondingly quite different
in the sense that the mean ionization is lower at a given value of the ionization parameter
in the steep spectrum case.

Our choice of spectrum is deficient in photons between $\sim$1 and 100 Ry, relative to higher energies.
This leads the gas to be thermally unstable, as originally described by \cite{Buff74}.  This manifests 
itself as a region of ionization parameter where the temperature and ionization balance of the gas 
can be multi-valued.  In the case of the spectrum we have chosen, this occurs at an ionization parameter 
log($\xi)\simeq$ 2.  This is less than the range of ionization parameters which can produce 
the ions observed from GRO J1655, and so it will not affect our results.  In what follows we will 
not discuss this further.

\begin{figure}[p]
\includegraphics*[angle=270, scale=0.5]{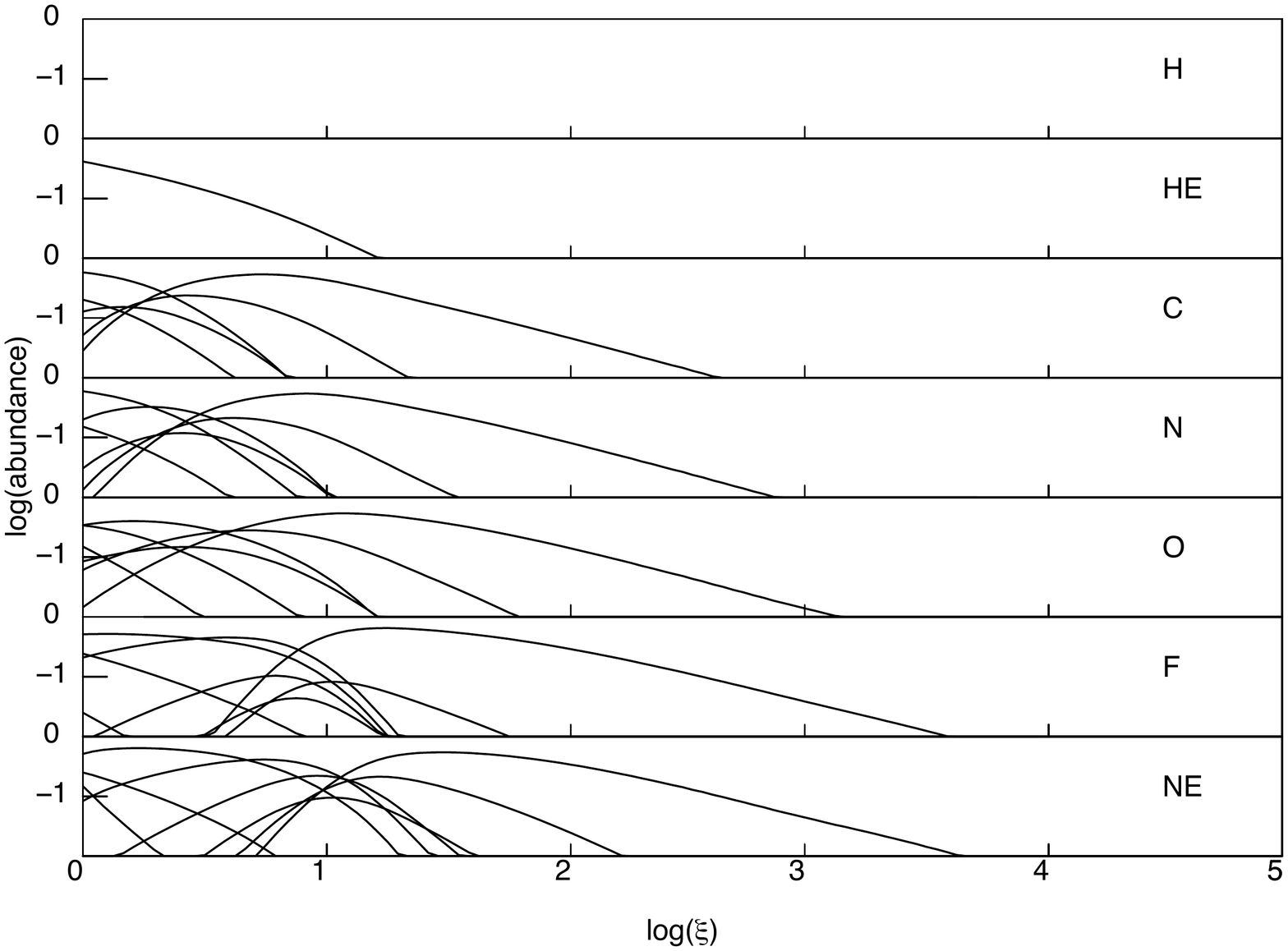}
\caption{\label{ionbala}ionization balance: H-Ne}
\end{figure}

\begin{figure}[p]
\includegraphics*[angle=270, scale=0.5]{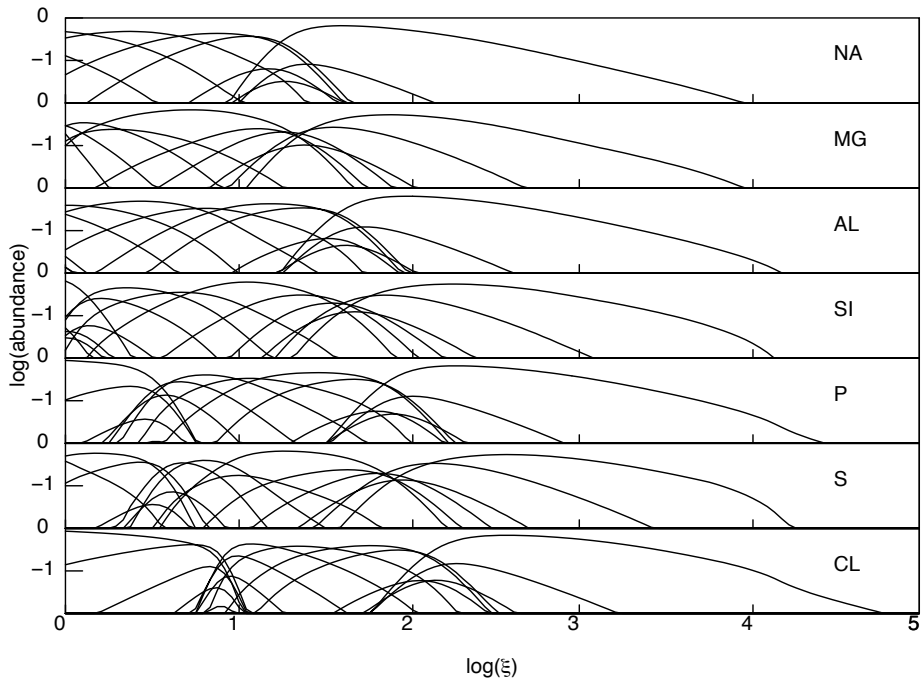}
\caption{\label{ionbalb}ionization balance: Na-Cl}
\end{figure}

\begin{figure}[p]
\includegraphics*[angle=270, scale=0.5]{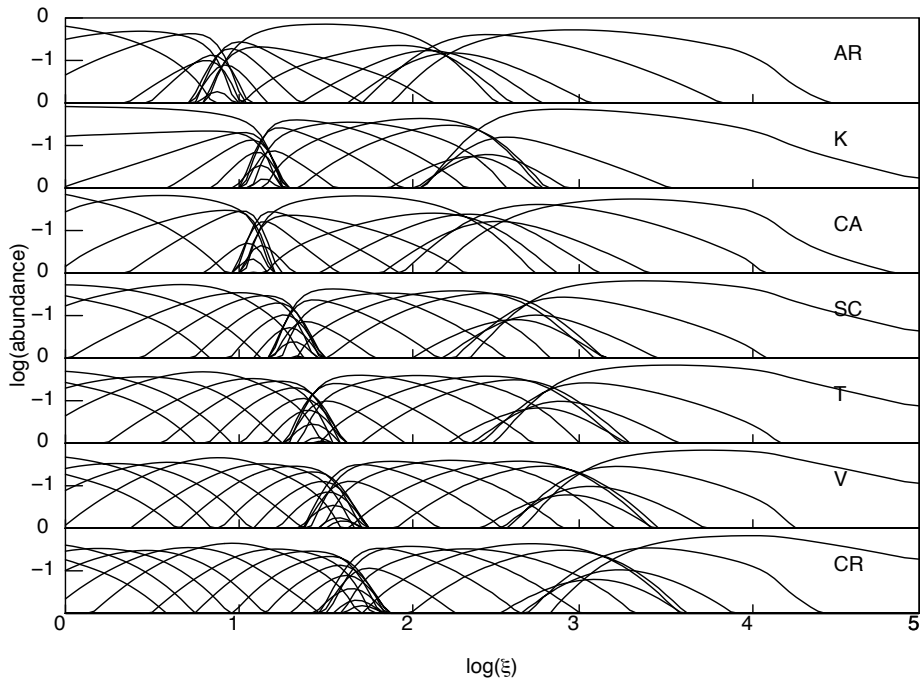}
\caption{\label{ionbalc}ionization balance: Ar-Cr}
\end{figure}

\begin{figure}[p]
\includegraphics*[angle=270, scale=0.5]{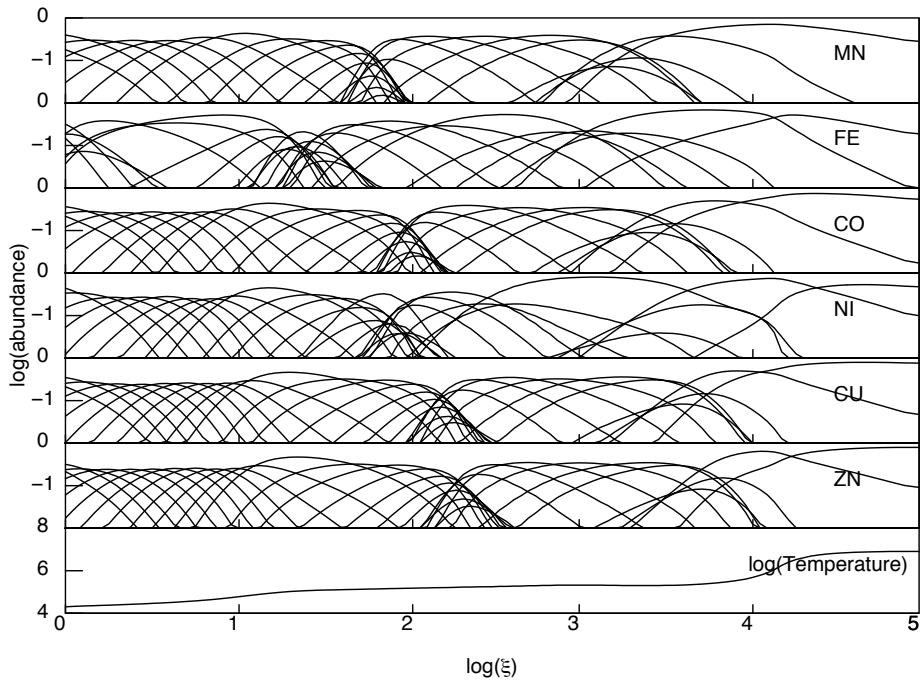}
\caption{\label{ionbald}ionization balance: Mn-Zn, and temperature.}
\end{figure}

Figures \ref{ionbala}-\ref{ionbald} show the ionization balance for our choice of incident spectrum, 
as a function of ionization parameter $\xi=L/(nR^2)$. In our case, we adopt cgs units, 
so $\xi$ has units erg cm s$^{-1}$. It is clear from this figure that the dominant observed ions
require ionization parameters log($\xi)\geq$4, but that production of Li-like and B-like iron and Ni 
requires ionization parameter values at the low end of this range.

\clearpage





\begin{thebibliography}{}

\bibitem[Aldrovandi and Pequignot(1973)]{Aldr73} Aldrovandi, S.~M.~V., \& Pequignot, D.\ 1973, Ast. Ap., 25, 137 

\bibitem[Allen(1973)]{Alle73} Allen, C.~W.\ 1973, London: University of London, Athlone Press, |c1973, 3rd ed.,  

\bibitem[Arnaud \& Raymond(1992)]{Arna92} Arnaud, M., \& Raymond, J.\ 1992, \apj, 398, 394 

\bibitem[Badnell et al.(2003)]{Badn03}Badnell, N. R.,  O'Mullane, M. G.,  Summers, H. P., Altun,  Z.,  Bautista, M. A., 
Colgan, J.,  Gorczyca, T. W.,  Mitnik, D. M.,  Pindzola, M. S. and Zatsarinny,  O., 2003, Astron. Astrophys. 406 1151-65

\bibitem[Badnell et al.(2005)]{Badn05}Badnell, N.,  2005, http://amdpp.phys.strath.ac.uk/tamoc/DATA/

\bibitem[Baluci{\'n}ska-Church(2001)]{Balu01} Baluci{\'n}ska-Church, M.\ 2001, Advances in Space Research, 28, 349 

\bibitem[Bautista and Kallman(2001)]{Baut01}Bautista, M., and Kallman, T., 2001 Ap. J. Supp. 134, 139

\bibitem[Begelman et al.(1983)]{Bege83} Begelman, M.~C., McKee, C.~F., \& Shields, G.~A.\ 1983, \apj, 271, 70

\bibitem[Behar et al.(2003)]{Beha03}Behar, E., et al., 2003, ApJ, 598, 232 

\bibitem[Buff \& McCray(1974)]{Buff74} Buff, J., \& McCray, R.\ 1974, \apj, 189, 147 

\bibitem[Calvet et al.(1993)]{Calv93} Calvet, N., Hartmann, L., \& Kenyon, S.~J.\ 1993, \apj, 402, 623 

\bibitem[Cash(1979)]{Cash79} Cash, W.\ 1979, \apj, 228, 939 

\bibitem[D{\'{\i}}az Trigo et al.(2006)]{Diaz06} D{\'{\i}}az Trigo, M., Parmar, A.~N., Boirin, L., M{\'e}ndez, M., \& Kaastra, J.~S.\ 2006, \aap, 445, 179 

\bibitem[Foellmi et al.(2007)]{Foel07} Foellmi, C., Dall, T.~H., \& Depagne, E.\ 2007, \aap, 464, L61

\bibitem[Gabel et al.(2005)]{Gabe05} Gabel, J.~R., et al.\ 2005, \apj, 623, 85

\bibitem[Grevesse et al.(1996)]{Grev96} Grevesse, N., Noels, A., \& Sauval, A.~J.\ 1996, Cosmic Abundances, 99, 117

\bibitem[Halpern(1984)]{Halp84}Halpern, J., 1984 Ap J 281 90.

\bibitem[Harmon et al.(1995)]{Harm95} Harmon, B.~A., et al.\ 1995, \nat, 374, 703 

\bibitem[Hartmann \& Calvet(1995)]{Hart95} Hartmann, L., \& Calvet, N.\ 1995, \aj, 109, 1846 

\bibitem[Hjellming \& Rupen(1995)]{Hjel95} Hjellming, R.~M., \& Rupen, M.~P.\ 1995, \nat, 375, 464 

\bibitem[Houck \& Denicola(2000)]{Houc00} Houck, J.~C., \& Denicola, L.~A.\ 2000, Astronomical Data Analysis Software and Systems IX, 216, 591

\bibitem[Israelian et al.(1999)]{Isra99} Israelian, G.,  et al.,\ 1999, \nat, 401, 142 

\bibitem[Juett et al.(2004)]{Juet04} Juett, A.~M., Schulz, N.~S., \& Chakrabarty, D.\ 2004, \apj, 612, 308

\bibitem[Kallman and Bautista(2001)]{Kall01}Kallman, T. R. and Bautista, M., 2001 Ap. J. Supp 133,221

\bibitem[Kallman and Palmeri(2007)]{Kall07}Jackman, T.R., and Palmeri, P.,  2007 Rev. Mod. Phys. 79, 79.

\bibitem[Kaspi et al.(2002)]{Kasp02}Kaspi, S., et al., 2002, Ap. J., 574, 643

\bibitem[Limongi et al(2000)]{Limo00} Limongi, M., Straniero, O., \& Chieffi, A.\ 2000, \apjs, 129, 625 

\bibitem[Masai \& Ishida(2004)]{Masa04} Masai, K., \& Ishida, M.\ 2004, \apj, 607, 76 

\bibitem[Miller et al.(2006)]{Mill06} Miller, J.~M., Raymond, J., Fabian, A., Steeghs, D., Homan, J., Reynolds, C., van der Klis, M., \& Wijnands, R.\ 2006, \nat, 441, 953 

\bibitem[Miller et al.(2008)]{Mill08} Miller, J.~M., Raymond, J., Reynolds, C.~S., Fabian, A.~C., Kallman, T.~R., \& Homan, J.\ 2008, \apj, 680, 1359

\bibitem[Netzer(2006)]{Netz06} Netzer, H.\ 2006, \apjl, 652, L117

\bibitem[Orosz \& Bailyn(1997)]{Oros97} Orosz, J.~A., \& Bailyn, C.~D.\ 1997, \apj, 477, 876 


\bibitem[Shabhaz et al.(2002)]{Shah02} Shahbas, T., van der Hooft, F., Casares, J., Charles, P. A., and van Paradijs, J., MNRAS 306, 89

\bibitem[Schultz, et al.(2000)]{Schu00}Schultz, D., et al.,  2000, http://www-cfadc.phy.ornl.gov/xbeam/xbmintro.html

\bibitem[Shirai et al.(2000)]{Shir00}Shirai, T., Sugar, J., Musgrove, A.,2000 J. Phys. Chem. Ref. Data, Monograph No. 8

\bibitem[Shull and Van Steenberg(1982)]{Shul82} Shull, J.~M., \& van Steenberg, M.\ 1982, \apjs, 48, 95 


\bibitem[Strohmayer(2001)]{Stro01} Strohmayer, T.~E.\ 2001, \apjl, 552, L49 

\bibitem[Tarter et al.(1969)]{Tart69} Tarter, C.~B., Tucker, W.~H., \& Salpeter, E.~E.\ 1969, \apj, 156, 943 


\bibitem[Ueda et al.(1998)]{Ueda98} Ueda, Y., et al., \ 1998, 
\apj, 492, 782 

\bibitem[Ueda et al.(2001)]{Ueda01} Ueda, Y., et al., \ 2001, \apjl, 556, L87 


\bibitem[Yamaoka et al.(2001)]{Yama01} Yamaoka, K., et al., \ 2001, \pasj, 53, 179 

\bibitem[van der Hooft et al.(1998)]{Vand98} van der Hooft, F.,  et al., \ 1998, \aap, 329, 538 


\bibitem[Verner \& Yakovlev(1995)]{Vern95} Verner, D.~A., \& Yakovlev, D.~G.\ 1995, \aaps, 109, 125 


\bibitem[Woods et al.(1996)]{Wood96} Woods, D.~T., Klein, R.~I., Castor, J.~I., McKee, C.~F., \& Bell, J.~B.\ 1996, \apj, 461, 767 

\end{thebibliography}
\end{document}